\documentclass{aa}  
\usepackage{graphicx}
\usepackage{subcaption}
\usepackage{multirow}
\usepackage{txfonts}
\usepackage{float}
\usepackage{tikz}
\usepackage{placeins}
\usepackage{threeparttable}
\usepackage{breqn}
\usepackage[hidelinks,colorlinks=true,linkcolor=blue,citecolor=blue]{hyperref}

\begin{document} 
   \title{Dynamical modelling and origin of gas turbulence in \textit{z} $\sim$ 4.5 galaxies}

   \author{F. Roman-Oliveira \inst{1} \and
          F. Rizzo \inst{2,3} \and
          F. Fraternali \inst{1}
          }

   \institute{Kapteyn Astronomical Institute, University of Groningen, Landleven 12, 9747 AD, Groningen, The Netherlands\\
              \email{romanoliveira@astro.rug.nl}
         \and
             Cosmic Dawn Center (DAWN), Jagtvej 128, DK2200, Copenhagen N, Denmark
            \and
            Niels Bohr Institute, University of Copenhagen, Lyngbyvej 2, DK-2100, Copenhagen Ø, Denmark
             }

   \date{}

% \abstract{}{}{}{}{} 
% 5 {} token are mandatory
  \abstract
   {In recent years, a growing number of regularly rotating galaxy discs have been found at z $\geq$ 4. Such systems provide us with the unique opportunity to study the properties of dark matter halos at these early epochs, the turbulence within the interstellar medium and the evolution of scaling relations.}
   {Here, we investigate the dynamics of four gas discs in galaxies at \textit{z} $\sim$ 4.5 observed with ALMA in the [CII] 158 $\mu$m fine-structure line. We aim to derive the structural properties of the gas, stars and dark matter halos of the galaxies and to study the mechanisms driving the turbulence in high-\textit{z} discs.}
   {We decompose the rotation curves into baryonic and dark matter components within the extent of the [CII] discs, i.e. 3 to 5 kpc. Furthermore, we use the gas velocity dispersion profiles as a diagnostic tool for the mechanisms driving the turbulence in the discs.}
   {We obtain total stellar, gas and dark matter masses in the ranges of $\log{(M/M_{\odot})} = 10.3 - 11.0$, $9.8 - 11.3$, and $11.2 - 13.3$, respectively. We find dynamical evidence in all four galaxies for the presence of compact stellar components, conceivably stellar bulges. The turbulence present in the galaxies appears to be primarily driven by stellar feedback, negating the necessity for large-scale gravitational instabilities.
   Finally, we investigate the position of our galaxies in the context of local scaling relations, in particular the stellar-to-halo mass and Tully-Fisher analogue relations.}
   {}

   \keywords{Galaxies: evolution -- Galaxies: high-redshift -- Galaxies: ISM -- Galaxies: kinematics and dynamics -- Submillimeter: galaxies
               }

   \maketitle

\section{Introduction}

%   We also find that these galaxies are in broad agreement with the expected stellar-to-halo mass relation for galaxies at this redshift and that they are situated similarly to local galaxies in the baryonic Tully Fisher relation, suggesting that they may have already fully formed their baryonic mass at \textit{z} $\sim$ 4.5.

%Observations of kinematically cold discs (ALMA) - gas, Discs JWST - stellar morphology

The advent of the Atacama Large Milimiter/submilimiter Array (ALMA) observatory has made possible to study the resolved kinematic properties of the gas in high-\textit{z} galaxies through the [CII] 158 $\mu$m line-emission, which is the brightest fine-structure line in the rest-frame far-infrared \citep{lagache18, casey14, carilli&walter13}.
Building upon the growing unprecedented data, many studies have reported the presence of disc galaxies at $z > 4$ with unexpectedly high levels of rotation support, with high ratios of rotation-to-random motion (V/$\sigma \sim 10$) similar to those of local galaxies \citep[e.g.][]{romanoliveira23, neeleman23, rizzo21, fraternali21, lelli21, tsukui21, rizzo20, neeleman20}.
These high V/$\sigma$ values indicate that early galaxy discs are kinematically colder than what it is typically predicted by cosmological simulations \citep[see detailed discussion in][]{romanoliveira23}.
The presence of such high-\textit{z} discs has been further supported by the discovery of both large fractions of disky morphologies in the rest-frame optical emission at $z > 4$ \citep{ferreira22a, ferreira22b, robertson23, kartaltepe23}, and barred spiral galaxies up to $z = 3$ \citep[e.g.][]{leconte23, costantin23}. The presence of bars at $z = 3$ indicates that dynamically settled disks are already present at a lookback time $\gtrsim$ 11 Gyr.

%Problem: cosmological simulations predict chaotic galaxies (accretion, mergers, feedback). Formation of discs, bulges, scaling relations, not understood
However, since galaxies are known to form hierarchically through the accretion of gas and mergers inside cold dark matter halos, there was a general expectation of a chaotic period in the early stages of matter build-up characterised by high levels of turbulence \citep{hopkins14, schaye15, vogelsberger20}.
The recent discoveries suggests instead a scenario in which disc galaxies can form earlier than previously expected and poses challenges to current models of galaxy formation as, in several simulations, disc galaxies at \textit{z} > 2 tend to be short-lived outliers \citep[e.g.][]{kretschmer22}.
Therefore, to better understand how galaxies can form persistent discs quickly and early in cosmic time, we need to gather more information on their dark matter halo content and the turbulence in their interstellar medium (ISM).
Fortunately, the regularity of these early gaseous discs provides the unique opportunity to obtain the gas rotation velocity that is linked to the total gravitational potential of the galaxy and consequently allows us to infer the dark matter halo properties \citep{cfn19} and the gas velocity dispersion that is linked to the local turbulence in the ISM and can be driven by different physical mechanisms \citep[e.g.][]{ejdetjarn22, yu21, bacchini20}.

%rotation velocity, mass decomposition and the TF
With an accurate measurement of the gas rotation velocities, we can derive the relative contribution of the different baryonic components, stars and gas, and of the dark matter \citep{lelli16b, mcgaugh04, vanalbada85, rubin80}.
There have been a few attempts at performing such dynamical modelling of galaxies at $z \sim 4$, which suggest that those galaxies already host stellar bulges, or at least a massive compact stellar component \citep{lelli21, rizzo20, tsukui21}.
Furthermore, the mass decomposition of rotation curves allows us to estimate the total stellar mass of the galaxies even without any observations of the stellar continuum, which are not available for most $z > 4$ galaxies.
With the complete information on the baryonic content of high-\textit{z} galaxies, we can study how they populate different scaling relations such as the stellar-to-halo mass relation (SHMR), which quantifies the efficiency of dark matter halo to transform baryonic matter into stars \citep{moster10, behroozi13}.
Moreover, we are able to study the evolution of one the tightest scaling relation in the study of galaxy evolution, the baryonic Tully-Fisher relation (BTFR) \citep{tullyfisher77, mcgaugh00}.
This is an empirical relation that correlates the baryonic mass and circular speed of galaxies over several orders of magnitude in mass \citep{lelli16, iorio17} and has been used as benchmark for galaxy formation models and cosmological simulations \citep[e.g.][]{governato07, posti19}.
At $z=1$ and $z=2$, much work has been done on the stellar Tully-Fisher relation \citep[e.g.][]{mcgaugh15}. However, no clear consensus has been reached regarding its evolution \citep{tiley16, diteodoro16, uebler17, harrison17}. Recently, \citet{fraternali21} investigated the location of two $z=4.5$ galaxies in the Early-Type-Galaxies (ETG) analogue version of the Tully-Fisher relation, which relates the rotational speed of inner CO discs of local ETGs with their stellar mass \citep{davis16}. They found that the two galaxies at $z=4.5$ lie nearly in the same relation as local ETGs if one assumes that all their available gas is turned into stars. 

%dispersion and turbulence in the ISM
The second key observable that we can extract through kinematic modelling is the gas velocity dispersion. Once all observational biases (instrumental spectral resolution and beam smearing) have been removed, we can considered the velocity dispersion as produced by thermal broadening and turbulence.

However, for [CII] observations at high-$z$ the thermal broadening can be considered negligible and even in HI observations of local galaxies, it is likely only important in the outer discs \citep[e.g.][]{tamburro09}.
Therefore, with an accurate measurement of the gas velocity dispersion we can attempt to study the origin of turbulence.
Turbulence plays a significant role in shaping the ISM of galaxies and influencing their evolution: it can arise from various factors such as mergers, gravitational instabilities, gas accretion, and feedback processes \citep{renaud14, tacchella16, kohandel20, orr20, nelson19}.
While it is not yet clear what mechanisms are responsible for the turbulence of the ISM of high-\textit{z}, stellar feedback and gravitational instabilities are usually considered the main drivers \citep{bournaud07, genzel11, krumholz16}.

Despite the importance of understanding the dynamical state of high-\textit{z} galaxies, there remains a paucity of detailed observations that allow for studies of their dark matter content and the drivers of turbulence. 
Although there have been many studies showing the presence of velocity gradients in high-\textit{z} galaxies \citep[e.g.][]{shao22, jones21, smit18, hodge12}, it is not clear how many of them are actually regularly rotating discs or perhaps poorly resolved disturbed systems, as the observations have varying levels of data quality. \citet{rizzo22} showed that both high angular resolution and high signal-to-noise ratio (SNR) are necessary to determine the dynamical state of galaxies (at least 6 resolution elements across the major axis). To date, the number of galaxies with such data quality at $z > 4$ is limited to a handful of cases.

In this paper, we focus on a sample of four disc galaxies at \textit{z} $\sim$ 4.5 with accurate [CII] kinematic measurements from \citet{romanoliveira23}. By modelling the mass contributions to the rotation curves, after correction for pressure support, we investigate the mass distribution of the dark matter, as well as the relative contributions of stars and gas to the total mass budget.
Additionally, we investigate the origin of the gas turbulence demonstrating that it can come from stellar feedback alone.
The paper is structured as follows: in Section~\ref{sec:data_methods}, we describe the data and methodology used in our analysis; in Section~\ref{sec:results}, we present our main results regarding the dynamical modelling of the rotation curves and the possible origins of the turbulence; in Section~\ref{sec:discussion}, we discuss the structure of the stellar mass component and where the galaxies fall into scaling relations within the context of the literature; in Section~\ref{sec:conclusions}, we summarise our main findings and implications. Throughout the paper, we adopt a $\Lambda$CDM cosmology from the 2018 Planck results \citep{planck20} and a Chabrier initial mass function \citep[IMF, ][]{chabrier03}.

\section{Data and methods}\label{sec:data_methods}
In this Section, we describe the data used and the methodologies we employ in the analysis throughout the paper.
We are analysing four galaxies at \textit{z} $\sim$ 4.5 with rotating gas discs to study the disc dynamics and turbulence.
The galaxies in this sample were selected due to the quality of the data available in the ALMA archive, in terms of SNR, angular and spectral resolution. This makes the sample heterogeneous in terms of physical properties: one of the galaxies is a quasar host (BRI1335-0417), two are in a group (SGP38326-1 and SGP38326-2) and the remaining galaxy was originally discovered as a damped Lyman alpha system \citep{wagg14, oteo16, neeleman19}.
To perform the analysis presented in the following Sections, in addition to the kinematic profiles from \citet{romanoliveira23}, we need estimates of the star formation rates and spatially resolved mass distributions of gas and stars. %In the following subsections, we describe in details the data we used for obtaining such quantities.

\subsection{Molecular gas mass}
The galaxies of our sample have previous estimates of the molecular gas masses \citep{wagg14, jones16, neeleman20, oteo16}. However, these were obtained from CO luminosities in different rotational transitions and assuming different $\alpha_{\mathrm{CO}}$ values \citep[see details in][]{romanoliveira23}. Ideally, the total molecular gas mass is best estimated from the CO (J=1-0), although at \textit{z} $\gtrsim 0.4$ this line is redshifted to frequency ranges not covered by current facilities. %z sim 0.37 out of band 3 of ALMA
In Table~\ref{tab:colum}, we report the CO luminosities for the lowest CO transitions available in the literature for the galaxies in our sample. We note that the measurement for SGP38326 is for the whole system and it is estimated that SGP38326-1 is 2.5 times more massive than SGP38326-2 \citep{oteo16}.
As described in Section~\ref{sec:dynmod}, we determine the total gas mass of the system independently from the gas emission through the mass decomposition of the rotation curves.
Hence, we use initial estimates of the gas mass in the Monte Carlo algorithm we use for the mass decomposition by scaling the values in Table~\ref{tab:colum} using a normalisation factor that accounts for conversions between CO luminosities and gas masses (details in Section~\ref{sec:dynmod}).

\begin{table}[H]
    \caption{CO luminosities of our sample. }
    \label{tab:colum}
    \centering
        \begin{tabular}{lll}
        \hline\hline \noalign{\vskip 1mm}
             \multirow{2}{*}{Object}   		& \multirow{2}{*}{CO transition}    & Luminosity	\\ \vspace{0.1cm}
                 & & 10$^{10}$ (K km s$^{-1}$ pc$^2$) \\  \hline \noalign{\vskip 1mm}
        BRI1335-0417  & J=2-1	& 10.9 $\pm$ 0.8 $^{a}$, 7.3 $\pm$ 0.6 $^{b}$ 	\\ \noalign{\vskip 1mm}
        %BRI1335-0417   & J=2-1	&       	\\ %wagg14
        J081740		& J=2-1	& 2.4 $\pm$ 0.7 $^{c}$        \\ \noalign{\vskip 1mm}
        SGP38326    & J=4-3	& 15.4 $\pm$ 1.6 $^{d, e}$	                 \\ \noalign{\vskip 1mm}
        \hline
        \end{tabular}
    \begin{tablenotes}
    \item \textbf{Notes.} References: $^{a}$ \citet{jones16}; $^{b}$ \citet{wagg14}; $^{c}$ \citet{neeleman20}; $^{d}$ \citet{oteo16}; $^{e}$ \citet{fudamoto17}.
    \end{tablenotes}
\end{table}

%SGP: from Fudamoto+17 following Solomon+97

\subsection{Star Formation Rates}\label{met:sfr}

The global star formation rate (SFR) of the galaxies J081740, SGP38326-1 and SGP38326-2 were derived from the total IR luminosity ($166 M_{\odot}$ yr$^{-1}$, $1830 M_{\odot}$ yr$^{-1}$, $882 M_{\odot}$ yr$^{-1}$, respectively), as reported by \citet{neeleman20} and \citet{oteo16}.
For BRI1335-0417, there have been studies deriving the SFR from fitting the spectral energy distribution (SED) and from the CO(J=7-6) luminosity, both resulting in a total SFR of $\sim$ 5000 $M_{\odot}$ yr$^{-1}$ \citep{lu18, wagg14}. However, more recently, \citet{tsukui23} rederived the SFR by taking into account the spatially resolved nuclear emission and thus excluding the contribution of the AGN to the dust heating, yielding a more accurate SFR of $1700^{+500}_{-400} M_{\odot}$ yr$^{-1}$.

\subsection{Dust and gas surface profiles}\label{sec:met_sb}
To perform the mass decomposition of the rotation curves, we need to consider the surface density profiles of the baryonic components in the galaxies.
Since there are no spatially-resolved data covering the rest-frame optical and near infrared emission, the stellar surface density profiles of the galaxies in our sample are unknown.
Therefore, for the stars we make the assumption that it should have an effective radius in between that of the dust and the gas, this will be discussed in further detail in Section~\ref{sec:dynmod}. For the gas, we assume that the [CII]-158 $\mu$m emission traces the total gas distribution of the galaxies and therefore the gas surface density is proportional to the [CII] surface brightness.

Considering that the galaxies in our sample have resolved ALMA data that comprise the dust continuum and the [CII]-158 $\mu$m line-emission, we can make use of the total map of the dust and the [CII] emission to derive surface brightness properties.
First, we separate the line-emission from the continuum and produce a dust-continuum only image and a [CII] line-emission only datacube, with the \texttt{mfs} and \texttt{cube} spectral definition modes in \texttt{CASA}, respectively \citep{mcmullin07}.
With the datacube, we obtain the [CII] total emission map by integrating the line emission along the velocity axis of the datacube.
For more details on the ALMA data reduction and imaging refer to \citet{romanoliveira23}.

We derive the surface brightness of the galaxies by fitting the [CII] and dust continuum maps with a two-dimensional S\'{e}rsic profile \citep{sersic68}, given by
\begin{equation}
    I(x,y) = I(R) = I_e \exp{\left\{ -b_n \left[ \left ( \frac{R}{R_{\mathrm{eff}}} \right) ^{(1/n)} -1 \right] \right\}},
\end{equation}
where $I_e$ is the surface brightness at the effective radius $R_{\mathrm{eff}}$, $n$ is the S\'{e}rsic index and $b_n$ is a constant defined as such that the profile contains half the total luminosity at $R_{\mathrm{eff}}$ and can be solved for numerically \citep{graham05}. We use the function \texttt{Sersic2D} as implemented in \texttt{Astropy} \citep{astropycollab13, astropycollab18, astropycollab22}, which takes into account the geometric parameters defining the 2D distributions, the ellipticity $\epsilon$, the position angle (PA) and the centre.
To account for the spatial resolution of the data, we convolve the model to the same resolution as the data. We then minimise the residuals to find the best-fit model using \texttt{DYNESTY}, a python package for estimating Bayesian posteriors and evidences using the Dynamic Nested Sampling algorithm \citep{speagle20, koposov22, skilling04, skilling06}.
In the fitting, we minimise the residuals using a chi-square likelihood, $(|\mathcal{D} - \mathcal{M}|)^2 / \sigma_{\mathrm{eff}}^2$. We note that, we weight the residuals by an effective uncertainty related to asymmetry in the galaxy \citep[see Section 2.5 in][]{marasco19}. This is because, given that our model is axi-symmetric, this provides a more realistic estimate of the parameter errors than only accounting for the background noise in the data. In this case, based on the methodology employed by \citet{marasco19}, we rotate the moment 0 map of the emission by 180$^\circ$ and subtract it from the original data. We then calculate the standard deviation of this residual map and use it as the $\sigma$ in the chi-square minimisation.
For both the gas and the dust we leave all the following parameters free: effective radius ($R_{\mathrm{eff}}$), surface brightness at the effective radius ($I_e$), galactic centre ($x_0$, $y_0$), ellipticity ($\epsilon$) and PA.
We use \texttt{DYNESTY} to compute the Bayesian posterior distribution of these parameters considering uniform priors.

As for the S\'ersic index, we perform a preliminary fit letting $n$ vary within 0.2 and 10.0. However, we notice that for the gas and dust emission of the galaxy J081740 and for the gas distribution of the SGP38326 system, the free \textit{n} fit does not converge, this is likely due to the data not having enough information to constrain the parameters. Moreover, for the free S\'ersic index fit of the dust emission of SGP38326-1 and SGP38326-2, we find an $n$ of 1.4 and 1.5, respectively. For the gas and dust emission of BRI1335-0417, we find $n = 1.5$ and $2.6$, respectively.
We conclude that the gas and dust emission of the galaxies are reasonably described with a S\'ersic index around 1, as for the majority of our tests that have converged we find an $n<2$.
Given that a S\'ersic index of 1 significantly simplifies our rotation curve decomposition (see Section~\ref{sec:dynmod}) we prefer to fix it to 1 in our fits for all galaxies, with the only exception of the dust emission of BRI for which we leave $n$ free.
We report our best-fit parameters in Section~\ref{sec:res_sb}; the maps are well reproduced with minimal residuals.

%Consequently, we conclude that the gas and the dust emission of the galaxies are compatible with an exponential profile which matches the assumption we use later in the mass decomposition of the rotation curves.

Finally, we assume the discs to be razor-thin such that the inclination is related to the ellipticity by $\cos^2{(i)} = 1 - \epsilon$. 
After deriving the [CII] surface brightness profiles as an exponential profile, we use the best-fit effective radius to obtain the gas disc scale-length as $R_{\rm gas} = R_{\rm eff, [CII]}/1.678$.

%This is a reasonable assumption given that the [CII] emission is a good tracer of the total H$_2$ content \citep{madden20}.
%We then compute the Bayes factor to decide which case is preferred.

\subsection{Gas kinematics}
The gas kinematic properties of these galaxies were derived using the [CII] 158 $\mu$m emission line, observed with ALMA. These observations have an average SNR of 3 - 5 per pixel and velocity channel, and spatial resolutions in the range of 0.13" - 0.24", which correspond to 6 - 10 independent radial elements across the major axis of the galaxies.
The kinematic modelling was performed with \texttt{$^{\text{3D}}$BAROLO} \citep{diteodoro15}, which fits the full emission line datacube considering an axisymmetric rotating disc model. For more details we refer to \citet{romanoliveira23}. Here we use the kinematic profiles obtained in our previous paper, namely the rotation curves to perform the dynamical modelling and the velocity dispersion profiles to study the turbulence in the discs.

\subsection{Circular speed of the gas}

For galaxy discs having a rotation support that largely dominate over the pressure support ($V/\sigma \gg 1$), the rotation velocity is essentially equal to the circular speed of a test particle that is moving in perfect circular obits under the influence of the galaxy gravitational potential,
\begin{equation}
    V_{\mathrm{c}}^2 = R \: \left( \frac{\partial \Phi}{\partial R} \right) \approx V_{\mathrm{rot}}^{2},
\end{equation}
where $\Phi$ is the gravitational potential, $V_{\mathrm{c}}$ is the circular speed and $V_{\mathrm{rot}}$ is the gas rotation velocity. For galaxies with $V_{\mathrm{rot}}$ more similar to $\sigma$, the pressure support should be considered.
Assuming that the velocity ellipsoid is isotropic and the disc thickness is constant with radius \citep{binney08, iorio17}, we can apply the asymmetric drift correction, $V_{\mathrm{A}}$, as follows
\begin{equation}\label{eq:adc}
    V_{\mathrm{c}}^2 = R \: \left( \frac{\partial \Phi}{\partial R} \right) = V_{\mathrm{rot}}^{2} + V_{\mathrm{A}}^{2} = V_{\mathrm{rot}}^{2} - R \sigma^2 \frac{\partial \ln (\sigma^2 \Sigma_{\rm gas})}{\partial R}.
\end{equation}
Therefore, to obtain $V_{\mathrm{A}}$, it is necessary to obtain the gas surface density $\Sigma_{\mathrm{gas}}$, the velocity dispersion $\sigma$ for each radial orbit $R$.
We apply this correction to the four galaxies, as described in Appendix~\ref{ap:adc}, but we observe that it is only significant for BRI1335-0417 given its low V/$\sigma$ compared to the other galaxies.

\subsection{Mass Decomposition}\label{sec:dynmod}

%we decompose the rotation curve into the different matter components: the stellar component, the gas component and the CDM halo component.
After obtaining the circular speed of the galaxies as a function of radius, we decompose it considering the contribution of stars, gas and DM as follows
\begin{equation}\label{eq:vc}
    V_{\mathrm{c}}(R) = \sqrt{-R \left(\frac{\partial \Phi_{\mathrm{star}}}{\partial R} + \frac{\partial \Phi_{\mathrm{gas}}}{\partial R} + \frac{\partial \Phi_{\mathrm{DM}}}{\partial R} \right) } = \sqrt{ V^2_{\mathrm{star}} + V^2_{\mathrm{gas}} + V^2_{\mathrm{DM}} },
\end{equation}
where $\Phi_{\rm i}$ is the gravitational potential contributed by the $i$-matter component (with $i$ star, gas and DM) and $V_{\mathrm{star}}$, $V_{\mathrm{gas}}$, $V_{\mathrm{DM}}$ are the contribution to the circular speed from the stars, gas and dark matter, respectively.
We note that we need to make a few major assumptions regarding the baryonic components. For the stars, we do not have any resolved information on the near-infrared stellar continuum of the galaxies and therefore no prior estimate of the stellar masses. For the gas, we do not have precise estimates of the total gas mass due to uncertainties in the CO line ratios and $\alpha_{\rm CO}$ conversion factor.
Below we describe how we model each of the three, while a summary of our assumptions is shown in Table~\ref{tab:priormm}.

\begin{enumerate}
    \item Stellar component: we assume that the projected light distribution follows a S\'ersic profile.
We use a density–potential relation for a deprojected spherical S\'ersic profile from \citet{terzic05} to describe the circular speed contributed by the stellar component as
\begin{equation}
    V_{\mathrm{star}}(R, M_{*}, R_{\mathrm{eff},*}, n) = \sqrt{\frac{G M_{*}}{R} \frac{\gamma(n(3-p)),b(R/R_{\mathrm{eff},*})^{1/n}}{\Gamma(n(3-p))}},
\end{equation}
where $M_{*}$ is the total stellar mass, $R_{\mathrm{eff},*}$ is the stellar effective radius and $n$ is the S\'ersic index. Moreover, the parameters $p$ and $b$ are a function of $n$, and $\gamma$ and $\Gamma$ are the incomplete and complete gamma functions, respectively. 
We let the log of the stellar mass vary uniformly between 9 and 12. The S\'ersic index is free to vary uniformly between 0.2 and 10. Lastly, the effective radius is free to vary uniformly between the effective radius of the dust and that of the [CII] emission. This is a reasonable assumption as different studies showed that the stellar continuum emission tends to be more extended than the dust continuum emission at high-\textit{z} \citep{gillman23, colina23, tadaki20b, kaasinen20}.
We note that some simulations suggest that the stellar continuum of high-\textit{z} galaxies may be up to four times less extended than the dust emission \citep[e.g.][]{popping22, cochrane19}. As a sanity check, we ran some tests with a larger prior in the stellar radius, covering a range between 0.2 times the effective radius of the dust and twice that of the [CII] emission. We find that it does not impact the estimates of the other parameters significantly nor does it lead the effective radius parameter to converge to a preferred value.
We highlight that our sample have not (yet) been observed with the James Webb Space Telescope (JWST) and that the reddest band available of the Hubble Space Telescope (HST) only covers the rest-frame far-UV emission of these galaxies. Given that these galaxies are also very dusty, we have no significant prior knowledge on the stellar component of our sample.

\item Gas component: we assume this component to be a razor-thin disc following an exponential disc profile so that the circular speed contributed by the gas is 
\begin{equation}
    V_{\mathrm{gas}}(R, M_{\mathrm{gas}}, R_{\mathrm{gas}}) = \sqrt{\frac{ 2 G M_{\mathrm{gas}} y^2}{R_{\mathrm{gas}}} (I_{0}(y)K_{0}(y) - I_{1}(y)K_{1}(y) ) },
\end{equation}
where $R_{\rm gas}$ is the scale length of the disc, $M_{\rm gas}$ is the total gas mass, $I_{0}, I_{1}, K_{0}, K_{1}$ are the modified Bessel functions and y = $R/(2 R_{\mathrm{gas}})$ \citep{binney08}.
We assume that the gas disc has the same distribution of the [CII] emission and therefore the same disc scale length $R_{\mathrm{gas}}$.
For the mass, we assume that the total gas mass is given by the CO luminosities reported in Table~\ref{tab:colum} times a gas mass normalisation ($M_{\mathrm{gas}}$ = $L_{\mathrm{CO}}$ $\times$ gas\_norm).
%We note that this gas normalisation affects the $\Sigma_0$ of the exponential profile of the gas surface density described in Section~\ref{sec:met_sb}.
This normalisation encompasses the conversion factor between the CO luminosity and the molecular gas mass ($\alpha_{\mathrm{CO}}$), as well as the line ratio between the CO(J=1-0) and the CO transition observed (r$_l$).
We use a uniform prior ranging from 0 to 10, where the former represents the complete lack of gas in the galaxy and 10 is a sufficiently high number to cover the range of expected $\alpha_{\mathrm{CO}}$ \citep{dunne22, gong20, papadopoulos18, bolatto13, narayanan11} and r$_l$ \citep{boogaard20, yang17, kamenetzky16, dacunha13}. We highlight the fact that both of these numbers are highly uncertain even in the local Universe, therefore we choose a conservative uniform prior that can flexibly cover many values of r$_l$ and $\alpha_{\mathrm{CO}}$. The $\alpha_{\mathrm{CO}}$ factor is usually taken to be either 0.8 for starburst galaxies or 4.0 for normal star-forming galaxies, however these values are highly sensitive to metallicity, gas surface density other ISM conditions \citep{dunne22, gong20, papadopoulos18, bolatto13, narayanan11}. Similarly, the CO line ratios are uncertain, especially at high-\textit{z}, as the CO spectral-line energy distributions (SLEDs) vary within a wide range of values and can greatly differ among different galaxies \citep{boogaard20, yang17, kamenetzky16, dacunha13}.

\item Dark matter halo component: we assume this component to be a spherical NFW halo profile \citep{navarro97}, whose circular speed is given by
\begin{equation}
\begin{split}
V_{\mathrm{DM}}(R, R_{200}, M_{200}, c_{200}) = & \\
= \sqrt{\frac{G M_{200}}{R_{200}} \frac{\ln{(1+ c_{200} x)} - c_{200} x / (1 + c_{200} x)}{x[\ln{(1+c_{200})}- c_{200} / (1+c_{200})]}},
\end{split}
\end{equation}
where $c_{200}$ is the concentration of the NFW dark matter halo, $M_{200}$ and $R_{200}$ are the dark matter halo virial mass and radius, respectively, and $x = R/ R_{200}$.
We assume a concentration ($c_{200}$) fixed at 3.4, which is the typical value of concentration for a wide range of CDM halo masses at \textit{z} $\sim$ 4.5 \citep{dutton14, correa15, child18}.
The virial CDM halo mass ($M_{200}$) is incorporated in the fit through the baryon fraction ($f_{\mathrm{bar}}$), which is the baryonic (star + gas) mass divided by the virial mass of the dark matter halo. We let $f_{\mathrm{bar}}$ vary between the cosmological baryon fraction of 0.187 \citep{planck20} and a baryon fraction of 10$^{-6}$, as an extreme limit of a highly DM dominated galaxy.
If we take the literature estimates for the gas mass of the galaxies in our sample as the total baryonic mass ($\sim 10^{10} M_{\odot}$), then the limits in $f_{\rm bar}$ would result in a CDM halo mass in the range of $\sim 5 \times 10^{10} - 10^{16} M_{\odot}$. We let the baryon fraction, $f_{\mathrm{bar}}$, vary uniformly in a log space.
\end{enumerate}

To obtain the best-fit model of the mass decomposition, we use Equation~\ref{eq:vc} to reproduce the observed circular speed of the galaxy using \texttt{DYNESTY}.
We compute Bayesian posterior distribution of the following free parameters: stellar mass ($M_*$), stellar effective radius ($R_{\mathrm{eff,*}}$), stellar S\'ersic index ($n$), gas mass normalisation (gas\_norm), baryon fraction ($f_{\mathrm{bar}}$).
These priors and assumptions were used to obtain the fiducial models discussed in Section~\ref{sec:res_dy}. We have also tested several different models considering different assumptions and priors, which are not used for the fiducial models although we discuss the implications of these different tests in Section~\ref{sec:res_dy}.

%The main setback in our modelling is that we need to fit the components taking into account these 5 free parameters while being limited by the spatial resolution of 3 to 5 independent resolution elements over the radial profile.

\begin{table}[H]
    \caption{Priors used in the rotation curve decomposition.}
    \label{tab:priormm}
    \centering
        \begin{tabular}{c l l l}
        \hline\hline \noalign{\vskip 1mm}
        Component & Parameter & Prior Type & Value Range  \\ \noalign{\vskip 1mm}
        \hline

\noalign{\vskip 1mm}
        \multirow{3}{*}{Stars} & $\log{(M_*/M_{\odot})}$              & Uniform log   & 9 -- 12  \\
        & $R_{\mathrm{eff},*}$ (kpc)    & Uniform    & $R_{\mathrm{eff}, \mathrm{dust}}$ -- $R_{\mathrm{eff}, \mathrm{[CII]}}$  \\ %from SB fits
        & $n$                       & Uniform    & 0.2 -- 10    \\

\noalign{\vskip 1mm}
        \multirow{3}{*}{Gas}  & gas\_norm   & Uniform   & 0 -- 10  \\
        & $R_{\mathrm{gas}}$ (kpc) & Fixed & $R_{\mathrm{[CII]}}$ \\

\noalign{\vskip 1mm}
        \multirow{2}{*}{Dark Matter} & $\log{(f_{\mathrm{bar}})}$      & Uniform log & -6 -- -0.73    \\ %fbar
        & $c_{200}$               & Fixed      & 3.4           \\ \noalign{\vskip 1mm}
        \hline
        \end{tabular}
    \begin{tablenotes}
    \item {Notes.} The stellar component is assumed to be spherical, with a corresponding projected light distribution described by a S\'{e}rsic profile, defined by a stellar mass ($M_*$), an effective radius ($R_{\mathrm{eff},*}$) and a S\'ersic index ($n$). The gas is described by an exponential disc defined by a disc scale length ($R_{\mathrm{gas}}$) fixed to the scale length of the [CII] emission and a gas mass ($M_{\mathrm{gas}}$) that is defined as as the CO luminosity times a gas normalisation (gas\_norm). The CDM halo follows an NFW profile with a concentration fixed at 3.4. The CDM halo virial mass is calculated from the baryon fraction which varies between the cosmological baryon fraction and a baryon-dominated halo (see main text for further explanation).
    \end{tablenotes}
\end{table}

\subsection{Supernova explosion driven turbulence}\label{met:snii}

To study whether the stellar feedback can be considered as a valid candidate for explaining the measured turbulence, we use a simple model in which we assume that the turbulence of the gas is fed by the energy injected in the ISM by supernova explosions \citep[for more information see][]{bacchini20}. Here, we neglect the contribution from supernova type Ia (SNIa) as high-\textit{z} galaxies tend to be dominated by supernova type II due to their high star formation rate and young ages \citep[SNII,][]{rodney14}. Therefore, the energy input from supernovae is given by
$\dot{E}_{\mathrm{SN}} = {\eta}  \: E_{\mathrm{SNII}}  \:  \xi$,
where $\eta$ is the SNII efficiency defined as the percentage of the total supernova energy that is injected in the ISM as kinetic energy; $E_{\mathrm{SNII}}$ is the total energy per SNII explosion and it is equal to $10^{51}$ erg; and $\xi$ is the expected number rate of SNII given the SFR of each galaxy.
We note that hydrodynamical simulations of supernova remnant evolution tend to find values of $\eta$ around $\sim~0.1$ or less \citep{fierlinger16, ohlin19}. Assuming a Chabrier IMF, the rate of SNII is given by $\xi = 0.01 \: M_{\odot}^{-1} \: \times \mathrm{SFR} $ \citep{mo10}. %page
The turbulence due to the supernova energy input is dissipated with a timescale $\tau = h / \sigma$, where h is the disc scale height  \citep{maclow99}. Therefore, the total supernova energy input can be rewritten as
\begin{equation}\label{eq:snii}
     E_{\mathrm{SN}} = {\eta}  \: E_{\mathrm{SNII}}  \:  \xi \: h \: \sigma^{-1}.
\end{equation}
If one assumes $\eta$ to be 1, then Equation~\ref{eq:snii} represents the maximum injectable energy from SNII into the ISM.
Then, we can compare this to the observed turbulent energy in the ISM given the measured gas velocity dispersion in a gas disc, given by
\begin{equation}\label{eq:turb}
E_{\mathrm{turb}} = 3/2 \: M_{\mathrm{gas}} \: \sigma^2,    
\end{equation}
where $M_{\mathrm{gas}}$ is the total molecular gas mass of the disc and $\sigma$ is the line-of-sight observed velocity dispersion of the gas that we assume to be isotropic.

%Lastly, equating $E_{\mathrm{turb}}$ to $E_{\mathrm{SN}}$ we can rewrite this expression in terms of the velocity dispersion
%\begin{equation}\label{eq:sigma_sn}
%	\sigma = \left( \: \frac{2}{3} \: \eta \: E_{\mathrm{SNII}} \: \xi \: h \: M_{\mathrm{gas}}^{-1} \: \right)^{1/3}
%\end{equation}
%Using Equation~\ref{eq:sigma_sn} we can estimate the gas velocity dispersion expected from a turbulent ISM driven solely by supernova feedback and then compare to the observed values reported in \citet{romanoliveira23}.

\section{Results}\label{sec:results}
In this Section, we present the main findings of our analysis, namely the surface brightness of the dust and [CII] 158 $\mu$m emission, the mass decomposition, the stellar masses and star formation rates of the galaxies.

\subsection{Surface Brightness}\label{sec:res_sb}

\begin{figure*}
    \centering
    \includegraphics[width=\textwidth]{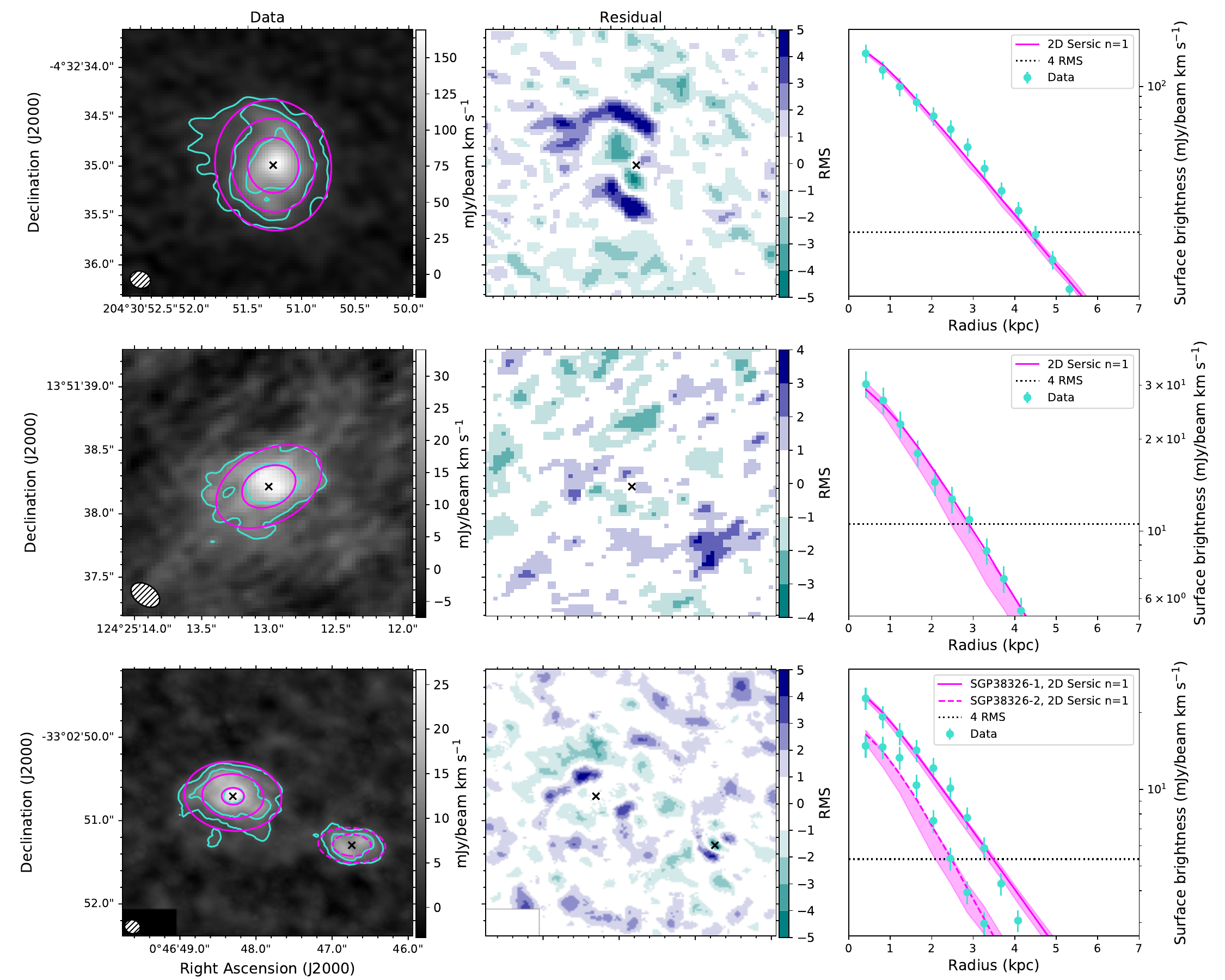}
    \caption{Surface brightness profiles of the gas distribution of the galaxies BRI1335-0417, J081740 and the SGP38326 system, respectively from top to bottom. Left panels: surface brightness map of the [CII] emission. The emission is shown as contours in turquoise for the data and magenta for the model. The contour levels follow levels of emission of 4$\sigma$, 8$\sigma$ and 16$\sigma$, where $\sigma$ is the RMS noise in the total map. We show the beam of the observations in the bottom left. Middle panel: residual map normalised by the RMS noise according to the colourbar shown. We show the centre of the best-fit model of each galaxy with a small cross. Right panel: 1D profile of the average gas emission in concentric ellipses (turquoise) and the best-fit S\'ersic model (magenta), the dotted line shows the 4$\sigma$ level above the noise. In the bottom panel we show the galaxies SGP38326-1 and SGP38326-2 together, as they have to be fitted simultaneously, and we mask the emission of SGP38326-3 in the bottom left corner.}
    \label{fig:sb_gas}
\end{figure*}

\begin{figure*}
    \centering
    \includegraphics[width=\textwidth]{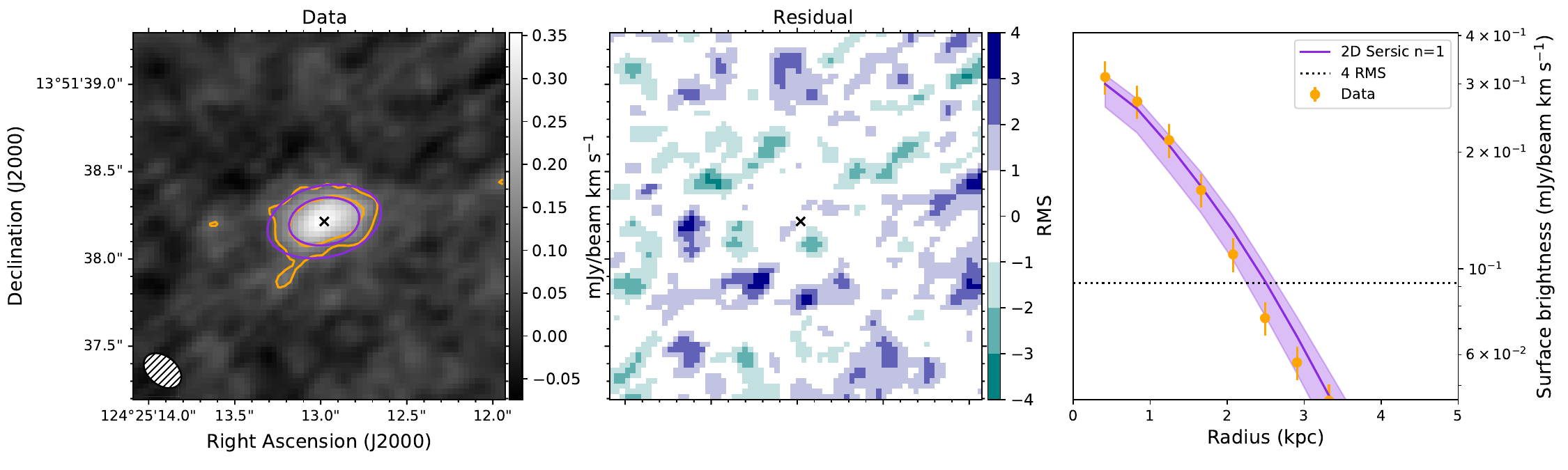}
    \caption{Surface brightness profile of the dust distribution of the galaxy J081740, the other galaxies are shown in Figure~\ref{fig:a_dustsb}. Left panel: surface brightness map of the dust continuum. The emission is shown as contours in orange for the data and in violet for the best-fit model. The contour levels follow levels of emission of 4$\sigma$, 8$\sigma$ and 16$\sigma$, where $\sigma$ is the RMS noise in the total map. We show the beam of the observations in the bottom left. Middle panel: residual map normalised by the RMS noise according to the colourbar shown. We show the centre of the best-fit model of the galaxy with a small cross. Right panel: 1D profile of the average gas emission in concentric ellipses (orange) and the best-fit S\'ersic model (violet), the black dotted line shows the 4$\sigma$ level above the noise.}
    \label{fig:sb_dust}
\end{figure*}

Using the method described in Section~\ref{sec:data_methods}, we found the best-fit parameters defining the surface brightness profiles for the [CII] and dust emission. In Figure~\ref{fig:sb_gas}, we show the comparison between [CII] data and model in 2D and 1D for the 4 galaxies of our sample. The 1D surface brightness is obtained after averaging the emission along rings with a separation of one third of the beam of the observations; these profiles are not deprojected.
Figure~\ref{fig:sb_dust} shows the surface brightness of the dust emission for J081740, while equivalent visualisation for the rest of sample is shown in Figure~\ref{fig:a_dustsb}.
We note that for the SGP38326 group we had to fit the emission of the two main galaxies simultaneously in order to avoid biases due to the diffuse emission of each galaxy influencing the fit of the other. In this case, we also masked the emission of the small companion (SGP38326-3).
In Appendix~\ref{ap:corner_sb}, we show the corner plots of the best-fits shown in Figure~\ref{fig:sb_gas} and~\ref{fig:sb_dust}. 
We summarise the properties of the best-fit models of the surface brightness in Table~\ref{tab:surfbrightness}. We note that, as discussed in Section~\ref{sec:met_sb}, for both the dust and gas we fit a 2D S\'ersic profile with $n$ fixed to 1 (equivalent to an exponential disc), given that the free S\'ersic index model does not significantly improves the fits. The only exception is for the dust emission of BRI1335-0417, which we keep free as it is significantly better described by a S\'ersic index of $n = 2.6 \pm 0.1$.
We note that for some of the galaxies, the surface brightness in the outer regions tends to be slightly overestimated by the model. This is due to the fact that the emission of the data in the outer parts is less axisymmetric than the model and that the residuals are minimised in a way that gives more importance to the bright central regions of the galaxies. Given that we only use the effective radii of these fits in the following analysis, we conclude that our best-fit models are sufficiently good representations of the shape and extent of the dust and gas emission.

\begin{table*}
    \caption{Gas and dust continuum surface brightness.}\label{tab:surfbrightness}
    \centering
        \begin{tabular}{llccccc}
        \hline \hline \noalign{\vskip 1mm}
\multirow{2}{*}{Object} &	& $I_{\mathrm{e, gas}}$	&	$I_{\mathrm{e, dust}}$ & $R_{\mathrm{eff}}$	&  $\epsilon$	&  PA	\\
&	& (mJy/beam km/s) &	(mJy/beam) & (kpc)			& 			& (deg)	\\ \noalign{\vskip 1mm} \hline
\noalign{\vskip 1mm}
\multirow{2}{*}{BRI1335-0417} & gas & 44.9 $^{+0.7}_{-0.6}$ & - & 3.06 $^{+0.03}_{-0.04}$ & 0.20 $^{+0.01}_{-0.02}$	& 6 $^{+2}_{-3}$ \\

\noalign{\vskip 1mm}
& dust	& - & 0.97 $\pm 0.07$ & 1.69 $^{+0.08}_{-0.06}$ 	& 0.22 $\pm$ 0.02 & -2 $^{+3}_{-2}$ \\

\noalign{\vskip 1mm}
\multirow{2}{*}{J081740}	& gas	& 9.9 $^{+0.5}_{-0.4}$ & -	& 3.2 $^{+0.2}_{-0.1}$  & 0.42 $\pm 0.03$	& 125 $\pm$ 3 \\

\noalign{\vskip 1mm}
& dust	& - & 0.17 $\pm 0.01$ & 1.79 $^{+0.09}_{-0.10}$ & 0.56 $\pm 0.04$ & 109 $^{+3}_{-2}$	\\

\noalign{\vskip 1mm}
\multirow{2}{*}{SGP38326-1}	& gas	& 6.8 $\pm 0.2$ &  - &	3.2 $\pm 0.1$  & 0.24 $\pm$ 0.02 & 83 $\pm$ 3 \\

\noalign{\vskip 1mm}
& dust	& - & 0.75 $\pm 0.01$ & 1.29 $\pm 0.01$	&  0.04 $\pm$ 0.01		& -49 $\pm$ 8	\\

\noalign{\vskip 1mm}
\multirow{2}{*}{SGP38326-2}	& gas	& 6.3 $\pm 0.3$	& 	- &	2.4 $\pm 0.1$	 & 0.50 $\pm 0.03$	& 85 $\pm 2$	\\

\noalign{\vskip 1mm}
& dust	& - & 0.57 $\pm 0.01$  & 1.08 $\pm 0.02$	& 0.25 $\pm$ 0.02		& 112 $\pm$ 3 \\ \noalign{\vskip 1mm}
\hline
        \end{tabular}
        
    \begin{tablenotes}
    \item \textbf{Notes.} Best-fit parameters of the 2D S\'ersic fits of the gas and dust distributions: surface brightness at the effective radius ($I_{\mathrm{e}}$); effective radius ($R_{\mathrm{eff}}$); ellipticity ($\epsilon$); and position angle (PA). All maps were fitted with a S\'ersic index ($n$) fixed to 1, except for the dust emission of BRI1335-0417 where $n$ was left free and led to $n = 2.6 \pm 0.1$.
    \end{tablenotes}
\end{table*}

\subsection{Mass decomposition of the rotation curves}\label{sec:res_dy}

Here, we present the fiducial dynamical models of the galaxies in our sample. Following the methods described in Section~\ref{sec:dynmod} we decompose the rotation curves of the galaxies into three mass components: stars, gas and a CDM halo.
The procedure employed for the dynamical modelling of the galaxies J081740, SGP38326-1 and SGP38326-2 is the same: a single spherical S\'ersic profile for the stars with free $n$, an exponential profile considering a razor-thin disc for the gas following the [CII] surface brightness and a spherical NFW CDM halo with the priors described in Table~\ref{tab:priormm}. However, for the galaxy BRI1335-0417, a different approach was taken, wherein the baryon fraction was fixed to the cosmological baryon fraction. This is due to the fact that the literature estimates for the gas mass of BRI1335-0417 are too high for its rotation curve. Therefore, as our fiducial model we assume a scenario that maximises the gas mass while still containing a marginally realistic CDM halo which provides an upper limit on the estimate of the total baryonic mass. We discuss this further in Appendix~\ref{ap:dynmod}.

In Figure~\ref{fig:dy_rotcur}, we show the rotation curves derived from the [CII] kinematics. These rotation curves are corrected for pressure support following Equation~\ref{eq:adc}, so that we are plotting the circular speeds (further description of this is provided in Appendix~\ref{ap:adc}). Similarly to local galaxies, the shapes of the rotation curves of $z \sim 4$ galaxies are diverse: BRI1335-0417 has a bump in the inner region that resembles the dynamical signature of a bulge; the rotation curves of J081470 and SGP38326-1 seem to flatten at large radii; and the rotation curve of SGP38326-2 has a shallow gradient. 
In Figure~\ref{fig:dy_rotcur}, we overlap the contributions to the circular speeds by the stellar component, the gas disc and the CDM halo, as well as the total circular speed of our model (in black) that matches the observed rotation curve (in blue).
In Table~\ref{tab:dy_fid}, we display the best-fit parameters that define the different components shown in Figure~\ref{fig:dy_rotcur}. Additionally, in Appendix~\ref{ap:dynmod} we show the posterior distribution of the best-fit parameters and discuss some extra tests to probe the limitations of our data.

\begin{figure*}
    \centering
    \includegraphics[width=0.98\textwidth]{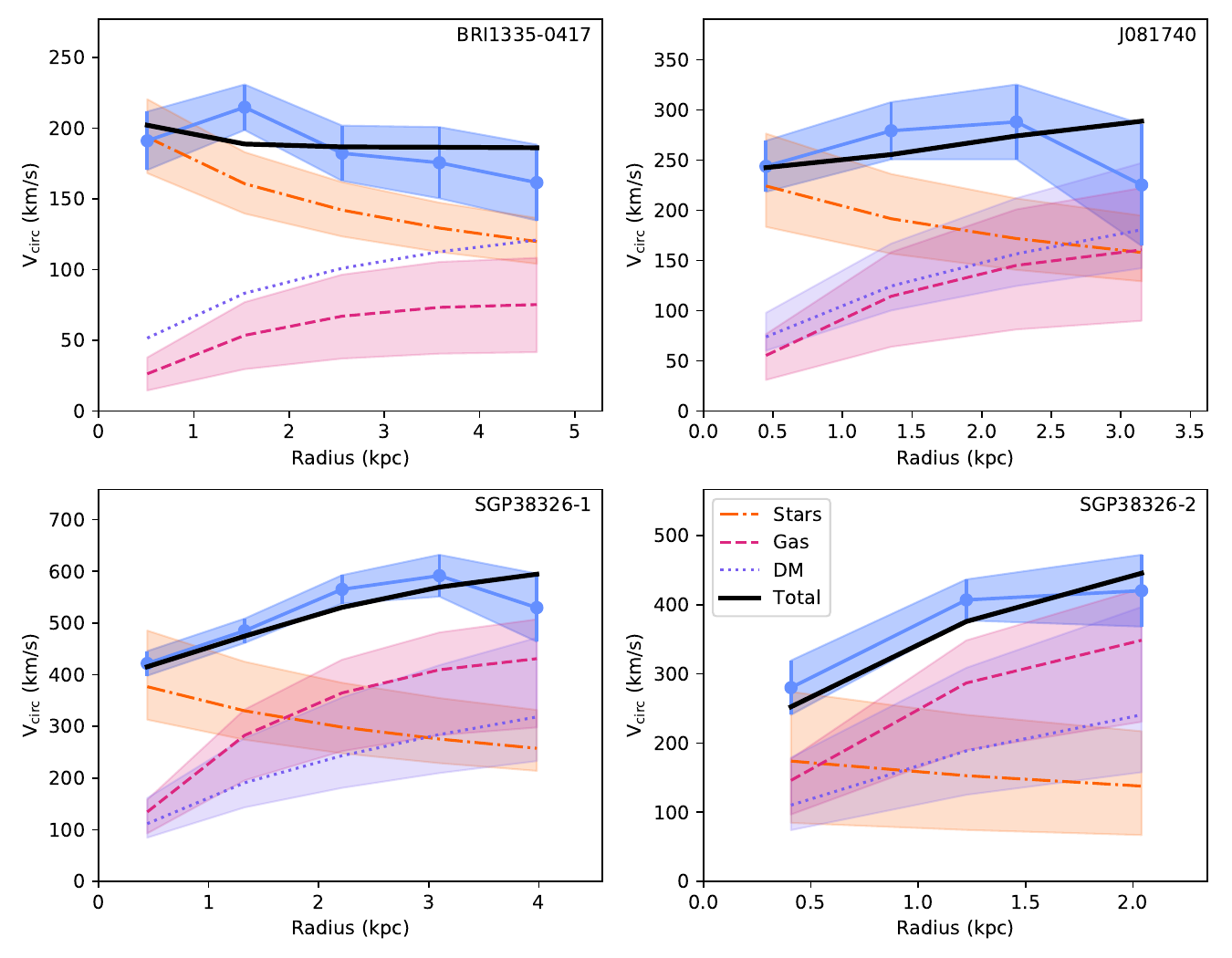}
    \caption{Fiducial dynamical model for the galaxies. We show the measured circular speed, i.e. the rotation curve corrected for pressure support, as the blue shaded region. We also show the circular speed predicted by our model in black. The circular speed of the different mass components in the galaxies are shown as different shaded regions: stellar component (dash-dot line, orange), gas disc (dashed line, pink) and NFW CDM halo (dotted line, blue). In the case of BRI1335-0417, we fix the CDM halo to the cosmological baryon fraction. As explained in the main text, due to difficulties in breaking the degeneracy of the gas and DM contribution to the rotation curve we keep the CDM halo fixed to provide an upper limit on the baryon components.}
    \label{fig:dy_rotcur}
\end{figure*}

\begin{table*}
\caption{Results of the fiducial mass decomposition models.} \label{tab:dy_fid}
\centering
\begin{tabular}{lccccccc}
\hline \hline \noalign{\vskip 1mm}
Object & $\log{(M_*/M_{\odot})}$ & $R_{\mathrm{eff, *}}$ (kpc) & $n$ & gas\_norm & $\log{(f_{\rm bar})}$ & $\log{(M_{\rm gas}/M_{\odot})}$ & $\log{(M_{200}/M_{\odot})}$ \\ \noalign{\vskip 1mm} \hline \noalign{\vskip 1mm}
BRI1335-0417 & 10.4 $^{+0.1}_{-0.1}$ & 2.3 $^{+0.5}_{-0.4}$ & 6.9 $^{+2.1}_{-2.1} $ & 0.06 $^{+0.06}_{-0.04}$ & \textit{-0.728} & 9.8 $^{+0.3}_{-0.5}$ & \textit{11.2} \\
\noalign{\vskip 1.5mm}

J081740      & 10.6 $^{+0.2}_{-0.2}$  & 2.4 $^{+0.5}_{-0.5}$ & 6.7 $^{+2.3}_{-2.4}$ & 1.4 $^{+1.2}_{-0.9}$  & -1.5 $^{+0.5}_{-0.7}$ & 10.5 $^{+0.3}_{-0.5}$ & 12.3 $^{+0.7}_{-0.5}$  \\ \noalign{\vskip 1.5mm}

SGP38326-1   & 11.0 $^{+0.2}_{-0.2}$ & 2.2 $^{+0.7}_{-0.6}$ & 6.3 $^{+2.5}_{-2.4}$ & 1.9 $^{+0.6}_{-0.9}$ & -1.7 $^{+0.7}_{-0.9}$ &  11.3 $^{+0.1}_{-0.3}$ &  13.2 $^{+0.9}_{-0.7}$  \\ \noalign{\vskip 1.5mm}

SGP38326-2   & 10.3 $^{+0.4}_{-0.6}$ & 1.7 $^{+0.4}_{-0.4}$ & 5.3 $^{+3.1}_{-2.9}$  & 2.5 $^{+1.1}_{-1.4}$ & -2.2 $^{+1.0}_{-1.3}$  & 11.1 $^{+0.2}_{-0.3}$ & 13.3 $^{+1.3}_{-1.0}$  \\ \noalign{\vskip 1mm} \hline \noalign{\vskip 1mm}
\end{tabular}
\begin{tablenotes}
\item \textbf{Notes.} Best-fit parameters of our dynamical model: stellar mass ($M_{*}$); stellar effective radius ($R_{\mathrm{eff,*}}$); stellar S\'ersic index ($n$); normalisation of the gas mass (gas\_norm); baryon fraction ($f_{\mathrm{bar}}$). We also show the total gas mass ($M_{\mathrm{gas}}$) derived as gas$\_$norm $\times$ L$_{\rm CO}$ (see Table~\ref{tab:colum}) and the CDM halo mass ($M_{200}$) derived as $M_{\rm bar} (1 - f_{\rm bar}) / f_{\rm bar}$. For BRI1335-0417, we report the values of $f_{\rm bar}$ and $M_{200}$ in italic as they are kept fixed in our fiducial model.
\end{tablenotes}
\end{table*}

\begin{table}
\caption{Other properties derived from the results of the mass decomposition.} \label{tab:dy_more}
\centering
\begin{tabular}{lccc}
\hline \hline \noalign{\vskip 1mm}
Object & $f_{\mathrm{gas}}$ & $R_{\rm eff, bar}$ (kpc) & $f_{\rm disc}$ \\ \noalign{\vskip 1mm} \hline \noalign{\vskip 1mm}
BRI1335-0417 &  0.16 $\pm$ 0.12  & 2.9 & 0.69 \\ \noalign{\vskip 1.5mm}
J081740      &  0.52 $\pm$ 0.24  & 2.8 & 0.63 \\ \noalign{\vskip 1.5mm}
SGP38326-1   &  0.68 $\pm$ 0.17  & 2.8 & 0.78 \\ \noalign{\vskip 1.5mm}
SGP38326-2   &  0.87 $\pm$ 0.22  & 2.0 & 0.72 \\ \noalign{\vskip 1mm} \hline 
\end{tabular}
\begin{tablenotes}
\item \textbf{Notes.} Parameters derived from the results of the mass decomposition: the gas fraction ($f_{\mathrm{gas}}$); the effective baryonic radius as the radius that comprises half of the baryonic mass ($R_{\rm eff, bar}$); and the disc fraction, as the fraction of baryons to dark matter inside $R_{\rm eff, bar}$ ($f_{\rm disc}$).
\end{tablenotes}
\end{table}

%explain the need for assumptions: free parameters

We find that our rotation curve decomposition can constrain the stellar mass for the four galaxies reasonably well, placing our sample in a considerably massive regime, with stellar masses in between $\log(M_{*}/M_{\odot}) = 10.3 - 11.0$. 
The accuracy of these estimates is due to the fact that the stellar component dominates the central region of the rotation curve, as a consequence, the inner shape of the rotation curve can only be reproduced by a massive stellar component. We find that this is only possible with stellar profiles with a high S\'ersic index, indicating that a massive compact stellar distribution, such as a bulge, is present in all of the four galaxies. We note that the effective radius of such a component cannot be constrained with precision with the current data, although, we use physically motivated priors that provide plausible limits. Moreover, in spite of all the assumptions involved, we obtain effective radii for the stellar component compatible with values recently found with JWST observations of $z > 2$ galaxies \citep{ormerod24, tadaki23}.

The gas mass is constrained for J081740, SGP38326-1 and SGP38326-2, while for BRI1335-0417 we find an upper limit that is driven by the imposition of the cosmological baryon fraction. In general, the values of the gas masses have larger uncertainties than those for the stellar masses. This is because at small radii the stellar component is dominant, while at larger radii the main contribution to the rotation curve comes from both the gas and the CDM halo.
Given the low concentration of DM at \textit{z} $\sim$ 4 and that we are still only probing the relatively inner regions of these galaxies, the contribution of the CDM halo is hard to discern from the contribution of the gas. 
For example, considering a concentration of 3.4, an NFW halo with a virial mass of $10^{12} M_{\odot}$ produces a circular speed of about only 90 km s$^{-1}$ lower than a halo with a virial mass of $10^{13} M_{\odot}$ at 3 kpc from the galactic centre.
Combining this with the large uncertainties regarding the conversion between the CO luminosity and the gas mass then it becomes difficult to disentangle and constrain both the gas mass and the CDM halo mass with the current data.
%that it is virtually impossible to constrain both the gas mass and the CDM halo mass with the current data.
However, we can clearly see a pattern regarding the DM content of these galaxies in which they all prefer baryon dominated models as seen by the posterior distributions that has a maximum close to the lower edge of the prior (see Figure~\ref{fig:a_dy}).
Finally, we find that the gas mass of the galaxies range between $\log{(M_{\rm gas}/M_{\odot})} = 9.8 - 11.3$ and we find CDM halos masses ranging between $\log{(M_{200}/M_{\odot})}  = 11.2 - 13.3$ with baryon fractions of 0.006 up to the cosmological baryon fraction of 0.187.

In Table~\ref{tab:dy_more}, we report other relevant properties that can be derived by the direct results of the mass decomposition: we show the gas fractions, defined as $f_{\mathrm{gas}} = M_{\mathrm{gas}}/M_{\mathrm{bar}}$; the baryonic effective radius, defined as the radius that comprises half of the baryonic mass ($R_{\rm eff, bar}$); and the disc fraction, defined as the fraction of baryons to dark matter inside the baryonic effective radius ($f_{\rm disc}$). Recently, \citet{blandhawthorn23} showed that these parameters can be used to predict the capability of a galaxy to form a stable bar. The four galaxies in our sample are above the thresholds set in \citet{blandhawthorn23}, indicating that all of these discs have high enough baryon concentrations that could enable the quick formation of a stable bar. 
To further support this prediction, \citet{tsukui24} performed a thorough analysis of the [CII] and dust emissions in BRI1335-0417 suggesting the presence of a strong bar.
Moreover, in general, our sample shows a diverse range of gas fractions, compatible to those found at $z > 1$ \citep{tacconi20, dessauges20, aravena16, decarli16}.

\subsection{Star Formation Rates and Stellar Masses}\label{sec:sfr}

\begin{figure}
    \centering
    \includegraphics[width=0.48\textwidth]{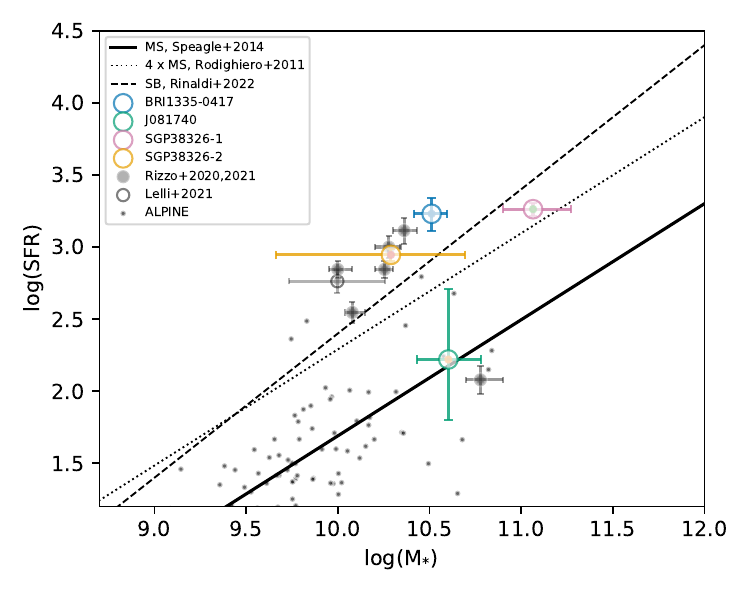}
    \caption{Stellar mass vs. star formation rate. Our galaxies are represented by coloured circles according to the legend. We also show some other galaxies in the literature at similar redshift \citep{rizzo20, faisst20, rizzo21, lelli21}. The black line shows the main-sequence relation at $z  = 4.5$ from \citet{speagle14}, while the dashed and dotted lines show the starburst/main-sequence boundary according to \citet{rodighiero11} and \citet{rinaldi22}, respectively.}
    \label{fig:ms}
\end{figure}

In Figure~\ref{fig:ms}, we show the position of our galaxies in the SFR vs stellar mass plane, compared to the main-sequence relation at $z =  4.5$ from \citet{speagle14}. For our galaxies, we use the stellar masses derived from the mass decomposition of the rotation curves. The SFRs are derived from IR luminosities and converted to a Chabrier IMF as described in Section~\ref{met:sfr}. We also show two thresholds for starburst galaxies: 4 times above the main-sequence \citep{rodighiero11} and the starburst bi-modality \citep{rinaldi22}, which is an empirical division between main-sequence and starburst galaxies at high-\textit{z} in the specific SFRs first proposed by \citet{caputi17}.
We conclude that our sample is diverse in terms of SFR: J081740 is a main-sequence galaxy, BRI1335-0417 and SGP38326-2 are both starburst and SGP38326-1 is borderline into the starburst region.

\subsection{Origin of Turbulence}
As was pointed out in the introduction to this paper, the velocity dispersion measured in cold gas components, in particular CO and [CII], is mostly contributed by turbulence as thermal broadening can be considered negligible. An accurate estimation of the velocity dispersion allows us to investigate the potential drivers of turbulence within the ISM.

In high-\textit{z} galaxies, gravitational instabilities and star formation feedback are believed to be the primary drivers of gas turbulence \citep{bournaud07, genzel11, krumholz16}. These galaxies have higher gas fractions and densities compared to those in the local Universe \citep{tacconi10, tacconi13, decarli16, scoville17}. This abundance of gas not only promotes elevated star formation rates but can also cause more significant disruptions to the equilibrium of the gas disc \citep{tacconi13}.
Here, we investigate which of these two possible scenarios are more likely: gas turbulence driven by gravitational instabilities in the disc and turbulence driven by stellar feedback.

\subsubsection{Gravitational Instabilities}

To test the hypothesis that the gas turbulence in our sample is driven by gravitational instabilities, we compare the observed gas velocity dispersions to those expected by the analytical model from \citet{krumholz18}. This model considers an axisymmetric disc of gas and stars in a dark matter halo, where the gas and stars have isotropic velocity dispersions and the disc is in both vertical hydrostatic and energy equilibrium. The model accounts for the energy contributed by star formation feedback and gravitational potential energy from inward gas flow. One main assumption of this model is that the gas in the disc self-regulates the inflow rate to maintain marginal stability. 
%The model considers an orbit within a steady gravitational potential.

\begin{figure*}
    \centering
    \includegraphics[width=0.9\textwidth]{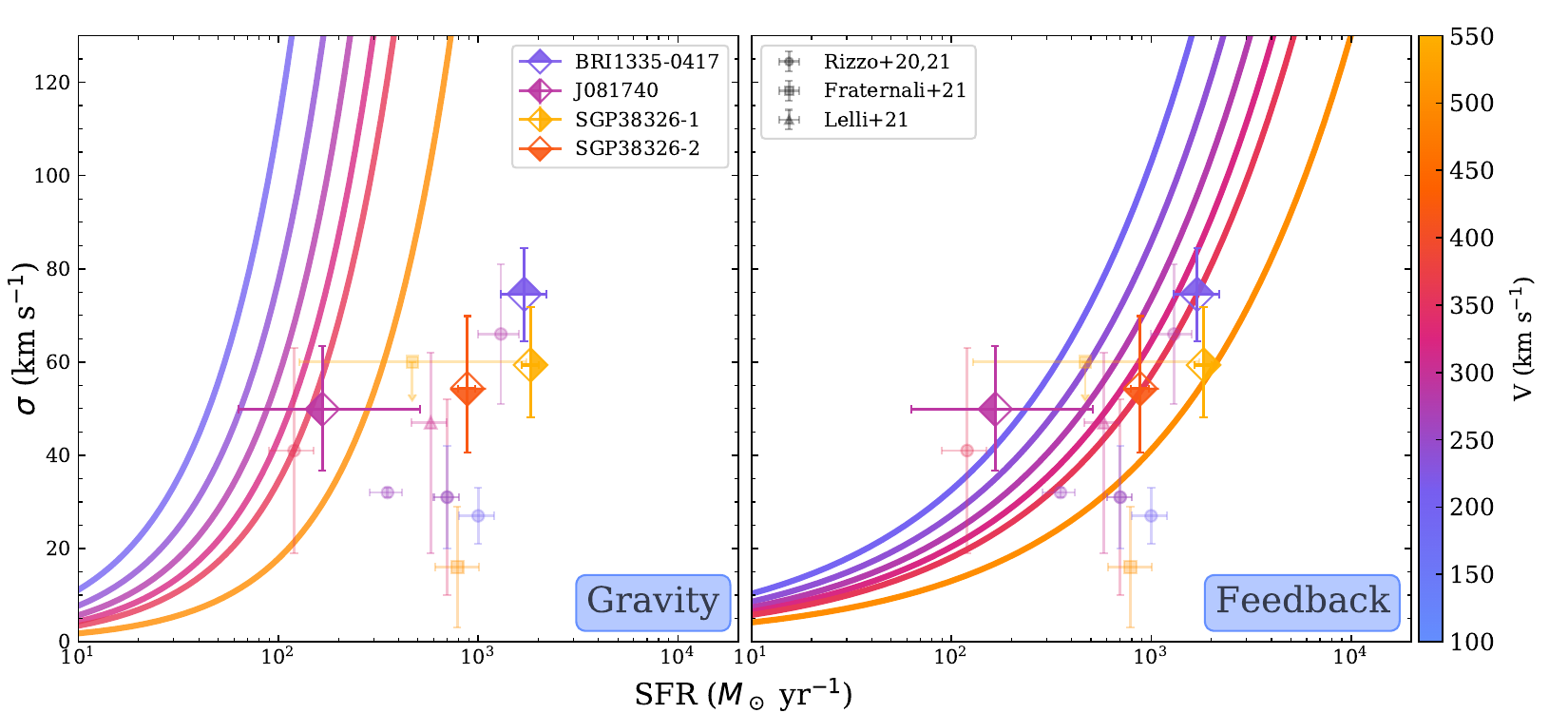}
    \caption{Comparison of the observed gas velocity dispersion and star formation rate to expectations from analytical models from \citet{krumholz18}. Left panel: The solid lines show different expectations from the model considering only gravitational instabilities as the driver of turbulence assuming fiducial parameters for high-\textit{z} galaxies. Right panel: The solid lines show different expectations from the model considering only stellar feedback as the driver of turbulence. The models (solid-lines) are colour-coded according to the maximum circular speed of the galaxies. We show our galaxies \citep{romanoliveira23} and other resolved observations from the literature \citep{fraternali21, rizzo21, lelli21} according to the legend.}
    \label{fig:krumholz}
\end{figure*}

We start by comparing our data to two models from \citet{krumholz18}: the scenario with only gravitational instabilities as the driver of turbulence and the scenario with only stellar feedback shown in Figure~\ref{fig:krumholz}. Although in reality both scenarios could be present simultaneously, we choose to analyse these two cases individually to understand which one gets closer to reproduce the data. In both scenarios we use the fiducial parameters for high-\textit{z} galaxies as reported in \citet{krumholz18} with different circular speeds that cover the range of velocities observed in our sample.
On the left panel, we explore the $\sigma$-SFR relation for the gravity driven scenario assuming that the stellar feedback does not contribute to the turbulence.
On the right panel, we explore the scenario of stellar feedback driven turbulence, in which there is no mass transport in the disc and the star formation efficiency is free to vary according to physical properties of the galaxy.
Both assume that the Toomre Q parameter, which related to the local stability of the disc \citep{toomre64}, is always equal to a marginally stable system.

Figure~\ref{fig:krumholz} suggests that, for the typical range of SFR of \textit{z}$ \sim 4.5$ galaxies, the analytical model including gravitational instabilities overestimates the level of turbulence in the gas discs. Instead, the stellar feedback only model is significantly better at reproducing the data trends, predicting velocity dispersions more similar to those observed. Moreover, despite the low-number statistics, we do not find any clear gradient in the rotation velocities of the observed dataset that follows the ones expected from either models, indicating that the turbulence might not be related to the total halo mass.

\subsubsection{Stellar Feedback}
Following up on the evidence that the stellar feedback only models perform better than the gravity only models from \citet{krumholz18}, we now compare our data with the simple model described in Section~\ref{met:snii}. Here we estimate the maximum possible energy that can be injected by SNII into the ISM of the galaxies given their current SFRs and compare to the turbulent energy present in the discs estimated from the observed velocity dispersions.

\begin{figure}
    \centering
    \includegraphics[width=0.48\textwidth]{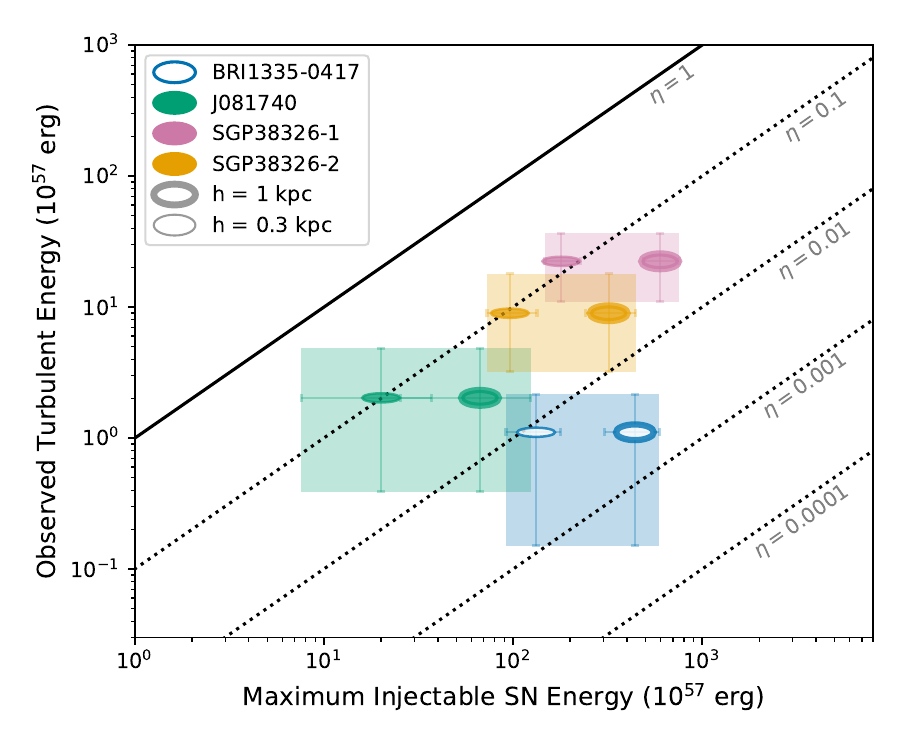}
    \caption{Observed turbulent energy in the discs (Equation~\ref{eq:turb}) vs. the maximum possible energy that can be injected in the ISM from SNII (Equation~\ref{eq:snii}). We explore the possibility of a thin (0.3 kpc, thin ellipse) and a thick disc (1 kpc, thick ellipse). BRI1335-0417 is shown as open marker due to the uncertainties on the dynamical estimate of the gas mass.}
    \label{fig:expsigma}
\end{figure}

We show this comparison in Figure~\ref{fig:expsigma}, where the x-axis shows the maximum injectable energy expected from SNII calculated with Equation~\ref{eq:snii} and assuming a supernova kinetic efficiency of 1. In the y-axis, we show the observed turbulent energy as calculated with Equation~\ref{eq:turb} using the estimated total gas mass of each galaxy (see Table~\ref{tab:dy_fid}) and the observed velocity dispersion averaged across the radial extent of the disc as reported in \citet{romanoliveira23}. Since we do not have reliable information about the thickness of gas discs, we analyse two different assumptions: one with a relatively thin disc ($h = 0.3$ kpc, thin ellipse) and a thick disc ($h = 1.0$ kpc, thick ellipse). Figure~\ref{fig:expsigma} also has identity lines that represent the SNII efficiency $\eta$ that is usually expected to be of the order of 10\%.
We find that the three of the galaxies are consistent with $\eta \approx 0.1$ while BRI1335-0417 is more consistent with a lower efficiency of $\eta \approx 0.001 - 0.01$. These are realistic values and indicate that the level of turbulence in the galaxies can be easily explained by the current SFR present in their discs, without the need of other drivers of turbulence.
We note here that we do not have a strong constrain in the gas mass of BRI1335-0417, since we had to keep the baryon fraction fixed to the cosmological fraction, effectively forcing the dark matter halo mass to its minimum realistic value.

\section{Discussion}\label{sec:discussion}
In this Section, we discuss the main outcomes of the dynamical modelling of our sample, the origin of turbulence in the galaxies and a comparison of our sample with known scaling relations.

\subsection{Compact and massive stellar component}

One of the main outcomes of the dynamical models described in Section~\ref{sec:res_dy} is that the stellar component is described by a compact S\'ersic profile with $n \gtrsim 5$.
This indicates that a large portion of the stellar mass has to be concentrated in a compact region (< 2 kpc) in the centre of the galaxies. This could be a compact stellar disc or, more likely, a stellar bulge. This is in line with other recent works that report dynamical signatures of bulges in $z > 4$ galaxies \citep{rizzo21, lelli21, tripodi23}.
In particular, \citet{rizzo21} reported steep rises in the rotation curve similar to those found in local galaxies with significant bulge components. In \citet{lelli21} and \citet{tripodi23}, there is a significant increase in the rotation velocity in the innermost resolution element that can only be reproduced in the dynamical modelling by a massive bulge component.
These are similar to the rotation curve of BRI1335-0417 shown in Figure~\ref{fig:dy_rotcur} which shows an increased velocity at the innermost radius compared to the outskirts.
For BRI1335-0417 in particular, we also observe a high concentration of dust in the centre as the best-fit surface brightness profile suggests a S\'ersic index of $\sim 4$. This could be related to the presence of a concentrated star formation region and perhaps a bulge still in formation.
It is also a possibility that the concentrated dust emission could be related to the fact that BRI1335-0417 is a QSO host, however we attempt to fit the surface brightness of the dust emission with a S\'ersic + PSF components and we find no evidence that it provides a better fit.
As for the other galaxies in our sample, even though they do not show such a pronounced inner rise in the rotation curves, all of them require a stellar component with a S\'ersic index higher than 4 to be able to reproduce the high inner circular speeds.

One may wonder whether the high rotation velocities that we observe in the inner regions could be explained by another component different from the stellar one.
However, a CDM halo cannot explain velocities of such a magnitude considering that there is essentially no leverage in the concentration parameter at $z \sim 4$ \citep{child18, correa15, dutton14}.
Moreover, a central supermassive black hole of $10^9 M_{\odot}$ can only give velocities of $\sim 200$ km s$^{-1}$ in the inner $\sim 100$ pc, while we observe these velocities typically within the inner 2 kpc from the centre (see inner points of our rotation curves in Figure~\ref{fig:dy_rotcur}).
The only alternative possibility would be an accumulation of gas in a phase that is completely missed by the [CII] emission. This seems quite unlikely especially considering that [CII] is expected to trace all major phases of the cold ISM. Thus we conclude that the presence of very compact stellar components in our galaxies is very likely.

The above conclusion opens questions as to the which of the pathways of bulge formation can be so efficient in forming bulges so early. There are several processes that may form bulges, e.g. mergers or gravitational instabilities in the stellar discs \citep{kormendy04, kormendy13, ishibashi14}.
Another possibility is that bulges can be formed in discs with high fraction of baryonic mass with respect to the dark matter halo within the baryonic effective radius ($f_{\rm disc}$ in Table~\ref{tab:dy_more}). In a recent work by \citet{blandhawthorn23}, they predict that such high baryonic systems are very efficient at forming persistent bars, but also that in the presence of high gas fractions a bulge can be formed almost instantaneously in the centre of the galaxies.
Independently on the specific mechanism, our findings suggest that the process of bulge formation has to happen rapidly and in the early phases of formation of massive galaxies \citep{ferreira22a, jacobs23, kartaltepe23}.
In fact, the bulges in galaxies at \textit{z} $ \sim 4.5$ may indicate that these galaxies are the progenitors of massive quenched galaxies that host mature bulges at $z = 2 - 4$ \citep{valentino23, ito23, lustig21, tacchella15}. Moreover, the discs we observe could also be possible progenitors of red discs recently observed with JWST at $z \sim 2 - 3$ \citep{huang23, fudamoto22, wu23, nelson23}.

\subsection{Scaling relations}

%The presence of galaxy scaling relations, and their cosmic evolution, provide crucial information in galaxy evolution and an important comparison with theory.
In this Section, we investigate how our sample of high-\textit{z} discs are situated in Tully-Fisher-like relations and in the stellar-to-halo mass relation (SHMR).

\subsubsection{Tully-Fisher relations}
%introduction of the BTFR
%The baryonic Tully-Fisher relation (BTFR) is a tight relation coupling the circular speed of galaxies to their total baryonic mass \citep{mcgaugh00}. Its existence indicates a tight link between the mass of the DM halos and the amount of cold gas and stars that reside in them. The normalisation and slope of the BTFR have been extensively studied at $z=0$ \citep{lelli16} and have been used as benchmarks for cosmological simulations and models \citep[e.g.][]{governato07, posti19}. However, due to observational difficulties, the evolution of the BTFR is uncertain and has been explored only up to $z \sim 2$ \citep{diteodoro16, harrison17, tiley16, uebler17}. Only recently, an analogous version of the Tully-Fisher relation applied nuclear CO discs found in early-type galaxies (ETGs) in the local Universe was used as comparison to dusty star-forming galaxies at $z \sim 4.5$ in \citet{fraternali21}. This is a relation of the stellar mass only and it was considered that the gas mass in the dusty star-forming galaxies in \citet{fraternali21} would be completely converted into stars by $z = 0$.

%plot
\begin{figure*}
    \centering
    \includegraphics[width=0.99\textwidth]{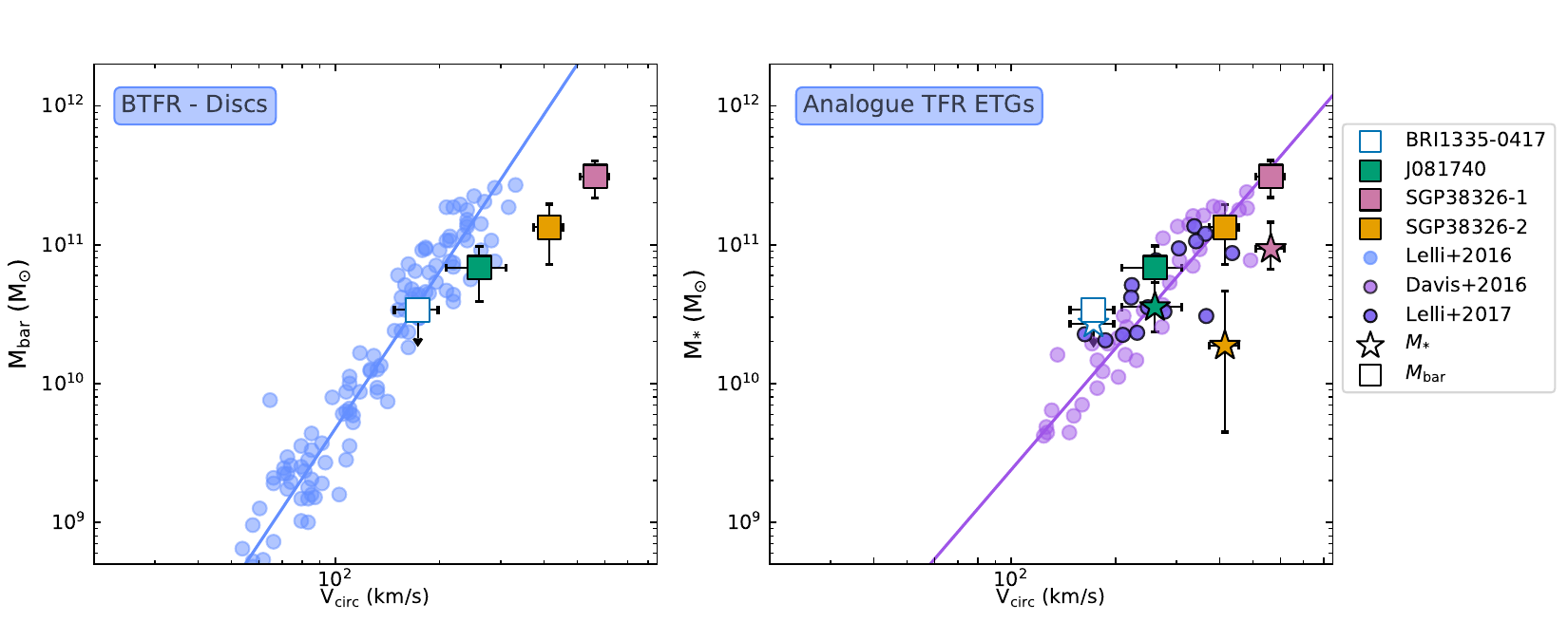}
    \caption{The baryonic Tully-Fisher relation (left panel) and the analogue of the Tully-Fisher relation for ETGs (right panel). Left panel: we show the baryonic mass and the circular speed on the flat region of the rotation curves of local disc galaxies and the best-fit BTFR relation in blue \citep{lelli16}. Right panel: we show the stellar masses and inner rotation of local ETGs with stellar kinematics in purple \citep{lelli17} and with inner CO rotating discs in violet \citep{davis16}. We show the baryonic mass and external circular speed of our sample according to the legend, in the right panel, we also show their stellar mass (stars).}
    \label{fig:btfr}
\end{figure*}

%what we show here
In Figure~\ref{fig:btfr}, we show the distribution of our sample with respect to the local BTFR (left panel) and the ETG analogue of the Tully Fisher relation (right panel). 
The latter, as mentioned in the Introduction, is derived using the rotation velocities of the inner CO discs in local ETGs.
We discuss both these panels separately below.

%left panel
On the left panel of Figure~\ref{fig:btfr}, we compare our sample to the BTFR for local disc galaxies where we show the disc galaxies from the Spitzer Photometry and Accurate Rotation Curves survey \citep[SPARC, blue markers,][]{lelli16b} and the BTFR best-fit relation from \citet{lelli16}. Here we show the baryonic masses of these galaxies and the circular speed estimated in the flat part of their rotation curves as obtained from kinematic models of the discs observed in HI.
For our galaxies, we plot the total baryonic mass derived from the mass decomposition of the rotation curves and the external circular speeds defined as the average of the circular speed from the outer two resolution independent elements.
We note that, like we do in the previous plots, we display BRI1335-0417 with an open symbol as we only have upper limits in the baryonic mass and therefore its placement in the BTFR is more uncertain.
We find that, for the galaxies with a more reliable baryonic mass, there is a shift from the relation towards higher velocities for a fixed mass. This trend goes in the same direction as the cosmic evolution predicted by simulations and theoretical models \citep{posti19, torrey14}.
However, we highlight that this is a small sample and that these galaxies are unlikely to be the progenitors of local disc galaxies. Therefore, it is difficult to draw firm conclusions and further comparison with theoretical models is needed.

%right panel
On the right panel of Figure~\ref{fig:btfr}, we compare our sample to local ETGs, both elliptical and lenticular galaxies, from the MASSIVE survey \citep[violet markers,][]{ma14}. This subset of galaxies have small molecular gas discs extending out to 1-2 kpc that are buried within the stellar component of the host galaxy \citep{davis13a, davis11}. We show the rotation derived from the velocity width of these small discs traced by the CO(J=1-0) emission, revised by \citet{davis16} and \citet{fraternali21} for better estimates of their inclinations. The total stellar masses of the local galaxies are derived from a $K-band$ luminosity assuming an ${\rm M/L} = 0.5 M_{\odot}/L_{\odot, K}$. We highlight that, unlike the disc Tully-Fisher shown in the left panel, we show the stellar mass instead of the baryonic mass here.
Moreover, we include a sample of ETGs from the ATLAS$^{\mathrm{3D}}$ survey \citep{cappellari11}. This comprises lenticular, disky elliptical and boxy elliptical galaxies with available stellar kinematics, therefore also tracing mostly the inner parts of the galaxies \citep[purple markers,][]{lelli17}.
We also plot the best-fit relation from \citet{davis16}, which has a shallower slope than the relation for discs shown on the left panel. We see that when we look at these inner rotations, which trace the baryon dominated inner potential of the galaxies, they deviate from the Tully-Fisher relation for disc galaxies from \citet{lelli16} as, in particular at high masses, they move towards a region of fast rotation with respect to their luminous mass content, as also noted by \citet{davis16}.
For our galaxies, we position them according to both their stellar mass (coloured stars in the panel) and their baryonic mass (stars + gas, squares) with the same external circular speed as shown in the left panel.
We see that our $z \sim 4.5$ galaxies tend to be shifted from the local ETG relation when comparing their stellar masses, but this shift disappears once we assume that their gas content will be fully converted into stars by $z=0$, similar to what was observed in \citet{fraternali21}.
This suggests that, these $z \sim 4.5$ galaxies are largely formed in the inner regions as they have gravitational potentials similar to local ETGs and do not need to accrete more gas in order to be consistent with the local relation.
%already accreted the majority of their baryonic mass during at only 10\% of the age of the Universe.
%convert their gas supply into stars, their potential becomes more similar to those of local ETGs. In fact, this suggests that these galaxies are functionally fully 
%
Given that it is rare to find spiral galaxies rotating at or above $\sim 300$ km s$^{-1}$ \citep{diteodoro23, ogle19, posti18}, we can expect that as the gas is converted into stars, the galaxies should undergo morphological changes and become more pressure supported \citep[potentially through a series of minor or major mergers, see ][]{hilz13, naab09a, naab09b}. Overall these findings give dynamical support to the scenario in which these massive $z \sim 4.5$ galaxies are likely progenitors of local ETGs \citep{fraternali21, rizzo20, toft14, barro14}.
%
%Moreover, if we account for possible mergers, the assumption that all gas is converted into stars does not need to be met. In which case, some of the gas could be lost or heated up and the stellar mass would still increase.

\subsubsection{The Stellar-to-Halo Mass Relation}
The SHMR is crucial in understanding the relationship between baryonic matter and the dark matter halos of galaxies as it traces the efficiency of the conversion of baryons into stars as a function of the halo mass, providing a proxy of the efficiency of galaxy formation over cosmic time \citep{moster10, behroozi13}.
Unlike the BTFR, the SHMR has been extended up to \textit{z} $\sim 6$ with the use of abundance-matching techniques. However, to date, the SMHR has been estimated only up to $z \sim 1$ with direct dynamical estimates of CDM halo masses of individual galaxies \citep{bouche22}.
Recent studies suggest that the SHMR may be dependent on the morphology of galaxies, where the relation breaks at the massive end with massive discs containing higher stellar mass fractions than elliptical galaxies \citep{correa20, marasco20, rodriguez-gomez22}.
Moreover, this was supported by \citet{posti19} and \citet{diteodoro23} that showed empirically that local massive disc galaxies deviate from the relation obtained from abundance-matching techniques.
It was suggested that these massive discs must have evolved at near isolation, without going through disruptive events such as AGN feedback or major mergers.

\begin{figure}
    \centering
    \includegraphics[width=0.48\textwidth]{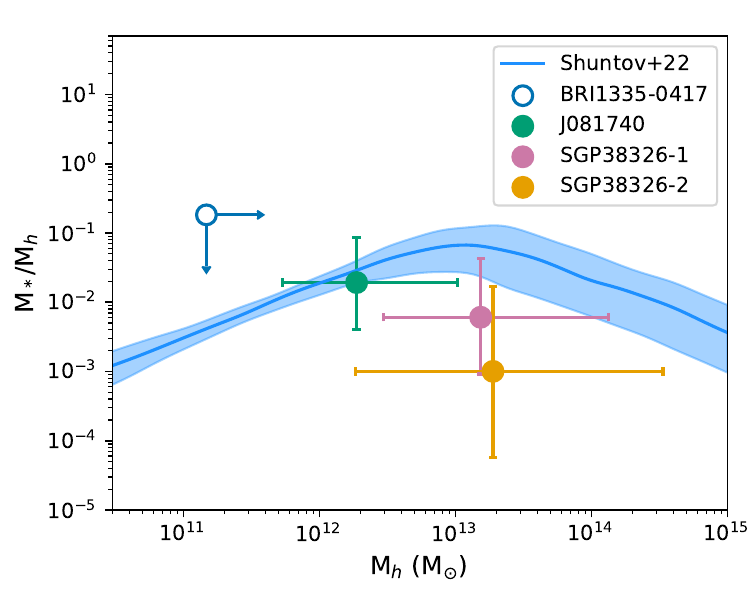}
    \caption{Stellar-to-halo mass relation (SHMR). We show the stellar-to-halo mass fraction on the y-axis versus the halo mass in the x axis. Our galaxies are shown as coloured circle markers according to the legend and the expected SHMR relation for central halos at 4.5 < \textit{z} < 5.5 as a curve with 16th, 50th and 84th percentiles from \citet{shuntov22}.}
    \label{fig:shmr}
\end{figure}

In Figure~\ref{fig:shmr}, we present the SHMR in the form of the stellar-to-halo mass fraction ($M_*/M_h$) as a function of the halo mass ($M_h$). We compare this to the expected relation from abundance-matching techniques at 4.5 < \textit{z} < 5.5 from \citet{shuntov22}. We see that the curve is reasonably compatible with the data, with the exception of BRI1335-0417, which has only a lower limits on the CDM halo mass. 
We note that, at least for this sample of \textit{z} $\sim$ 4.5 massive disc galaxies, we do not find any significant deviation towards higher stellar fractions at the massive end.

\section{Conclusions}\label{sec:conclusions}

In this paper, we provided mass decomposition of the rotation curves of four galaxies at \textit{z} $\sim 4.5$ using resolved kinematics traced by the [CII] 158 $\mu \mathrm{m}$ emission-line observed with ALMA to determine the main properties of the various matter components in galaxies in the Early Universe.
We also investigated the possible origin of turbulence in the cold gas component of these galaxies to gain insights into the main contributions to the observed velocity dispersions.
Our main results are the following:

%We find that the gas disc of the galaxies is well represented by exponential profiles with effective radii ranging between 2.57 and 3.87 kpc. 

\begin{itemize}
    %mass models
    \item From the rotation curve decomposition, we constrain the stellar masses (ranging between $\log(M_{\rm *}/M_{\odot}) = 10.3 - 11.0$ and gas masses (ranging between $\log(M_{\rm gas}/M_{\odot}) = 9.8 - 11.3$). We estimate the dark matter halo mass of the galaxies (ranging between $\log(M_{200}/M_{\odot}) = 11.2 - 13.3$), albeit with larger uncertainties. We found that our galaxies are strongly baryon dominated in the regions probed (inner 3 to 5 kpc) and that they have a variety of gas-to-baryon fractions, from 0.16 to 0.87.
    % turbulence
    \item We investigated the origin of the turbulence in the gas discs by comparing the observed [CII] velocity dispersion to analytical models of turbulence driven by either gravitational instabilities or stellar feedback. We find that the observed gas turbulence in the discs can be fully reproduced by the stellar feedback alone, without the need for large-scale gravitational instabilities or other mechanisms. We established that the observed gas turbulence can be easily maintained by supernova explosions with efficiencies of only 10\% or less.
    %bulges
    \item Our rotation curve decompositions indicate that there must be a massive compact stellar component dominating in the inner 2 kpc of the galaxies in our sample, compatible with what would be expected from a stellar bulge. This dynamical evidence, already suggested by other similar studies of galaxies at $z = 4 - 5$, can be validated by upcoming JWST/NIRCam observations of the stellar continuum of these galaxies.
    %scaling relations
    \item With the properties derived from the rotation curve decomposition, we place the galaxies in our sample into two scaling relations: the stellar-to-halo mass relation, where our data is in broad agreement with the relation obtained from abundance-matching techniques; and two versions of the Tully-Fisher relation. Compared to the local baryonic Tully-Fisher relations our galaxies show a shift towards higher rotation velocities as expected from theory. Instead in ETG- analogue of the Tully-Fisher relation our galaxies overlap with local ETGs once their gas mass is added to the stellar mass suggesting that, at least in the inner regions, they have gathered the majority of their baryonic mass already at \textit{z} $\sim$ 4.5.
\end{itemize}

Finally, we reiterate that spatially-resolved kinematics of galaxies is a powerful tool for studying the physical mechanisms involved in galaxy formation and evolution. It is crucial to obtain constrains on the internal structure and morphology of galaxies, providing insights into the complex interactions between DM, gas, stars, and feedback processes in the ISM.
We emphasise that the detailed analysis carried out in this paper, in particular the mass decomposition of the rotation curves, requires at least 6 independent resolution elements along the major axis in any single galaxy.
ALMA is currently the only instrument that can provide high enough SNR and angular resolution to achieve these requirements. Such observations are time expensive, but critical, and in the future we should push for even higher data quality to be able to investigate the shape of the dark matter halos.
We can achieve this with  with JWST data, that will give us spatially resolved information for the stellar component, combined with higher sensitivity ALMA data to probe the outer regions of the rotation curve that are more dark matter dominated.
%
%However, the detailed analysis done in this paper is only possible with high SNR and high angular resolution data which is time expensive to obtain and, currently, only ALMA is capable of providing.
%
%We emphasise that we are highly limited by the resolution of the current data and the physical extent of the [CII] emission, as we have about 3 to 5 independent resolution elements in our rotation curves probing only the inner 3-5 kpc. 
%
%Such resolutions and the SNRs used in this paper are minimum requirements for the type of work presented here to be done reliably. 
%
%In the future, we need to increase the number of independent resolution elements in the rotation curves to be able to also increase the number of free parameters that describe the mass components.

%Resolved observations of the stellar continuum with the JWST/NIRCam together with higher sensitivity [CII] data will constitute a new source of information able to independently pin down the properties of the stellar components.

\section*{Data Availability}

The data underlying this article are public and available in the ALMA Science Archive, at \url{https://almascience.nrao.edu/asax/}.
The processed data (i.e. the rotation curves corrected for ADC and the 0th-moment maps for the dust continuum and [CII] emission) along with the scripts to produce the Figures in this paper are made available at Zenodo (DOI: 10.5281/zenodo.10707348).

\begin{acknowledgements}
We thank Vasily Kokorev and Takafumi Tsukui for the discussion on the SFR of BRI1335-0417, Marko Shuntov for the SHMR and Joss Bland-Hawthorn for the discussion on the disc fraction and the formation of bars.
FRO and FF acknowledge support from The Dutch Research Council (NWO) through the Klein-1 Grant code OCEN2.KLEIN.088.
FR acknowledges support from the European Union’s Horizon 2020 research and innovation program under the Marie Sklodowska-Curie grant agreement No. 847523 ‘INTERACTIONS’ and the Cosmic Dawn Center that is funded by the Danish National Research Foundation under grant No. 140.
The project leading to this publication has received support from ORP, that is funded by the European Union’s Horizon 2020 research and innovation programme under grant agreement No 101004719 [ORP].
We acknowledge extensive assistance from Allegro, the European ALMA Regional Center node in the Netherlands.
The Joint ALMA Observatory is operated by ESO, AUI/NRAO and NAOJ.
This work makes use of the following ALMA data: ADS/JAO.ALMA\#2015.1.00330.S, ADS/JAO.ALMA\#2017.1.00127.S, \#2017.1.00394.S, \#2017.1.01052.S, \#2018.1.00001.S. ALMA is a partnership of ESO (representing its member states), NSF (USA) and NINS (Japan), together with NRC (Canada), MOST and ASIAA (Taiwan), and KASI (Republic of Korea), in cooperation with the Republic of Chile.
This publication makes use of data products from the Wide-field Infrared Survey Explorer, which is a joint project of the University of California, Los Angeles, and the Jet Propulsion Laboratory/California Institute of Technology, funded by the National Aeronautics and Space Administration.
\end{acknowledgements}

\bibliographystyle{aa}
\bibliography{references.bib}

\begin{thebibliography}{154}
\expandafter\ifx\csname natexlab\endcsname\relax\def\natexlab#1{#1}\fi

\bibitem[{{Aravena} {et~al.}(2016){Aravena}, {Spilker}, {Bethermin},
  {Bothwell}, {Chapman}, {de Breuck}, {Furstenau}, {G{\'o}nzalez-L{\'o}pez},
  {Greve}, {Litke}, {Ma}, {Malkan}, {Marrone}, {Murphy}, {Stark}, {Strandet},
  {Vieira}, {Weiss}, {Welikala}, {Wong}, \& {Collier}}]{aravena16}
{Aravena}, M., {Spilker}, J.~S., {Bethermin}, M., {et~al.} 2016, \mnras, 457,
  4406

\bibitem[{{Astropy Collaboration} {et~al.}(2022){Astropy Collaboration},
  {Price-Whelan}, {Lim}, {Earl}, {Starkman}, {Bradley}, {Shupe}, {Patil},
  {Corrales}, {Brasseur}, {N{"o}the}, {Donath}, {Tollerud}, {Morris},
  {Ginsburg}, {Vaher}, {Weaver}, {Tocknell}, {Jamieson}, {van Kerkwijk},
  {Robitaille}, {Merry}, {Bachetti}, {G{"u}nther}, {Aldcroft},
  {Alvarado-Montes}, {Archibald}, {B{'o}di}, {Bapat}, {Barentsen}, {Baz{'a}n},
  {Biswas}, {Boquien}, {Burke}, {Cara}, {Cara}, {Conroy}, {Conseil}, {Craig},
  {Cross}, {Cruz}, {D'Eugenio}, {Dencheva}, {Devillepoix}, {Dietrich},
  {Eigenbrot}, {Erben}, {Ferreira}, {Foreman-Mackey}, {Fox}, {Freij}, {Garg},
  {Geda}, {Glattly}, {Gondhalekar}, {Gordon}, {Grant}, {Greenfield}, {Groener},
  {Guest}, {Gurovich}, {Handberg}, {Hart}, {Hatfield-Dodds}, {Homeier},
  {Hosseinzadeh}, {Jenness}, {Jones}, {Joseph}, {Kalmbach}, {Karamehmetoglu},
  {Ka{l}uszy{'n}ski}, {Kelley}, {Kern}, {Kerzendorf}, {Koch}, {Kulumani},
  {Lee}, {Ly}, {Ma}, {MacBride}, {Maljaars}, {Muna}, {Murphy}, {Norman},
  {O'Steen}, {Oman}, {Pacifici}, {Pascual}, {Pascual-Granado}, {Patil},
  {Perren}, {Pickering}, {Rastogi}, {Roulston}, {Ryan}, {Rykoff}, {Sabater},
  {Sakurikar}, {Salgado}, {Sanghi}, {Saunders}, {Savchenko}, {Schwardt},
  {Seifert-Eckert}, {Shih}, {Jain}, {Shukla}, {Sick}, {Simpson},
  {Singanamalla}, {Singer}, {Singhal}, {Sinha}, {Sip{H{o}}cz}, {Spitler},
  {Stansby}, {Streicher}, {{{S}}umak}, {Swinbank}, {Taranu}, {Tewary},
  {Tremblay}, {Val-Borro}, {Van Kooten}, {Vasovi{'c}}, {Verma}, {de Miranda
  Cardoso}, {Williams}, {Wilson}, {Winkel}, {Wood-Vasey}, {Xue}, {Yoachim},
  {Zhang}, {Zonca}, \& {Astropy Project Contributors}}]{astropycollab22}
{Astropy Collaboration}, {Price-Whelan}, A.~M., {Lim}, P.~L., {et~al.} 2022,
  \apj, 935, 167

\bibitem[{{Astropy Collaboration} {et~al.}(2018){Astropy Collaboration},
  {Price-Whelan}, {Sip{\H{o}}cz}, {G{\"u}nther}, {Lim}, {Crawford}, {Conseil},
  {Shupe}, {Craig}, {Dencheva}, {Ginsburg}, {VanderPlas}, {Bradley},
  {P{\'e}rez-Su{\'a}rez}, {de Val-Borro}, {Aldcroft}, {Cruz}, {Robitaille},
  {Tollerud}, {Ardelean}, {Babej}, {Bach}, {Bachetti}, {Bakanov}, {Bamford},
  {Barentsen}, {Barmby}, {Baumbach}, {Berry}, {Biscani}, {Boquien}, {Bostroem},
  {Bouma}, {Brammer}, {Bray}, {Breytenbach}, {Buddelmeijer}, {Burke},
  {Calderone}, {Cano Rodr{\'\i}guez}, {Cara}, {Cardoso}, {Cheedella}, {Copin},
  {Corrales}, {Crichton}, {D'Avella}, {Deil}, {Depagne}, {Dietrich}, {Donath},
  {Droettboom}, {Earl}, {Erben}, {Fabbro}, {Ferreira}, {Finethy}, {Fox},
  {Garrison}, {Gibbons}, {Goldstein}, {Gommers}, {Greco}, {Greenfield},
  {Groener}, {Grollier}, {Hagen}, {Hirst}, {Homeier}, {Horton}, {Hosseinzadeh},
  {Hu}, {Hunkeler}, {Ivezi{\'c}}, {Jain}, {Jenness}, {Kanarek}, {Kendrew},
  {Kern}, {Kerzendorf}, {Khvalko}, {King}, {Kirkby}, {Kulkarni}, {Kumar},
  {Lee}, {Lenz}, {Littlefair}, {Ma}, {Macleod}, {Mastropietro}, {McCully},
  {Montagnac}, {Morris}, {Mueller}, {Mumford}, {Muna}, {Murphy}, {Nelson},
  {Nguyen}, {Ninan}, {N{\"o}the}, {Ogaz}, {Oh}, {Parejko}, {Parley}, {Pascual},
  {Patil}, {Patil}, {Plunkett}, {Prochaska}, {Rastogi}, {Reddy Janga},
  {Sabater}, {Sakurikar}, {Seifert}, {Sherbert}, {Sherwood-Taylor}, {Shih},
  {Sick}, {Silbiger}, {Singanamalla}, {Singer}, {Sladen}, {Sooley},
  {Sornarajah}, {Streicher}, {Teuben}, {Thomas}, {Tremblay}, {Turner},
  {Terr{\'o}n}, {van Kerkwijk}, {de la Vega}, {Watkins}, {Weaver}, {Whitmore},
  {Woillez}, {Zabalza}, \& {Astropy Contributors}}]{astropycollab18}
{Astropy Collaboration}, {Price-Whelan}, A.~M., {Sip{\H{o}}cz}, B.~M., {et~al.}
  2018, \aj, 156, 123

\bibitem[{{Astropy Collaboration} {et~al.}(2013){Astropy Collaboration},
  {Robitaille}, {Tollerud}, {Greenfield}, {Droettboom}, {Bray}, {Aldcroft},
  {Davis}, {Ginsburg}, {Price-Whelan}, {Kerzendorf}, {Conley}, {Crighton},
  {Barbary}, {Muna}, {Ferguson}, {Grollier}, {Parikh}, {Nair}, {Unther},
  {Deil}, {Woillez}, {Conseil}, {Kramer}, {Turner}, {Singer}, {Fox}, {Weaver},
  {Zabalza}, {Edwards}, {Azalee Bostroem}, {Burke}, {Casey}, {Crawford},
  {Dencheva}, {Ely}, {Jenness}, {Labrie}, {Lim}, {Pierfederici}, {Pontzen},
  {Ptak}, {Refsdal}, {Servillat}, \& {Streicher}}]{astropycollab13}
{Astropy Collaboration}, {Robitaille}, T.~P., {Tollerud}, E.~J., {et~al.} 2013,
  \aap, 558, A33

\bibitem[{{Bacchini} {et~al.}(2020){Bacchini}, {Fraternali}, {Iorio},
  {Pezzulli}, {Marasco}, \& {Nipoti}}]{bacchini20}
{Bacchini}, C., {Fraternali}, F., {Iorio}, G., {et~al.} 2020, \aap, 641, A70

\bibitem[{{Barro} {et~al.}(2014){Barro}, {Faber}, {P{\'e}rez-Gonz{\'a}lez},
  {Pacifici}, {Trump}, {Koo}, {Wuyts}, {Guo}, {Bell}, {Dekel}, {Porter},
  {Primack}, {Ferguson}, {Ashby}, {Caputi}, {Ceverino}, {Croton}, {Fazio},
  {Giavalisco}, {Hsu}, {Kocevski}, {Koekemoer}, {Kurczynski}, {Kollipara},
  {Lee}, {McIntosh}, {McGrath}, {Moody}, {Somerville}, {Papovich}, {Salvato},
  {Santini}, {Tal}, {van der Wel}, {Williams}, {Willner}, \&
  {Zolotov}}]{barro14}
{Barro}, G., {Faber}, S.~M., {P{\'e}rez-Gonz{\'a}lez}, P.~G., {et~al.} 2014,
  \apj, 791, 52

\bibitem[{{Behroozi} {et~al.}(2013){Behroozi}, {Wechsler}, \&
  {Conroy}}]{behroozi13}
{Behroozi}, P.~S., {Wechsler}, R.~H., \& {Conroy}, C. 2013, \apjl, 762, L31

\bibitem[{{Binney} \& {Tremaine}(2008)}]{binney08}
{Binney}, J. \& {Tremaine}, S. 2008, {Galactic Dynamics: Second Edition}

\bibitem[{{Bland-Hawthorn} {et~al.}(2023){Bland-Hawthorn}, {Tepper-Garcia},
  {Agertz}, \& {Freeman}}]{blandhawthorn23}
{Bland-Hawthorn}, J., {Tepper-Garcia}, T., {Agertz}, O., \& {Freeman}, K. 2023,
  \apj, 947, 80

\bibitem[{{Bolatto} {et~al.}(2013){Bolatto}, {Wolfire}, \& {Leroy}}]{bolatto13}
{Bolatto}, A.~D., {Wolfire}, M., \& {Leroy}, A.~K. 2013, \araa, 51, 207

\bibitem[{{Boogaard} {et~al.}(2020){Boogaard}, {van der Werf}, {Weiss},
  {Popping}, {Decarli}, {Walter}, {Aravena}, {Bouwens}, {Riechers},
  {Gonz{\'a}lez-L{\'o}pez}, {Smail}, {Carilli}, {Kaasinen}, {Daddi}, {Cox},
  {D{\'\i}az-Santos}, {Inami}, {Cortes}, \& {Wagg}}]{boogaard20}
{Boogaard}, L.~A., {van der Werf}, P., {Weiss}, A., {et~al.} 2020, \apj, 902,
  109

\bibitem[{{Bouch{\'e}} {et~al.}(2022){Bouch{\'e}}, {Bera}, {Krajnovi{\'c}},
  {Emsellem}, {Mercier}, {Schaye}, {Epinat}, {Richard}, {Zoutendijk},
  {Abril-Melgarejo}, {Brinchmann}, {Bacon}, {Contini}, {Boogaard}, {Wisotzki},
  {Maseda}, \& {Steinmetz}}]{bouche22}
{Bouch{\'e}}, N.~F., {Bera}, S., {Krajnovi{\'c}}, D., {et~al.} 2022, \aap, 658,
  A76

\bibitem[{{Bournaud} {et~al.}(2007){Bournaud}, {Elmegreen}, \&
  {Elmegreen}}]{bournaud07}
{Bournaud}, F., {Elmegreen}, B.~G., \& {Elmegreen}, D.~M. 2007, \apj, 670, 237

\bibitem[{{Cappellari} {et~al.}(2011){Cappellari}, {Emsellem}, {Krajnovi{\'c}},
  {McDermid}, {Scott}, {Verdoes Kleijn}, {Young}, {Alatalo}, {Bacon}, {Blitz},
  {Bois}, {Bournaud}, {Bureau}, {Davies}, {Davis}, {de Zeeuw}, {Duc},
  {Khochfar}, {Kuntschner}, {Lablanche}, {Morganti}, {Naab}, {Oosterloo},
  {Sarzi}, {Serra}, \& {Weijmans}}]{cappellari11}
{Cappellari}, M., {Emsellem}, E., {Krajnovi{\'c}}, D., {et~al.} 2011, \mnras,
  413, 813

\bibitem[{{Caputi} {et~al.}(2017){Caputi}, {Deshmukh}, {Ashby}, {Cowley},
  {Bisigello}, {Fazio}, {Fynbo}, {Le F{\`e}vre}, {Milvang-Jensen}, \&
  {Ilbert}}]{caputi17}
{Caputi}, K.~I., {Deshmukh}, S., {Ashby}, M.~L.~N., {et~al.} 2017, \apj, 849,
  45

\bibitem[{{Carilli} \& {Walter}(2013)}]{carilli&walter13}
{Carilli}, C.~L. \& {Walter}, F. 2013, \araa, 51, 105

\bibitem[{{Casey} {et~al.}(2014){Casey}, {Narayanan}, \& {Cooray}}]{casey14}
{Casey}, C.~M., {Narayanan}, D., \& {Cooray}, A. 2014, \physrep, 541, 45

\bibitem[{{Chabrier}(2003)}]{chabrier03}
{Chabrier}, G. 2003, \pasp, 115, 763

\bibitem[{{Child} {et~al.}(2018){Child}, {Habib}, {Heitmann}, {Frontiere},
  {Finkel}, {Pope}, \& {Morozov}}]{child18}
{Child}, H.~L., {Habib}, S., {Heitmann}, K., {et~al.} 2018, \apj, 859, 55

\bibitem[{{Cimatti} {et~al.}(2019){Cimatti}, {Fraternali}, \& {Nipoti}}]{cfn19}
{Cimatti}, A., {Fraternali}, F., \& {Nipoti}, C. 2019, {Introduction to Galaxy
  Formation and Evolution: From Primordial Gas to Present-Day Galaxies}

\bibitem[{{Cochrane} {et~al.}(2019){Cochrane}, {Hayward},
  {Angl{\'e}s-Alc{\'a}zar}, {Lotz}, {Parsotan}, {Ma}, {Kere{\v{s}}},
  {Feldmann}, {Faucher-Gigu{\`e}re}, \& {Hopkins}}]{cochrane19}
{Cochrane}, R.~K., {Hayward}, C.~C., {Angl{\'e}s-Alc{\'a}zar}, D., {et~al.}
  2019, \mnras, 488, 1779

\bibitem[{{Colina} {et~al.}(2023){Colina}, {Crespo G{\'o}mez},
  {{\'A}lvarez-M{\'a}rquez}, {Bik}, {Walter}, {Boogaard}, {Labiano},
  {Peissker}, {P{\'e}rez-Gonz{\'a}lez}, {{\"O}stlin}, {Greve},
  {N{\o}rgaard-Nielsen}, {Wright}, {Alonso-Herrero}, {Azollini}, {Caputi},
  {Dicken}, {Garc{\'\i}a-Mar{\'\i}n}, {Hjorth}, {Ilbert}, {Kendrew}, {Pye},
  {Tikkanen}, {van der Werf}, {Costantin}, {Iani}, {Gillman}, {Jermann},
  {Langeroodi}, {Moutard}, {Rinaldi}, {Topinka}, {van Dishoeck}, {G{\"u}del},
  {Henning}, {Lagage}, {Ray}, \& {Vandenbussche}}]{colina23}
{Colina}, L., {Crespo G{\'o}mez}, A., {{\'A}lvarez-M{\'a}rquez}, J., {et~al.}
  2023, \aap, 673, L6

\bibitem[{{Correa} \& {Schaye}(2020)}]{correa20}
{Correa}, C.~A. \& {Schaye}, J. 2020, \mnras, 499, 3578

\bibitem[{{Correa} {et~al.}(2015){Correa}, {Wyithe}, {Schaye}, \&
  {Duffy}}]{correa15}
{Correa}, C.~A., {Wyithe}, J. S.~B., {Schaye}, J., \& {Duffy}, A.~R. 2015,
  \mnras, 452, 1217

\bibitem[{{Costantin} {et~al.}(2023){Costantin}, {P{\'e}rez-Gonz{\'a}lez},
  {Guo}, {Buttitta}, {Jogee}, {Bagley}, {Barro}, {Kartaltepe}, {Koekemoer},
  {Cabello}, {Corsini}, {M{\'e}ndez-Abreu}, {de la Vega}, {Iyer}, {Bisigello},
  {Cheng}, {Morelli}, {Arrabal Haro}, {Buitrago}, {Cooper}, {Dekel},
  {Dickinson}, {Finkelstein}, {Giavalisco}, {Holwerda}, {Huertas-Company},
  {Lucas}, {Papovich}, {Pirzkal}, {Seill{\'e}}, {Vega-Ferrero}, {Wuyts}, \&
  {Yung}}]{costantin23}
{Costantin}, L., {P{\'e}rez-Gonz{\'a}lez}, P.~G., {Guo}, Y., {et~al.} 2023,
  \nat, 623, 499

\bibitem[{{da Cunha} {et~al.}(2013){da Cunha}, {Groves}, {Walter}, {Decarli},
  {Weiss}, {Bertoldi}, {Carilli}, {Daddi}, {Elbaz}, {Ivison}, {Maiolino},
  {Riechers}, {Rix}, {Sargent}, \& {Smail}}]{dacunha13}
{da Cunha}, E., {Groves}, B., {Walter}, F., {et~al.} 2013, \apj, 766, 13

\bibitem[{{Davis} {et~al.}(2013){Davis}, {Alatalo}, {Bureau}, {Cappellari},
  {Scott}, {Young}, {Blitz}, {Crocker}, {Bayet}, {Bois}, {Bournaud}, {Davies},
  {de Zeeuw}, {Duc}, {Emsellem}, {Khochfar}, {Krajnovi{\'c}}, {Kuntschner},
  {Lablanche}, {McDermid}, {Morganti}, {Naab}, {Oosterloo}, {Sarzi}, {Serra},
  \& {Weijmans}}]{davis13a}
{Davis}, T.~A., {Alatalo}, K., {Bureau}, M., {et~al.} 2013, \mnras, 429, 534

\bibitem[{{Davis} {et~al.}(2011){Davis}, {Bureau}, {Young}, {Alatalo}, {Blitz},
  {Cappellari}, {Scott}, {Bois}, {Bournaud}, {Davies}, {de Zeeuw}, {Emsellem},
  {Khochfar}, {Krajnovi{\'c}}, {Kuntschner}, {Lablanche}, {McDermid},
  {Morganti}, {Naab}, {Oosterloo}, {Sarzi}, {Serra}, \& {Weijmans}}]{davis11}
{Davis}, T.~A., {Bureau}, M., {Young}, L.~M., {et~al.} 2011, \mnras, 414, 968

\bibitem[{{Davis} {et~al.}(2016){Davis}, {Greene}, {Ma}, {Pandya}, {Blakeslee},
  {McConnell}, \& {Thomas}}]{davis16}
{Davis}, T.~A., {Greene}, J., {Ma}, C.-P., {et~al.} 2016, \mnras, 455, 214

\bibitem[{{Decarli} {et~al.}(2016){Decarli}, {Walter}, {Aravena}, {Carilli},
  {Bouwens}, {da Cunha}, {Daddi}, {Ivison}, {Popping}, {Riechers}, {Smail},
  {Swinbank}, {Weiss}, {Anguita}, {Assef}, {Bauer}, {Bell}, {Bertoldi},
  {Chapman}, {Colina}, {Cortes}, {Cox}, {Dickinson}, {Elbaz},
  {G{\'o}nzalez-L{\'o}pez}, {Ibar}, {Infante}, {Hodge}, {Karim}, {Le Fevre},
  {Magnelli}, {Neri}, {Oesch}, {Ota}, {Rix}, {Sargent}, {Sheth}, {van der Wel},
  {van der Werf}, \& {Wagg}}]{decarli16}
{Decarli}, R., {Walter}, F., {Aravena}, M., {et~al.} 2016, \apj, 833, 69

\bibitem[{{Dessauges-Zavadsky} {et~al.}(2020){Dessauges-Zavadsky}, {Ginolfi},
  {Pozzi}, {B{\'e}thermin}, {Le F{\`e}vre}, {Fujimoto}, {Silverman}, {Jones},
  {Vallini}, {Schaerer}, {Faisst}, {Khusanova}, {Fudamoto}, {Cassata},
  {Loiacono}, {Capak}, {Yan}, {Amorin}, {Bardelli}, {Boquien}, {Cimatti},
  {Gruppioni}, {Hathi}, {Ibar}, {Koekemoer}, {Lemaux}, {Narayanan}, {Oesch},
  {Rodighiero}, {Romano}, {Talia}, {Toft}, {Vergani}, {Zamorani}, \&
  {Zucca}}]{dessauges20}
{Dessauges-Zavadsky}, M., {Ginolfi}, M., {Pozzi}, F., {et~al.} 2020, \aap, 643,
  A5

\bibitem[{{Di Teodoro} \& {Fraternali}(2015)}]{diteodoro15}
{Di Teodoro}, E.~M. \& {Fraternali}, F. 2015, \mnras, 451, 3021

\bibitem[{{Di Teodoro} {et~al.}(2016){Di Teodoro}, {Fraternali}, \&
  {Miller}}]{diteodoro16}
{Di Teodoro}, E.~M., {Fraternali}, F., \& {Miller}, S.~H. 2016, \aap, 594, A77

\bibitem[{{Di Teodoro} {et~al.}(2023){Di Teodoro}, {Posti}, {Fall}, {Ogle},
  {Jarrett}, {Appleton}, {Cluver}, {Haynes}, \& {Lisenfeld}}]{diteodoro23}
{Di Teodoro}, E.~M., {Posti}, L., {Fall}, S.~M., {et~al.} 2023, \mnras, 518,
  6340

\bibitem[{{Dunne} {et~al.}(2022){Dunne}, {Maddox}, {Papadopoulos}, {Ivison}, \&
  {Gomez}}]{dunne22}
{Dunne}, L., {Maddox}, S.~J., {Papadopoulos}, P.~P., {Ivison}, R.~J., \&
  {Gomez}, H.~L. 2022, \mnras, 517, 962

\bibitem[{{Dutton} \& {Macci{\`o}}(2014)}]{dutton14}
{Dutton}, A.~A. \& {Macci{\`o}}, A.~V. 2014, \mnras, 441, 3359

\bibitem[{{Ejdetj{\"a}rn} {et~al.}(2022){Ejdetj{\"a}rn}, {Agertz},
  {{\"O}stlin}, {Renaud}, \& {Romeo}}]{ejdetjarn22}
{Ejdetj{\"a}rn}, T., {Agertz}, O., {{\"O}stlin}, G., {Renaud}, F., \& {Romeo},
  A.~B. 2022, \mnras, 514, 480

\bibitem[{{Faisst} {et~al.}(2020){Faisst}, {Schaerer}, {Lemaux}, {Oesch},
  {Fudamoto}, {Cassata}, {B{\'e}thermin}, {Capak}, {Le F{\`e}vre}, {Silverman},
  {Yan}, {Ginolfi}, {Koekemoer}, {Morselli}, {Amor{\'\i}n}, {Bardelli},
  {Boquien}, {Brammer}, {Cimatti}, {Dessauges-Zavadsky}, {Fujimoto},
  {Gruppioni}, {Hathi}, {Hemmati}, {Ibar}, {Jones}, {Khusanova}, {Loiacono},
  {Pozzi}, {Talia}, {Tasca}, {Riechers}, {Rodighiero}, {Romano}, {Scoville},
  {Toft}, {Vallini}, {Vergani}, {Zamorani}, \& {Zucca}}]{faisst20}
{Faisst}, A.~L., {Schaerer}, D., {Lemaux}, B.~C., {et~al.} 2020, \apjs, 247, 61

\bibitem[{{Ferreira} {et~al.}(2022){Ferreira}, {Adams}, {Conselice},
  {Sazonova}, {Austin}, {Caruana}, {Ferrari}, {Verma}, {Trussler},
  {Broadhurst}, {Diego}, {Frye}, {Pascale}, {Wilkins}, {Windhorst}, \&
  {Zitrin}}]{ferreira22a}
{Ferreira}, L., {Adams}, N., {Conselice}, C.~J., {et~al.} 2022, \apjl, 938, L2

\bibitem[{{Ferreira} {et~al.}(2023){Ferreira}, {Conselice}, {Sazonova},
  {Ferrari}, {Caruana}, {Tohill}, {Lucatelli}, {Adams}, {Irodotou}, {Marshall},
  {Roper}, {Lovell}, {Verma}, {Austin}, {Trussler}, \& {Wilkins}}]{ferreira22b}
{Ferreira}, L., {Conselice}, C.~J., {Sazonova}, E., {et~al.} 2023, \apj, 955,
  94

\bibitem[{{Fierlinger} {et~al.}(2016){Fierlinger}, {Burkert}, {Ntormousi},
  {Fierlinger}, {Schartmann}, {Ballone}, {Krause}, \& {Diehl}}]{fierlinger16}
{Fierlinger}, K.~M., {Burkert}, A., {Ntormousi}, E., {et~al.} 2016, \mnras,
  456, 710

\bibitem[{{Fraternali} {et~al.}(2021){Fraternali}, {Karim}, {Magnelli},
  {G{\'o}mez-Guijarro}, {Jim{\'e}nez-Andrade}, \& {Posses}}]{fraternali21}
{Fraternali}, F., {Karim}, A., {Magnelli}, B., {et~al.} 2021, \aap, 647, A194

\bibitem[{{Fudamoto} {et~al.}(2022){Fudamoto}, {Inoue}, \&
  {Sugahara}}]{fudamoto22}
{Fudamoto}, Y., {Inoue}, A.~K., \& {Sugahara}, Y. 2022, \apjl, 938, L24

\bibitem[{{Fudamoto} {et~al.}(2017){Fudamoto}, {Ivison}, {Oteo}, {Krips},
  {Zhang}, {Weiss}, {Dannerbauer}, {Omont}, {Chapman}, {Christensen},
  {Arumugam}, {Bertoldi}, {Bremer}, {Clements}, {Dunne}, {Eales}, {Greenslade},
  {Maddox}, {Martinez-Navajas}, {Michalowski}, {P{\'e}rez-Fournon}, {Riechers},
  {Simpson}, {Stalder}, {Valiante}, \& {van der Werf}}]{fudamoto17}
{Fudamoto}, Y., {Ivison}, R.~J., {Oteo}, I., {et~al.} 2017, \mnras, 472, 2028

\bibitem[{{Genzel} {et~al.}(2011){Genzel}, {Newman}, {Jones}, {F{\"o}rster
  Schreiber}, {Shapiro}, {Genel}, {Lilly}, {Renzini}, {Tacconi}, {Bouch{\'e}},
  {Burkert}, {Cresci}, {Buschkamp}, {Carollo}, {Ceverino}, {Davies}, {Dekel},
  {Eisenhauer}, {Hicks}, {Kurk}, {Lutz}, {Mancini}, {Naab}, {Peng},
  {Sternberg}, {Vergani}, \& {Zamorani}}]{genzel11}
{Genzel}, R., {Newman}, S., {Jones}, T., {et~al.} 2011, \apj, 733, 101

\bibitem[{{Gillman} {et~al.}(2023){Gillman}, {Gullberg}, {Brammer}, {Vijayan},
  {Lee}, {Bl{\'a}nquez}, {Brinch}, {Greve}, {Jermann}, {Jin}, {Kokorev}, {Liu},
  {Magdis}, {Rizzo}, \& {Valentino}}]{gillman23}
{Gillman}, S., {Gullberg}, B., {Brammer}, G., {et~al.} 2023, \aap, 676, A26

\bibitem[{{Gong} {et~al.}(2020){Gong}, {Ostriker}, {Kim}, \& {Kim}}]{gong20}
{Gong}, M., {Ostriker}, E.~C., {Kim}, C.-G., \& {Kim}, J.-G. 2020, \apj, 903,
  142

\bibitem[{{Governato} {et~al.}(2007){Governato}, {Willman}, {Mayer}, {Brooks},
  {Stinson}, {Valenzuela}, {Wadsley}, \& {Quinn}}]{governato07}
{Governato}, F., {Willman}, B., {Mayer}, L., {et~al.} 2007, \mnras, 374, 1479

\bibitem[{{Graham} \& {Driver}(2005)}]{graham05}
{Graham}, A.~W. \& {Driver}, S.~P. 2005, \pasa, 22, 118

\bibitem[{{Harrison} {et~al.}(2017){Harrison}, {Johnson}, {Swinbank}, {Stott},
  {Bower}, {Smail}, {Tiley}, {Bunker}, {Cirasuolo}, {Sobral}, {Sharples},
  {Best}, {Bureau}, {Jarvis}, \& {Magdis}}]{harrison17}
{Harrison}, C.~M., {Johnson}, H.~L., {Swinbank}, A.~M., {et~al.} 2017, \mnras,
  467, 1965

\bibitem[{{Hilz} {et~al.}(2013){Hilz}, {Naab}, \& {Ostriker}}]{hilz13}
{Hilz}, M., {Naab}, T., \& {Ostriker}, J.~P. 2013, \mnras, 429, 2924

\bibitem[{{Hodge} {et~al.}(2012){Hodge}, {Carilli}, {Walter}, {de Blok},
  {Riechers}, {Daddi}, \& {Lentati}}]{hodge12}
{Hodge}, J.~A., {Carilli}, C.~L., {Walter}, F., {et~al.} 2012, \apj, 760, 11

\bibitem[{{Hopkins} {et~al.}(2014){Hopkins}, {Kere{\v{s}}}, {O{\~n}orbe},
  {Faucher-Gigu{\`e}re}, {Quataert}, {Murray}, \& {Bullock}}]{hopkins14}
{Hopkins}, P.~F., {Kere{\v{s}}}, D., {O{\~n}orbe}, J., {et~al.} 2014, \mnras,
  445, 581

\bibitem[{{Huang} {et~al.}(2023){Huang}, {Kawabe}, {Kohno}, {Saito},
  {Mizukoshi}, {Iono}, {Michiyama}, {Tamura}, {Hayward}, \&
  {Umehata}}]{huang23}
{Huang}, S., {Kawabe}, R., {Kohno}, K., {et~al.} 2023, \apjl, 958, L26

\bibitem[{{Iorio} {et~al.}(2017){Iorio}, {Fraternali}, {Nipoti}, {Di Teodoro},
  {Read}, \& {Battaglia}}]{iorio17}
{Iorio}, G., {Fraternali}, F., {Nipoti}, C., {et~al.} 2017, \mnras, 466, 4159

\bibitem[{{Ishibashi} \& {Fabian}(2014)}]{ishibashi14}
{Ishibashi}, W. \& {Fabian}, A.~C. 2014, \mnras, 441, 1474

\bibitem[{{Ito} {et~al.}(2023){Ito}, {Valentino}, {Brammer}, {Faisst},
  {Gillman}, {Gomez-Guijarro}, {Gould}, {Heintz}, {Ilbert}, {Kragh Jespersen},
  {Kokorev}, {Kubo}, {Magdis}, {McPartland}, {Onodera}, {Rizzo}, {Tanaka},
  {Toft}, {Vijayan}, {Weaver}, {Whitaker}, \& {Wright}}]{ito23}
{Ito}, K., {Valentino}, F., {Brammer}, G., {et~al.} 2023, arXiv e-prints,
  arXiv:2307.06994

\bibitem[{{Jacobs} {et~al.}(2023){Jacobs}, {Glazebrook}, {Calabr{\`o}}, {Treu},
  {Nannayakkara}, {Jones}, {Merlin}, {Abraham}, {Stevens}, {Vulcani}, {Yang},
  {Bonchi}, {Boyett}, {Brada{\v{c}}}, {Castellano}, {Fontana}, {Marchesini},
  {Malkan}, {Mason}, {Morishita}, {Paris}, {Santini}, {Trenti}, \&
  {Wang}}]{jacobs23}
{Jacobs}, C., {Glazebrook}, K., {Calabr{\`o}}, A., {et~al.} 2023, \apjl, 948,
  L13

\bibitem[{{Jones} {et~al.}(2016){Jones}, {Carilli}, {Momjian}, {Wagg},
  {Riechers}, {Walter}, {Decarli}, {Ota}, \& {McMahon}}]{jones16}
{Jones}, G.~C., {Carilli}, C.~L., {Momjian}, E., {et~al.} 2016, \apj, 830, 63

\bibitem[{{Jones} {et~al.}(2021){Jones}, {Vergani}, {Romano}, {Ginolfi},
  {Fudamoto}, {B{\'e}thermin}, {Fujimoto}, {Lemaux}, {Morselli}, {Capak},
  {Cassata}, {Faisst}, {Le F{\`e}vre}, {Schaerer}, {Silverman}, {Yan},
  {Boquien}, {Cimatti}, {Dessauges-Zavadsky}, {Ibar}, {Maiolino}, {Rizzo},
  {Talia}, \& {Zamorani}}]{jones21}
{Jones}, G.~C., {Vergani}, D., {Romano}, M., {et~al.} 2021, \mnras, 507, 3540

\bibitem[{{Kaasinen} {et~al.}(2020){Kaasinen}, {Walter}, {Novak}, {Neeleman},
  {Smail}, {Boogaard}, {Cunha}, {Weiss}, {Liu}, {Decarli}, {Popping},
  {Diaz-Santos}, {Cort{\'e}s}, {Aravena}, {Werf}, {Riechers}, {Inami}, {Hodge},
  {Rix}, \& {Cox}}]{kaasinen20}
{Kaasinen}, M., {Walter}, F., {Novak}, M., {et~al.} 2020, \apj, 899, 37

\bibitem[{{Kamenetzky} {et~al.}(2016){Kamenetzky}, {Rangwala}, {Glenn},
  {Maloney}, \& {Conley}}]{kamenetzky16}
{Kamenetzky}, J., {Rangwala}, N., {Glenn}, J., {Maloney}, P.~R., \& {Conley},
  A. 2016, \apj, 829, 93

\bibitem[{{Kartaltepe} {et~al.}(2023){Kartaltepe}, {Rose}, {Vanderhoof},
  {McGrath}, {Costantin}, {Cox}, {Yung}, {Kocevski}, {Wuyts}, {Ferguson},
  {Bagley}, {Finkelstein}, {Amor{\'\i}n}, {Andrews}, {Arrabal Haro},
  {Backhaus}, {Behroozi}, {Bisigello}, {Calabr{\`o}}, {Casey}, {Coogan},
  {Cooper}, {Croton}, {de la Vega}, {Dickinson}, {Fontana}, {Franco},
  {Grazian}, {Grogin}, {Hathi}, {Holwerda}, {Huertas-Company}, {Iyer}, {Jogee},
  {Jung}, {Kewley}, {Kirkpatrick}, {Koekemoer}, {Liu}, {Lotz}, {Lucas},
  {Newman}, {Pacifici}, {Pandya}, {Papovich}, {Pentericci},
  {P{\'e}rez-Gonz{\'a}lez}, {Petersen}, {Pirzkal}, {Rafelski}, {Ravindranath},
  {Simons}, {Snyder}, {Somerville}, {Stanway}, {Straughn}, {Tacchella},
  {Trump}, {Vega-Ferrero}, {Wilkins}, {Yang}, \& {Zavala}}]{kartaltepe23}
{Kartaltepe}, J.~S., {Rose}, C., {Vanderhoof}, B.~N., {et~al.} 2023, \apjl,
  946, L15

\bibitem[{{Kohandel} {et~al.}(2020){Kohandel}, {Pallottini}, {Ferrara},
  {Carniani}, {Gallerani}, {Vallini}, {Zanella}, \& {Behrens}}]{kohandel20}
{Kohandel}, M., {Pallottini}, A., {Ferrara}, A., {et~al.} 2020, \mnras, 499,
  1250

\bibitem[{{Koposov} {et~al.}(2022){Koposov}, {Speagle}, {Barbary}, {Ashton},
  {Buchner}, {Scheffler}, {Cook}, {Talbot}, {Guillochon}, {Cubillos}, {Asensio
  Ramos}, {Johnson}, {Lang}, {Ilya}, {Dartiailh}, {Nitz}, {McCluskey},
  {Archibald}, {Deil}, {Foreman-Mackey}, {Goldstein}, {Tollerud}, {Leja},
  {Kirk}, {Pitkin}, {Sheehan}, {Cargile}, {Ruskin23}, {Angus}, \&
  {Daylan}}]{koposov22}
{Koposov}, S., {Speagle}, J., {Barbary}, K., {et~al.} 2022,
  {joshspeagle/dynesty: v1.2.3}, Zenodo

\bibitem[{{Kormendy} \& {Ho}(2013)}]{kormendy13}
{Kormendy}, J. \& {Ho}, L.~C. 2013, \araa, 51, 511

\bibitem[{{Kormendy} \& {Kennicutt}(2004)}]{kormendy04}
{Kormendy}, J. \& {Kennicutt}, Robert~C., J. 2004, \araa, 42, 603

\bibitem[{{Kretschmer} {et~al.}(2022){Kretschmer}, {Dekel}, \&
  {Teyssier}}]{kretschmer22}
{Kretschmer}, M., {Dekel}, A., \& {Teyssier}, R. 2022, \mnras, 510, 3266

\bibitem[{{Krumholz} \& {Burkhart}(2016)}]{krumholz16}
{Krumholz}, M.~R. \& {Burkhart}, B. 2016, \mnras, 458, 1671

\bibitem[{{Krumholz} {et~al.}(2018){Krumholz}, {Burkhart}, {Forbes}, \&
  {Crocker}}]{krumholz18}
{Krumholz}, M.~R., {Burkhart}, B., {Forbes}, J.~C., \& {Crocker}, R.~M. 2018,
  \mnras, 477, 2716

\bibitem[{{Lagache} {et~al.}(2018){Lagache}, {Cousin}, \&
  {Chatzikos}}]{lagache18}
{Lagache}, G., {Cousin}, M., \& {Chatzikos}, M. 2018, \aap, 609, A130

\bibitem[{{Le Conte} {et~al.}(2023){Le Conte}, {Gadotti}, {Ferreira},
  {Conselice}, {de S{\'a}-Freitas}, {Kim}, {Neumann}, {Fragkoudi},
  {Athanassoula}, \& {Adams}}]{leconte23}
{Le Conte}, Z.~A., {Gadotti}, D.~A., {Ferreira}, L., {et~al.} 2023, arXiv
  e-prints, arXiv:2309.10038

\bibitem[{{Lelli} {et~al.}(2021){Lelli}, {Di Teodoro}, {Fraternali}, {Man},
  {Zhang}, {De Breuck}, {Davis}, \& {Maiolino}}]{lelli21}
{Lelli}, F., {Di Teodoro}, E.~M., {Fraternali}, F., {et~al.} 2021, Science,
  371, 713

\bibitem[{{Lelli} {et~al.}(2016{\natexlab{a}}){Lelli}, {McGaugh}, \&
  {Schombert}}]{lelli16b}
{Lelli}, F., {McGaugh}, S.~S., \& {Schombert}, J.~M. 2016{\natexlab{a}}, \aj,
  152, 157

\bibitem[{{Lelli} {et~al.}(2016{\natexlab{b}}){Lelli}, {McGaugh}, \&
  {Schombert}}]{lelli16}
{Lelli}, F., {McGaugh}, S.~S., \& {Schombert}, J.~M. 2016{\natexlab{b}}, \apjl,
  816, L14

\bibitem[{{Lelli} {et~al.}(2017){Lelli}, {McGaugh}, {Schombert}, \&
  {Pawlowski}}]{lelli17}
{Lelli}, F., {McGaugh}, S.~S., {Schombert}, J.~M., \& {Pawlowski}, M.~S. 2017,
  \apj, 836, 152

\bibitem[{{Lu} {et~al.}(2018){Lu}, {Cao}, {D{\'\i}az-Santos}, {Zhao}, {Privon},
  {Cheng}, {Gao}, {Xu}, {Charmandaris}, {Rigopoulou}, {van der Werf}, {Huang},
  {Wang}, {Evans}, \& {Sanders}}]{lu18}
{Lu}, N., {Cao}, T., {D{\'\i}az-Santos}, T., {et~al.} 2018, \apj, 864, 38

\bibitem[{{Lustig} {et~al.}(2021){Lustig}, {Strazzullo}, {D'Eugenio}, {Daddi},
  {Pannella}, {Renzini}, {Cimatti}, {Gobat}, {Jin}, {Mohr}, \&
  {Onodera}}]{lustig21}
{Lustig}, P., {Strazzullo}, V., {D'Eugenio}, C., {et~al.} 2021, \mnras, 501,
  2659

\bibitem[{{Ma} {et~al.}(2014){Ma}, {Greene}, {McConnell}, {Janish},
  {Blakeslee}, {Thomas}, \& {Murphy}}]{ma14}
{Ma}, C.-P., {Greene}, J.~E., {McConnell}, N., {et~al.} 2014, \apj, 795, 158

\bibitem[{{Mac Low}(1999)}]{maclow99}
{Mac Low}, M.-M. 1999, \apj, 524, 169

\bibitem[{{Marasco} {et~al.}(2019){Marasco}, {Fraternali}, {Heald}, {de Blok},
  {Oosterloo}, {Kamphuis}, {J{\'o}zsa}, {Vargas}, {Winkel}, {Walterbos},
  {Dettmar}, \& {Juẗte}}]{marasco19}
{Marasco}, A., {Fraternali}, F., {Heald}, G., {et~al.} 2019, \aap, 631, A50

\bibitem[{{Marasco} {et~al.}(2020){Marasco}, {Posti}, {Oman}, {Famaey},
  {Cresci}, \& {Fraternali}}]{marasco20}
{Marasco}, A., {Posti}, L., {Oman}, K., {et~al.} 2020, \aap, 640, A70

\bibitem[{{McGaugh}(2004)}]{mcgaugh04}
{McGaugh}, S.~S. 2004, \apj, 609, 652

\bibitem[{{McGaugh} \& {Schombert}(2015)}]{mcgaugh15}
{McGaugh}, S.~S. \& {Schombert}, J.~M. 2015, \apj, 802, 18

\bibitem[{{McGaugh} {et~al.}(2000){McGaugh}, {Schombert}, {Bothun}, \& {de
  Blok}}]{mcgaugh00}
{McGaugh}, S.~S., {Schombert}, J.~M., {Bothun}, G.~D., \& {de Blok}, W.~J.~G.
  2000, \apjl, 533, L99

\bibitem[{{McMullin} {et~al.}(2007){McMullin}, {Waters}, {Schiebel}, {Young},
  \& {Golap}}]{mcmullin07}
{McMullin}, J.~P., {Waters}, B., {Schiebel}, D., {Young}, W., \& {Golap}, K.
  2007, in Astronomical Society of the Pacific Conference Series, Vol. 376,
  Astronomical Data Analysis Software and Systems XVI, ed. R.~A. {Shaw},
  F.~{Hill}, \& D.~J. {Bell}, 127

\bibitem[{{Mo} {et~al.}(2010){Mo}, {van den Bosch}, \& {White}}]{mo10}
{Mo}, H., {van den Bosch}, F.~C., \& {White}, S. 2010, {Galaxy Formation and
  Evolution}

\bibitem[{{Moster} {et~al.}(2010){Moster}, {Somerville}, {Maulbetsch}, {van den
  Bosch}, {Macci{\`o}}, {Naab}, \& {Oser}}]{moster10}
{Moster}, B.~P., {Somerville}, R.~S., {Maulbetsch}, C., {et~al.} 2010, \apj,
  710, 903

\bibitem[{{Naab} {et~al.}(2009){Naab}, {Johansson}, \& {Ostriker}}]{naab09b}
{Naab}, T., {Johansson}, P.~H., \& {Ostriker}, J.~P. 2009, \apjl, 699, L178

\bibitem[{{Naab} \& {Ostriker}(2009)}]{naab09a}
{Naab}, T. \& {Ostriker}, J.~P. 2009, \apj, 690, 1452

\bibitem[{{Narayanan} {et~al.}(2011){Narayanan}, {Krumholz}, {Ostriker}, \&
  {Hernquist}}]{narayanan11}
{Narayanan}, D., {Krumholz}, M., {Ostriker}, E.~C., \& {Hernquist}, L. 2011,
  \mnras, 418, 664

\bibitem[{{Navarro} {et~al.}(1997){Navarro}, {Frenk}, \& {White}}]{navarro97}
{Navarro}, J.~F., {Frenk}, C.~S., \& {White}, S. D.~M. 1997, \apj, 490, 493

\bibitem[{{Neeleman} {et~al.}(2019){Neeleman}, {Kanekar}, {Prochaska},
  {Rafelski}, \& {Carilli}}]{neeleman19}
{Neeleman}, M., {Kanekar}, N., {Prochaska}, J.~X., {Rafelski}, M.~A., \&
  {Carilli}, C.~L. 2019, \apjl, 870, L19

\bibitem[{{Neeleman} {et~al.}(2020){Neeleman}, {Prochaska}, {Kanekar}, \&
  {Rafelski}}]{neeleman20}
{Neeleman}, M., {Prochaska}, J.~X., {Kanekar}, N., \& {Rafelski}, M. 2020,
  \nat, 581, 269

\bibitem[{{Neeleman} {et~al.}(2023){Neeleman}, {Walter}, {Decarli}, {Drake},
  {Eilers}, {Meyer}, \& {Venemans}}]{neeleman23}
{Neeleman}, M., {Walter}, F., {Decarli}, R., {et~al.} 2023, \apj, 958, 132

\bibitem[{{Nelson} {et~al.}(2019){Nelson}, {Pillepich}, {Springel}, {Pakmor},
  {Weinberger}, {Genel}, {Torrey}, {Vogelsberger}, {Marinacci}, \&
  {Hernquist}}]{nelson19}
{Nelson}, D., {Pillepich}, A., {Springel}, V., {et~al.} 2019, \mnras, 490, 3234

\bibitem[{{Nelson} {et~al.}(2023){Nelson}, {Suess}, {Bezanson}, {Price}, {van
  Dokkum}, {Leja}, {Wang}, {Whitaker}, {Labb{\'e}}, {Barrufet}, {Brammer},
  {Eisenstein}, {Gibson}, {Hartley}, {Johnson}, {Heintz}, {Mathews}, {Miller},
  {Oesch}, {Sandles}, {Setton}, {Speagle}, {Tacchella}, {Tadaki}, {{\"U}bler},
  \& {Weaver}}]{nelson23}
{Nelson}, E.~J., {Suess}, K.~A., {Bezanson}, R., {et~al.} 2023, \apjl, 948, L18

\bibitem[{{Ogle} {et~al.}(2019){Ogle}, {Lanz}, {Appleton}, {Helou}, \&
  {Mazzarella}}]{ogle19}
{Ogle}, P.~M., {Lanz}, L., {Appleton}, P.~N., {Helou}, G., \& {Mazzarella}, J.
  2019, \apjs, 243, 14

\bibitem[{{Ohlin} {et~al.}(2019){Ohlin}, {Renaud}, \& {Agertz}}]{ohlin19}
{Ohlin}, L., {Renaud}, F., \& {Agertz}, O. 2019, \mnras, 485, 3887

\bibitem[{{Ormerod} {et~al.}(2024){Ormerod}, {Conselice}, {Adams}, {Harvey},
  {Austin}, {Trussler}, {Ferreira}, {Caruana}, {Lucatelli}, {Li}, \&
  {Roper}}]{ormerod24}
{Ormerod}, K., {Conselice}, C.~J., {Adams}, N.~J., {et~al.} 2024, \mnras, 527,
  6110

\bibitem[{{Orr} {et~al.}(2020){Orr}, {Hayward}, {Medling}, {Gurvich},
  {Hopkins}, {Murray}, {Pineda}, {Faucher-Gigu{\`e}re}, {Kere{\v{s}}},
  {Wetzel}, \& {Su}}]{orr20}
{Orr}, M.~E., {Hayward}, C.~C., {Medling}, A.~M., {et~al.} 2020, \mnras, 496,
  1620

\bibitem[{{Oteo} {et~al.}(2016){Oteo}, {Ivison}, {Dunne}, {Smail}, {Swinbank},
  {Zhang}, {Lewis}, {Maddox}, {Riechers}, {Serjeant}, {Van der Werf}, {Biggs},
  {Bremer}, {Cigan}, {Clements}, {Cooray}, {Dannerbauer}, {Eales}, {Ibar},
  {Messias}, {Micha{\l}owski}, {P{\'e}rez-Fournon}, \& {van Kampen}}]{oteo16}
{Oteo}, I., {Ivison}, R.~J., {Dunne}, L., {et~al.} 2016, \apj, 827, 34

\bibitem[{{Papadopoulos} {et~al.}(2018){Papadopoulos}, {Bisbas}, \&
  {Zhang}}]{papadopoulos18}
{Papadopoulos}, P.~P., {Bisbas}, T.~G., \& {Zhang}, Z.-Y. 2018, \mnras, 478,
  1716

\bibitem[{{Planck Collaboration} {et~al.}(2020){Planck Collaboration},
  {Aghanim}, {Akrami}, {Arroja}, {Ashdown}, {Aumont}, {Baccigalupi},
  {Ballardini}, {Banday}, {Barreiro}, {Bartolo}, {Basak}, {Battye}, {Benabed},
  {Bernard}, {Bersanelli}, {Bielewicz}, {Bock}, {Bond}, {Borrill}, {Bouchet},
  {Boulanger}, {Bucher}, {Burigana}, {Butler}, {Calabrese}, {Cardoso},
  {Carron}, {Casaponsa}, {Challinor}, {Chiang}, {Colombo}, {Combet},
  {Contreras}, {Crill}, {Cuttaia}, {de Bernardis}, {de Zotti}, {Delabrouille},
  {Delouis}, {D{\'e}sert}, {Di Valentino}, {Dickinson}, {Diego}, {Donzelli},
  {Dor{\'e}}, {Douspis}, {Ducout}, {Dupac}, {Efstathiou}, {Elsner},
  {En{\ss}lin}, {Eriksen}, {Falgarone}, {Fantaye}, {Fergusson},
  {Fernandez-Cobos}, {Finelli}, {Forastieri}, {Frailis}, {Franceschi},
  {Frolov}, {Galeotta}, {Galli}, {Ganga}, {G{\'e}nova-Santos}, {Gerbino},
  {Ghosh}, {Gonz{\'a}lez-Nuevo}, {G{\'o}rski}, {Gratton}, {Gruppuso},
  {Gudmundsson}, {Hamann}, {Handley}, {Hansen}, {Helou}, {Herranz},
  {Hildebrandt}, {Hivon}, {Huang}, {Jaffe}, {Jones}, {Karakci}, {Keih{\"a}nen},
  {Keskitalo}, {Kiiveri}, {Kim}, {Kisner}, {Knox}, {Krachmalnicoff}, {Kunz},
  {Kurki-Suonio}, {Lagache}, {Lamarre}, {Langer}, {Lasenby}, {Lattanzi},
  {Lawrence}, {Le Jeune}, {Leahy}, {Lesgourgues}, {Levrier}, {Lewis},
  {Liguori}, {Lilje}, {Lilley}, {Lindholm}, {L{\'o}pez-Caniego}, {Lubin}, {Ma},
  {Mac{\'\i}as-P{\'e}rez}, {Maggio}, {Maino}, {Mandolesi}, {Mangilli},
  {Marcos-Caballero}, {Maris}, {Martin}, {Martinelli},
  {Mart{\'\i}nez-Gonz{\'a}lez}, {Matarrese}, {Mauri}, {McEwen}, {Meerburg},
  {Meinhold}, {Melchiorri}, {Mennella}, {Migliaccio}, {Millea}, {Mitra},
  {Miville-Desch{\^e}nes}, {Molinari}, {Moneti}, {Montier}, {Morgante}, {Moss},
  {Mottet}, {M{\"u}nchmeyer}, {Natoli}, {N{\o}rgaard-Nielsen}, {Oxborrow},
  {Pagano}, {Paoletti}, {Partridge}, {Patanchon}, {Pearson}, {Peel}, {Peiris},
  {Perrotta}, {Pettorino}, {Piacentini}, {Polastri}, {Polenta}, {Puget},
  {Rachen}, {Reinecke}, {Remazeilles}, {Renault}, {Renzi}, {Rocha}, {Rosset},
  {Roudier}, {Rubi{\~n}o-Mart{\'\i}n}, {Ruiz-Granados}, {Salvati}, {Sandri},
  {Savelainen}, {Scott}, {Shellard}, {Shiraishi}, {Sirignano}, {Sirri},
  {Spencer}, {Sunyaev}, {Suur-Uski}, {Tauber}, {Tavagnacco}, {Tenti},
  {Terenzi}, {Toffolatti}, {Tomasi}, {Trombetti}, {Valiviita}, {Van Tent},
  {Vibert}, {Vielva}, {Villa}, {Vittorio}, {Wandelt}, {Wehus}, {White},
  {White}, {Zacchei}, \& {Zonca}}]{planck20}
{Planck Collaboration}, {Aghanim}, N., {Akrami}, Y., {et~al.} 2020, \aap, 641,
  A1

\bibitem[{{Popping} {et~al.}(2022){Popping}, {Pillepich}, {Calistro Rivera},
  {Schulz}, {Hernquist}, {Kaasinen}, {Marinacci}, {Nelson}, \&
  {Vogelsberger}}]{popping22}
{Popping}, G., {Pillepich}, A., {Calistro Rivera}, G., {et~al.} 2022, \mnras,
  510, 3321

\bibitem[{{Posti} {et~al.}(2018){Posti}, {Fraternali}, {Di Teodoro}, \&
  {Pezzulli}}]{posti18}
{Posti}, L., {Fraternali}, F., {Di Teodoro}, E.~M., \& {Pezzulli}, G. 2018,
  \aap, 612, L6

\bibitem[{{Posti} {et~al.}(2019){Posti}, {Marasco}, {Fraternali}, \&
  {Famaey}}]{posti19}
{Posti}, L., {Marasco}, A., {Fraternali}, F., \& {Famaey}, B. 2019, \aap, 629,
  A59

\bibitem[{{Renaud} {et~al.}(2014){Renaud}, {Bournaud}, {Kraljic}, \&
  {Duc}}]{renaud14}
{Renaud}, F., {Bournaud}, F., {Kraljic}, K., \& {Duc}, P.~A. 2014, \mnras, 442,
  L33

\bibitem[{{Rinaldi} {et~al.}(2022){Rinaldi}, {Caputi}, {van Mierlo}, {Ashby},
  {Caminha}, \& {Iani}}]{rinaldi22}
{Rinaldi}, P., {Caputi}, K.~I., {van Mierlo}, S.~E., {et~al.} 2022, \apj, 930,
  128

\bibitem[{{Rizzo} {et~al.}(2022){Rizzo}, {Kohandel}, {Pallottini}, {Zanella},
  {Ferrara}, {Vallini}, \& {Toft}}]{rizzo22}
{Rizzo}, F., {Kohandel}, M., {Pallottini}, A., {et~al.} 2022, \aap, 667, A5

\bibitem[{{Rizzo} {et~al.}(2021){Rizzo}, {Vegetti}, {Fraternali}, {Stacey}, \&
  {Powell}}]{rizzo21}
{Rizzo}, F., {Vegetti}, S., {Fraternali}, F., {Stacey}, H.~R., \& {Powell}, D.
  2021, \mnras, 507, 3952

\bibitem[{{Rizzo} {et~al.}(2020){Rizzo}, {Vegetti}, {Powell}, {Fraternali},
  {McKean}, {Stacey}, \& {White}}]{rizzo20}
{Rizzo}, F., {Vegetti}, S., {Powell}, D., {et~al.} 2020, \nat, 584, 201

\bibitem[{{Robertson} {et~al.}(2023){Robertson}, {Tacchella}, {Johnson},
  {Hausen}, {Alabi}, {Boyett}, {Bunker}, {Carniani}, {Egami}, {Eisenstein},
  {Hainline}, {Helton}, {Ji}, {Kumari}, {Lyu}, {Maiolino}, {Nelson}, {Rieke},
  {Shivaei}, {Sun}, {{\"U}bler}, {Williams}, {Willmer}, \&
  {Witstok}}]{robertson23}
{Robertson}, B.~E., {Tacchella}, S., {Johnson}, B.~D., {et~al.} 2023, \apjl,
  942, L42

\bibitem[{{Rodighiero} {et~al.}(2011){Rodighiero}, {Daddi}, {Baronchelli},
  {Cimatti}, {Renzini}, {Aussel}, {Popesso}, {Lutz}, {Andreani}, {Berta},
  {Cava}, {Elbaz}, {Feltre}, {Fontana}, {F{\"o}rster Schreiber},
  {Franceschini}, {Genzel}, {Grazian}, {Gruppioni}, {Ilbert}, {Le Floch},
  {Magdis}, {Magliocchetti}, {Magnelli}, {Maiolino}, {McCracken}, {Nordon},
  {Poglitsch}, {Santini}, {Pozzi}, {Riguccini}, {Tacconi}, {Wuyts}, \&
  {Zamorani}}]{rodighiero11}
{Rodighiero}, G., {Daddi}, E., {Baronchelli}, I., {et~al.} 2011, \apjl, 739,
  L40

\bibitem[{{Rodney} {et~al.}(2014){Rodney}, {Riess}, {Strolger}, {Dahlen},
  {Graur}, {Casertano}, {Dickinson}, {Ferguson}, {Garnavich}, {Hayden}, {Jha},
  {Jones}, {Kirshner}, {Koekemoer}, {McCully}, {Mobasher}, {Patel}, {Weiner},
  {Cenko}, {Clubb}, {Cooper}, {Filippenko}, {Frederiksen}, {Hjorth},
  {Leibundgut}, {Matheson}, {Nayyeri}, {Penner}, {Trump}, {Silverman}, {U},
  {Azalee Bostroem}, {Challis}, {Rajan}, {Wolff}, {Faber}, {Grogin}, \&
  {Kocevski}}]{rodney14}
{Rodney}, S.~A., {Riess}, A.~G., {Strolger}, L.-G., {et~al.} 2014, \aj, 148, 13

\bibitem[{{Rodriguez-Gomez} {et~al.}(2022){Rodriguez-Gomez}, {Genel}, {Fall},
  {Pillepich}, {Huertas-Company}, {Nelson}, {P{\'e}rez-Monta{\~n}o},
  {Marinacci}, {Pakmor}, {Springel}, {Vogelsberger}, \&
  {Hernquist}}]{rodriguez-gomez22}
{Rodriguez-Gomez}, V., {Genel}, S., {Fall}, S.~M., {et~al.} 2022, \mnras, 512,
  5978

\bibitem[{{Roman-Oliveira} {et~al.}(2023){Roman-Oliveira}, {Fraternali}, \&
  {Rizzo}}]{romanoliveira23}
{Roman-Oliveira}, F., {Fraternali}, F., \& {Rizzo}, F. 2023, \mnras, 521, 1045

\bibitem[{{Rubin} {et~al.}(1980){Rubin}, {Ford}, \& {Thonnard}}]{rubin80}
{Rubin}, V.~C., {Ford}, W.~K., J., \& {Thonnard}, N. 1980, \apj, 238, 471

\bibitem[{{Schaye} {et~al.}(2015){Schaye}, {Crain}, {Bower}, {Furlong},
  {Schaller}, {Theuns}, {Dalla Vecchia}, {Frenk}, {McCarthy}, {Helly},
  {Jenkins}, {Rosas-Guevara}, {White}, {Baes}, {Booth}, {Camps}, {Navarro},
  {Qu}, {Rahmati}, {Sawala}, {Thomas}, \& {Trayford}}]{schaye15}
{Schaye}, J., {Crain}, R.~A., {Bower}, R.~G., {et~al.} 2015, \mnras, 446, 521

\bibitem[{{Scoville} {et~al.}(2017){Scoville}, {Lee}, {Vanden Bout},
  {Diaz-Santos}, {Sanders}, {Darvish}, {Bongiorno}, {Casey}, {Murchikova},
  {Koda}, {Capak}, {Vlahakis}, {Ilbert}, {Sheth}, {Morokuma-Matsui}, {Ivison},
  {Aussel}, {Laigle}, {McCracken}, {Armus}, {Pope}, {Toft}, \&
  {Masters}}]{scoville17}
{Scoville}, N., {Lee}, N., {Vanden Bout}, P., {et~al.} 2017, \apj, 837, 150

\bibitem[{{Sersic}(1968)}]{sersic68}
{Sersic}, J.~L. 1968, {Atlas de Galaxias Australes}

\bibitem[{{Shao} {et~al.}(2022){Shao}, {Wang}, {Weiss}, {Wagg}, {Carilli},
  {Strauss}, {Walter}, {Cox}, {Fan}, {Menten}, {Narayanan}, {Riechers},
  {Bertoldi}, {Omont}, \& {Jiang}}]{shao22}
{Shao}, Y., {Wang}, R., {Weiss}, A., {et~al.} 2022, \aap, 668, A121

\bibitem[{{Shuntov} {et~al.}(2022){Shuntov}, {McCracken}, {Gavazzi}, {Laigle},
  {Weaver}, {Davidzon}, {Ilbert}, {Kauffmann}, {Faisst}, {Dubois}, {Koekemoer},
  {Moneti}, {Milvang-Jensen}, {Mobasher}, {Sanders}, \& {Toft}}]{shuntov22}
{Shuntov}, M., {McCracken}, H.~J., {Gavazzi}, R., {et~al.} 2022, \aap, 664, A61

\bibitem[{{Skilling}(2004)}]{skilling04}
{Skilling}, J. 2004, in American Institute of Physics Conference Series, Vol.
  735, Bayesian Inference and Maximum Entropy Methods in Science and
  Engineering: 24th International Workshop on Bayesian Inference and Maximum
  Entropy Methods in Science and Engineering, ed. R.~{Fischer}, R.~{Preuss}, \&
  U.~V. {Toussaint}, 395--405

\bibitem[{Skilling(2006)}]{skilling06}
Skilling, J. 2006, Bayesian Analysis, 1, 833

\bibitem[{{Smit} {et~al.}(2018){Smit}, {Bouwens}, {Carniani}, {Oesch},
  {Labb{\'e}}, {Illingworth}, {van der Werf}, {Bradley}, {Gonzalez}, {Hodge},
  {Holwerda}, {Maiolino}, \& {Zheng}}]{smit18}
{Smit}, R., {Bouwens}, R.~J., {Carniani}, S., {et~al.} 2018, \nat, 553, 178

\bibitem[{{Speagle}(2020)}]{speagle20}
{Speagle}, J.~S. 2020, \mnras, 493, 3132

\bibitem[{{Speagle} {et~al.}(2014){Speagle}, {Steinhardt}, {Capak}, \&
  {Silverman}}]{speagle14}
{Speagle}, J.~S., {Steinhardt}, C.~L., {Capak}, P.~L., \& {Silverman}, J.~D.
  2014, \apjs, 214, 15

\bibitem[{{Tacchella} {et~al.}(2015){Tacchella}, {Carollo}, {Renzini},
  {F{\"o}rster Schreiber}, {Lang}, {Wuyts}, {Cresci}, {Dekel}, {Genzel},
  {Lilly}, {Mancini}, {Newman}, {Onodera}, {Shapley}, {Tacconi}, {Woo}, \&
  {Zamorani}}]{tacchella15}
{Tacchella}, S., {Carollo}, C.~M., {Renzini}, A., {et~al.} 2015, Science, 348,
  314

\bibitem[{{Tacchella} {et~al.}(2016){Tacchella}, {Dekel}, {Carollo},
  {Ceverino}, {DeGraf}, {Lapiner}, {Mandelker}, \& {Primack
  Joel}}]{tacchella16}
{Tacchella}, S., {Dekel}, A., {Carollo}, C.~M., {et~al.} 2016, \mnras, 457,
  2790

\bibitem[{{Tacconi} {et~al.}(2010){Tacconi}, {Genzel}, {Neri}, {Cox}, {Cooper},
  {Shapiro}, {Bolatto}, {Bouch{\'e}}, {Bournaud}, {Burkert}, {Combes},
  {Comerford}, {Davis}, {F{\"o}rster Schreiber}, {Garcia-Burillo},
  {Gracia-Carpio}, {Lutz}, {Naab}, {Omont}, {Shapley}, {Sternberg}, \&
  {Weiner}}]{tacconi10}
{Tacconi}, L.~J., {Genzel}, R., {Neri}, R., {et~al.} 2010, \nat, 463, 781

\bibitem[{{Tacconi} {et~al.}(2020){Tacconi}, {Genzel}, \&
  {Sternberg}}]{tacconi20}
{Tacconi}, L.~J., {Genzel}, R., \& {Sternberg}, A. 2020, \araa, 58, 157

\bibitem[{{Tacconi} {et~al.}(2013){Tacconi}, {Neri}, {Genzel}, {Combes},
  {Bolatto}, {Cooper}, {Wuyts}, {Bournaud}, {Burkert}, {Comerford}, {Cox},
  {Davis}, {F{\"o}rster Schreiber}, {Garc{\'\i}a-Burillo}, {Gracia-Carpio},
  {Lutz}, {Naab}, {Newman}, {Omont}, {Saintonge}, {Shapiro Griffin}, {Shapley},
  {Sternberg}, \& {Weiner}}]{tacconi13}
{Tacconi}, L.~J., {Neri}, R., {Genzel}, R., {et~al.} 2013, \apj, 768, 74

\bibitem[{{Tadaki} {et~al.}(2020){Tadaki}, {Belli}, {Burkert}, {Dekel},
  {F{\"o}rster Schreiber}, {Genzel}, {Hayashi}, {Herrera-Camus}, {Kodama},
  {Kohno}, {Koyama}, {Lee}, {Lutz}, {Mowla}, {Nelson}, {Renzini}, {Suzuki},
  {Tacconi}, {{\"U}bler}, {Wisnioski}, \& {Wuyts}}]{tadaki20b}
{Tadaki}, K.-i., {Belli}, S., {Burkert}, A., {et~al.} 2020, \apj, 901, 74

\bibitem[{{Tadaki} {et~al.}(2023){Tadaki}, {Kodama}, {Koyama}, {Suzuki},
  {Mitsuhashi}, \& {Ikeda}}]{tadaki23}
{Tadaki}, K.-i., {Kodama}, T., {Koyama}, Y., {et~al.} 2023, \apjl, 957, L15

\bibitem[{{Tamburro} {et~al.}(2009){Tamburro}, {Rix}, {Leroy}, {Mac Low},
  {Walter}, {Kennicutt}, {Brinks}, \& {de Blok}}]{tamburro09}
{Tamburro}, D., {Rix}, H.~W., {Leroy}, A.~K., {et~al.} 2009, \aj, 137, 4424

\bibitem[{{Terzi{\'c}} \& {Graham}(2005)}]{terzic05}
{Terzi{\'c}}, B. \& {Graham}, A.~W. 2005, \mnras, 362, 197

\bibitem[{{Tiley} {et~al.}(2016){Tiley}, {Stott}, {Swinbank}, {Bureau},
  {Harrison}, {Bower}, {Johnson}, {Bunker}, {Jarvis}, {Magdis}, {Sharples},
  {Smail}, {Sobral}, \& {Best}}]{tiley16}
{Tiley}, A.~L., {Stott}, J.~P., {Swinbank}, A.~M., {et~al.} 2016, \mnras, 460,
  103

\bibitem[{{Toft} {et~al.}(2014){Toft}, {Smol{\v{c}}i{\'c}}, {Magnelli},
  {Karim}, {Zirm}, {Michalowski}, {Capak}, {Sheth}, {Schawinski}, {Krogager},
  {Wuyts}, {Sanders}, {Man}, {Lutz}, {Staguhn}, {Berta}, {Mccracken}, {Krpan},
  \& {Riechers}}]{toft14}
{Toft}, S., {Smol{\v{c}}i{\'c}}, V., {Magnelli}, B., {et~al.} 2014, \apj, 782,
  68

\bibitem[{{Toomre}(1964)}]{toomre64}
{Toomre}, A. 1964, \apj, 139, 1217

\bibitem[{{Torrey} {et~al.}(2014){Torrey}, {Vogelsberger}, {Genel}, {Sijacki},
  {Springel}, \& {Hernquist}}]{torrey14}
{Torrey}, P., {Vogelsberger}, M., {Genel}, S., {et~al.} 2014, \mnras, 438, 1985

\bibitem[{{Tripodi} {et~al.}(2023){Tripodi}, {Lelli}, {Feruglio}, {Fiore},
  {Fontanot}, {Bischetti}, \& {Maiolino}}]{tripodi23}
{Tripodi}, R., {Lelli}, F., {Feruglio}, C., {et~al.} 2023, \aap, 671, A44

\bibitem[{{Tsukui} \& {Iguchi}(2021)}]{tsukui21}
{Tsukui}, T. \& {Iguchi}, S. 2021, Science, 372, 1201

\bibitem[{{Tsukui} {et~al.}(2024){Tsukui}, {Wisnioski}, {Bland-Hawthorn},
  {Mai}, {Iguchi}, {Baba}, \& {Freeman}}]{tsukui24}
{Tsukui}, T., {Wisnioski}, E., {Bland-Hawthorn}, J., {et~al.} 2024, \mnras,
  527, 8941

\bibitem[{{Tsukui} {et~al.}(2023){Tsukui}, {Wisnioski}, {Krumholz}, \&
  {Battisti}}]{tsukui23}
{Tsukui}, T., {Wisnioski}, E., {Krumholz}, M.~R., \& {Battisti}, A. 2023,
  \mnras, 523, 4654

\bibitem[{{Tully} \& {Fisher}(1977)}]{tullyfisher77}
{Tully}, R.~B. \& {Fisher}, J.~R. 1977, \aap, 54, 661

\bibitem[{{{\"U}bler} {et~al.}(2017){{\"U}bler}, {F{\"o}rster Schreiber},
  {Genzel}, {Wisnioski}, {Wuyts}, {Lang}, {Naab}, {Burkert}, {van Dokkum},
  {Tacconi}, {Wilman}, {Fossati}, {Mendel}, {Beifiori}, {Belli}, {Bender},
  {Brammer}, {Chan}, {Davies}, {Fabricius}, {Galametz}, {Lutz}, {Momcheva},
  {Nelson}, {Saglia}, {Seitz}, \& {Tadaki}}]{uebler17}
{{\"U}bler}, H., {F{\"o}rster Schreiber}, N.~M., {Genzel}, R., {et~al.} 2017,
  \apj, 842, 121

\bibitem[{{Valentino} {et~al.}(2023){Valentino}, {Brammer}, {Gould}, {Kokorev},
  {Fujimoto}, {Jespersen}, {Vijayan}, {Weaver}, {Ito}, {Tanaka}, {Ilbert},
  {Magdis}, {Whitaker}, {Faisst}, {Gallazzi}, {Gillman}, {Gim{\'e}nez-Arteaga},
  {G{\'o}mez-Guijarro}, {Kubo}, {Heintz}, {Hirschmann}, {Oesch}, {Onodera},
  {Rizzo}, {Lee}, {Strait}, \& {Toft}}]{valentino23}
{Valentino}, F., {Brammer}, G., {Gould}, K. M.~L., {et~al.} 2023, \apj, 947, 20

\bibitem[{{van Albada} {et~al.}(1985){van Albada}, {Bahcall}, {Begeman}, \&
  {Sancisi}}]{vanalbada85}
{van Albada}, T.~S., {Bahcall}, J.~N., {Begeman}, K., \& {Sancisi}, R. 1985,
  \apj, 295, 305

\bibitem[{{Vogelsberger} {et~al.}(2020){Vogelsberger}, {Marinacci}, {Torrey},
  \& {Puchwein}}]{vogelsberger20}
{Vogelsberger}, M., {Marinacci}, F., {Torrey}, P., \& {Puchwein}, E. 2020,
  Nature Reviews Physics, 2, 42

\bibitem[{{Wagg} {et~al.}(2014){Wagg}, {Carilli}, {Aravena}, {Cox}, {Lentati},
  {Maiolino}, {McMahon}, {Riechers}, {Walter}, {Andreani}, {Hills}, \&
  {Wolfe}}]{wagg14}
{Wagg}, J., {Carilli}, C.~L., {Aravena}, M., {et~al.} 2014, \apj, 783, 71

\bibitem[{{Wu} {et~al.}(2023){Wu}, {Cai}, {Sun}, {Bian}, {Lin}, {Li}, {Li},
  {Bauer}, {Egami}, {Fan}, {Gonz{\'a}lez-L{\'o}pez}, {Li}, {Wang}, {Yang},
  {Zhang}, \& {Zou}}]{wu23}
{Wu}, Y., {Cai}, Z., {Sun}, F., {et~al.} 2023, \apjl, 942, L1

\bibitem[{{Yang} {et~al.}(2017){Yang}, {Omont}, {Beelen}, {Gao}, {van der
  Werf}, {Gavazzi}, {Zhang}, {Ivison}, {Lehnert}, {Liu}, {Oteo},
  {Gonz{\'a}lez-Alfonso}, {Dannerbauer}, {Cox}, {Krips}, {Neri}, {Riechers},
  {Baker}, {Micha{\l}owski}, {Cooray}, \& {Smail}}]{yang17}
{Yang}, C., {Omont}, A., {Beelen}, A., {et~al.} 2017, \aap, 608, A144

\bibitem[{{Yu} {et~al.}(2021){Yu}, {Bian}, {Krumholz}, {Shi}, {Li}, \&
  {Chen}}]{yu21}
{Yu}, X., {Bian}, F., {Krumholz}, M.~R., {et~al.} 2021, \mnras, 505, 5075

\end{thebibliography}

\begin{appendix}

\section{Asymmetric Drift Correction}\label{ap:adc}

\begin{figure}
    \centering
    \includegraphics[width=0.46\textwidth]{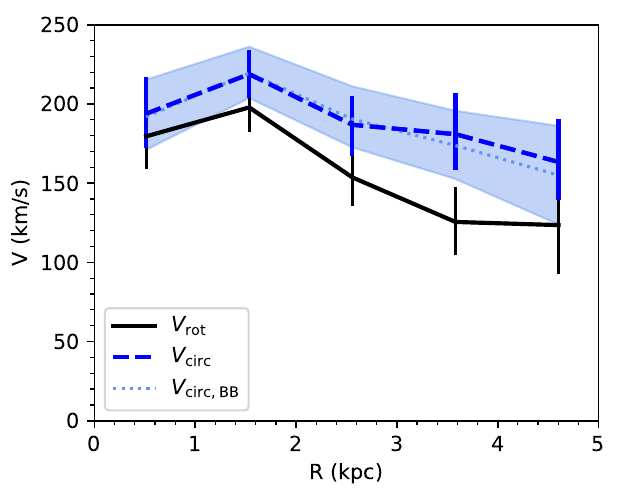}
    \caption{Comparison between the original rotation velocity obtained from the kinematic modelling ($V_{\rm rot}$) and the circular speed after the asymmetric drift correction was applied with the method described in this Appendix ($V_{\rm circ}$) and the one obtained with \texttt{$^{3D}$BAROLO} ($V_{\rm circ, BB}$).}
    \label{fig:adc}
\end{figure}

As discussed in Section~\ref{sec:dynmod}, to obtain the circular speed from the rotation velocity it is first necessary to correct the rotation curve for the gas pressure support by applying the asymmetric drift correction.
Although \texttt{$^{3D}$BAROLO} can apply this correction during the kinematic fitting, it does so without convolving the [CII] surface brightness with the beam. To properly account for the spatial resolution, we assume that the gas is distributed as an exponential disc and use the best-fit disc scale length reported in Table~\ref{tab:surfbrightness}. Therefore, we can rewrite the asymmetric drift correction from Equation~\ref{eq:adc} as
\begin{equation}
V_{\mathrm{A}}^2 = - R \sigma^2 \frac{\partial}{\partial R} \left[\ln{(\sigma^2  \exp{(-R/R_{\mathrm{gas}}}})) \right],
\end{equation}
and consequently,
\begin{equation}
V_{\mathrm{A}}^2 = - 2 R \sigma^2 \frac{\partial \ln \sigma}{\partial R}  + \frac{R \sigma^2}{R_{\mathrm{gas}}}.
\end{equation}
To apply this correction to the rotation velocity, we numerically calculate the slope of $\ln{\sigma}$ at each radial element. In Figure~\ref{fig:adc} we show the result of the asymmetric drift correction for BRI1335-0417, which, in spite of the difference in methodology, agrees with the circular speed derived with \texttt{$^{3D}$BAROLO}.

\section{Surface Brightness Models Best-Fit Posteriors}\label{ap:corner_sb}
Here we show the posterior distributions of the best-fit surface brightness models for the gas and dust emission obtained as described in Section~\ref{sec:met_sb}. We fit the surface brightness at the effective radius ($I_{\mathrm{e}}$), the effective radius ($R_{\mathrm{eff}}$), the centre (x0, y0), ellipticity ($\epsilon$) and position angle (PA). For the dust emission of BRI1335-0417 we let the S\'ersic index ($n$) free to vary as it is significantly different from 1.

\begin{figure*}
    \centering
    \includegraphics[width=\textwidth]{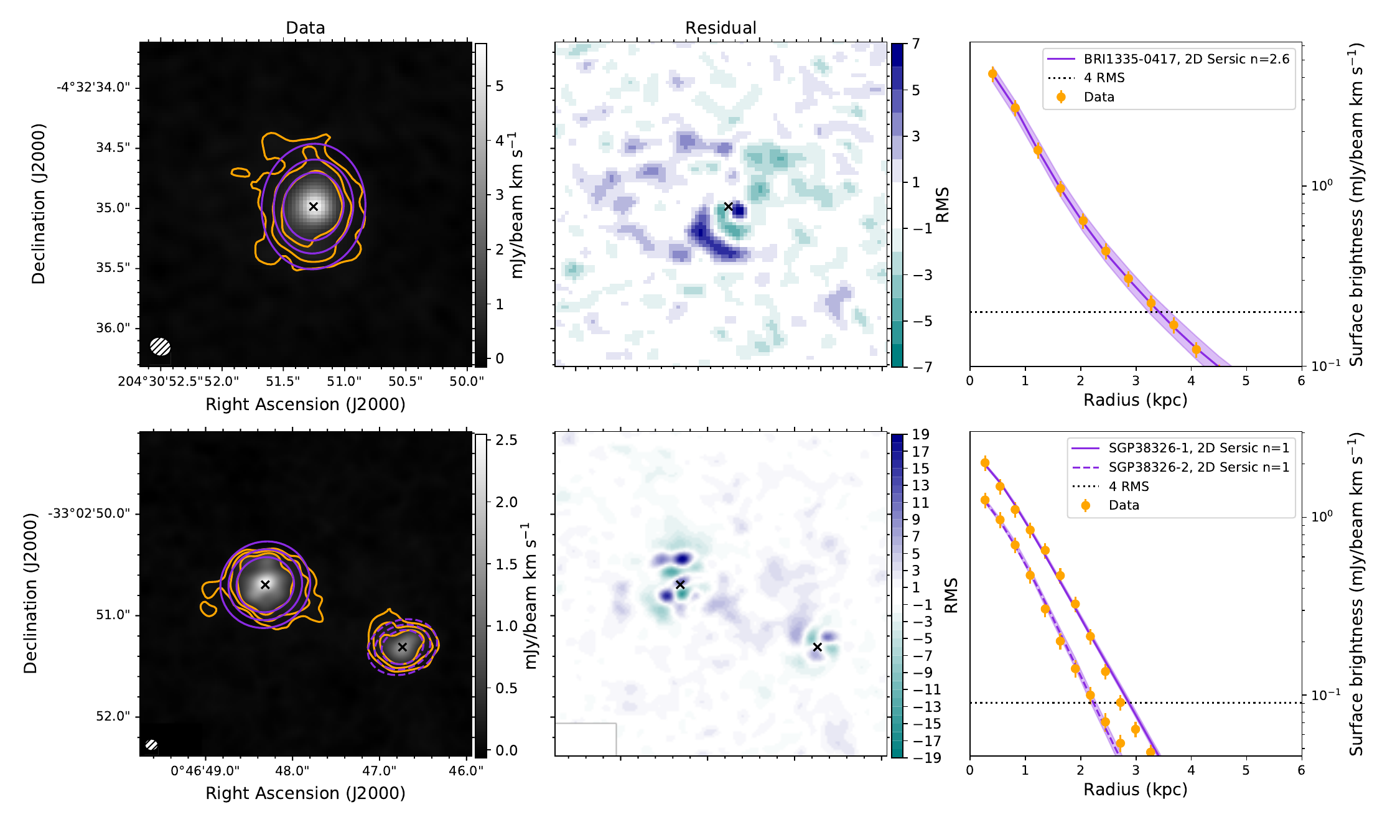}
    \caption{Surface brightness profiles of the dust distribution. Details are as described in the caption of Figure~\ref{fig:sb_gas}.}
    \label{fig:a_dustsb}
\end{figure*}

\begin{figure*}
\begin{subfigure}{0.5\textwidth}
  \centering
    \includegraphics[width=\textwidth]{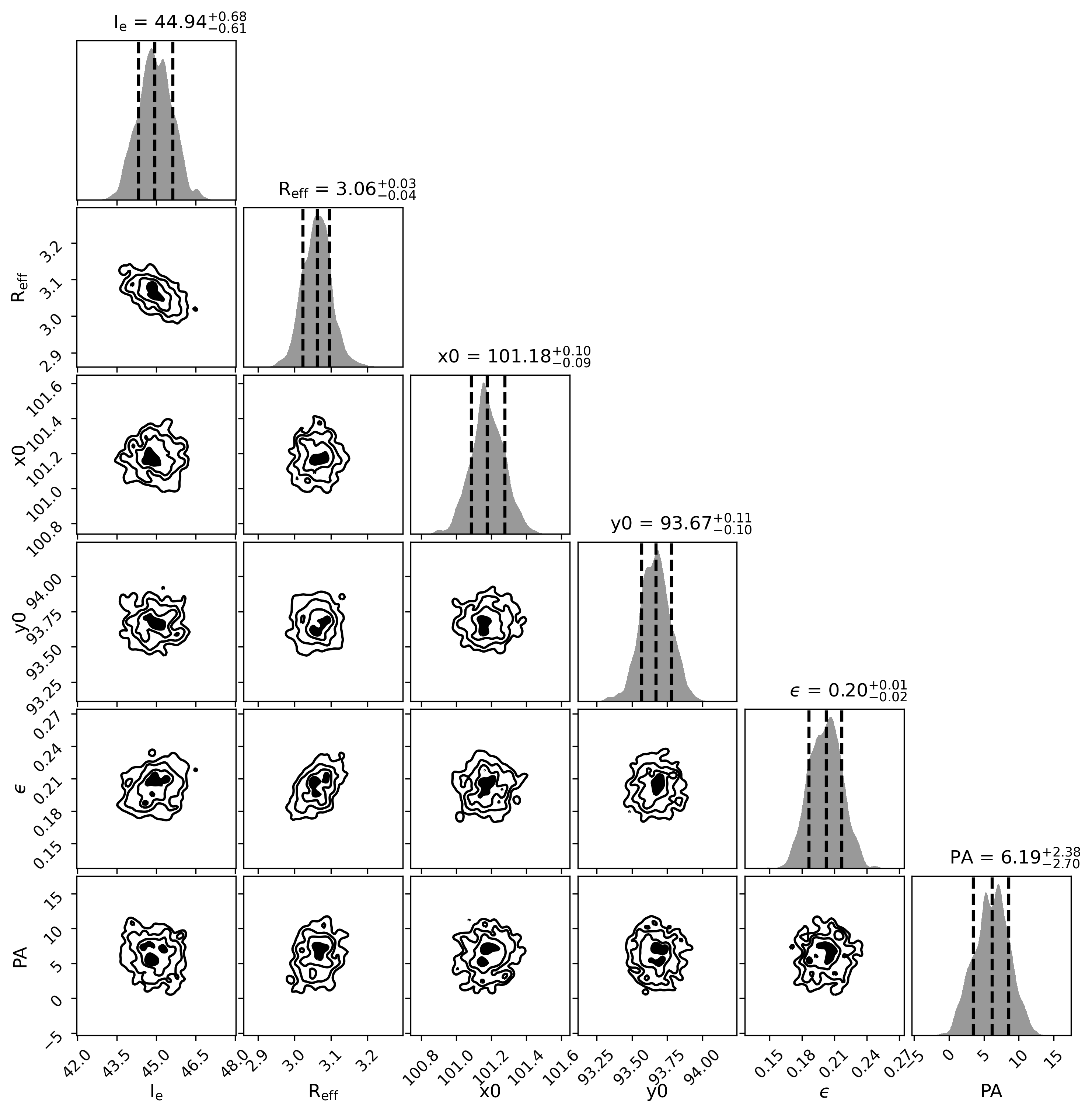}
\end{subfigure}\hfill
\begin{subfigure}{0.5\textwidth}
  \centering
    \includegraphics[width=\textwidth]{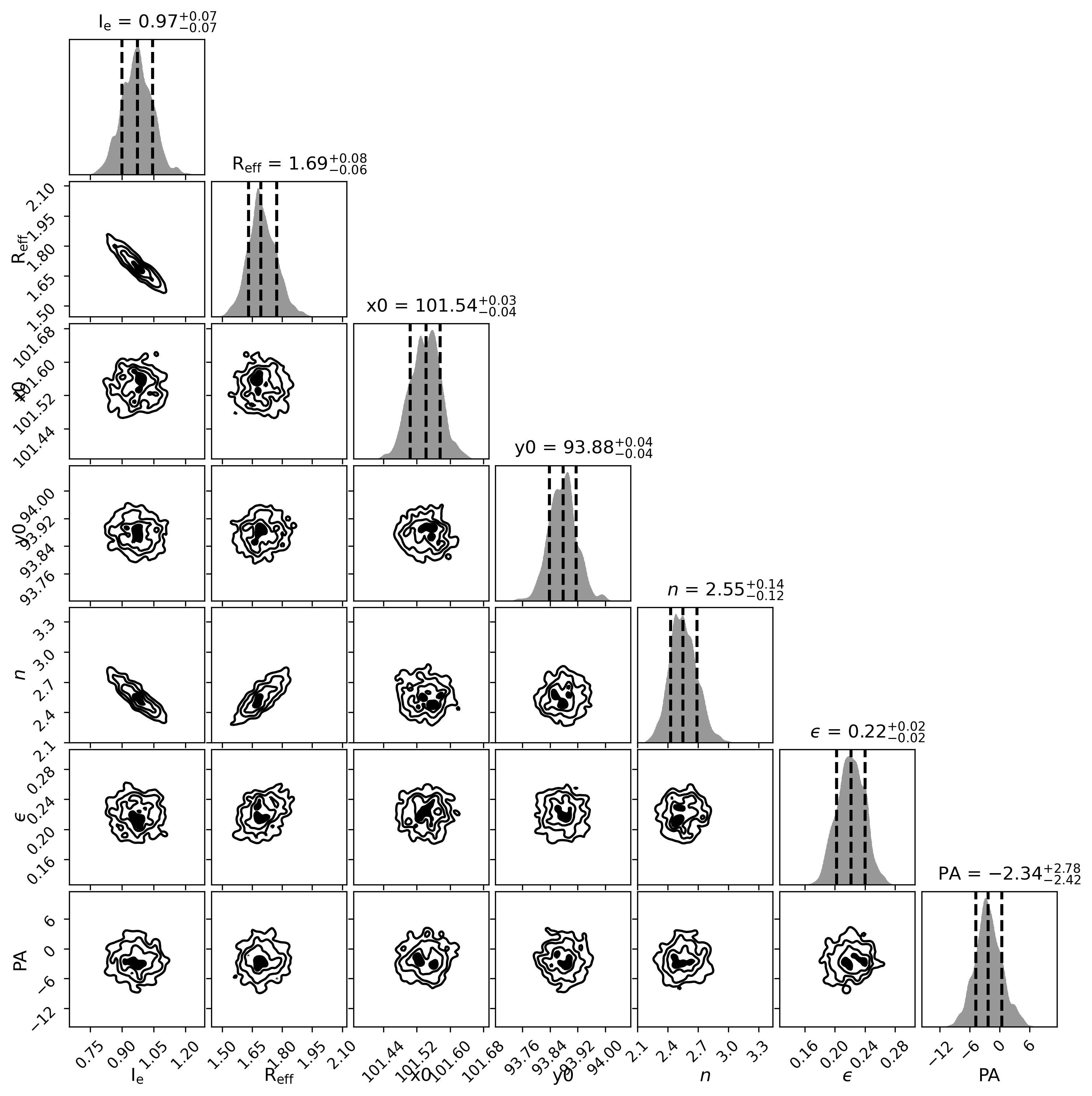}
\end{subfigure}

\begin{subfigure}{0.5\textwidth}
  \centering
    \includegraphics[width=\textwidth]{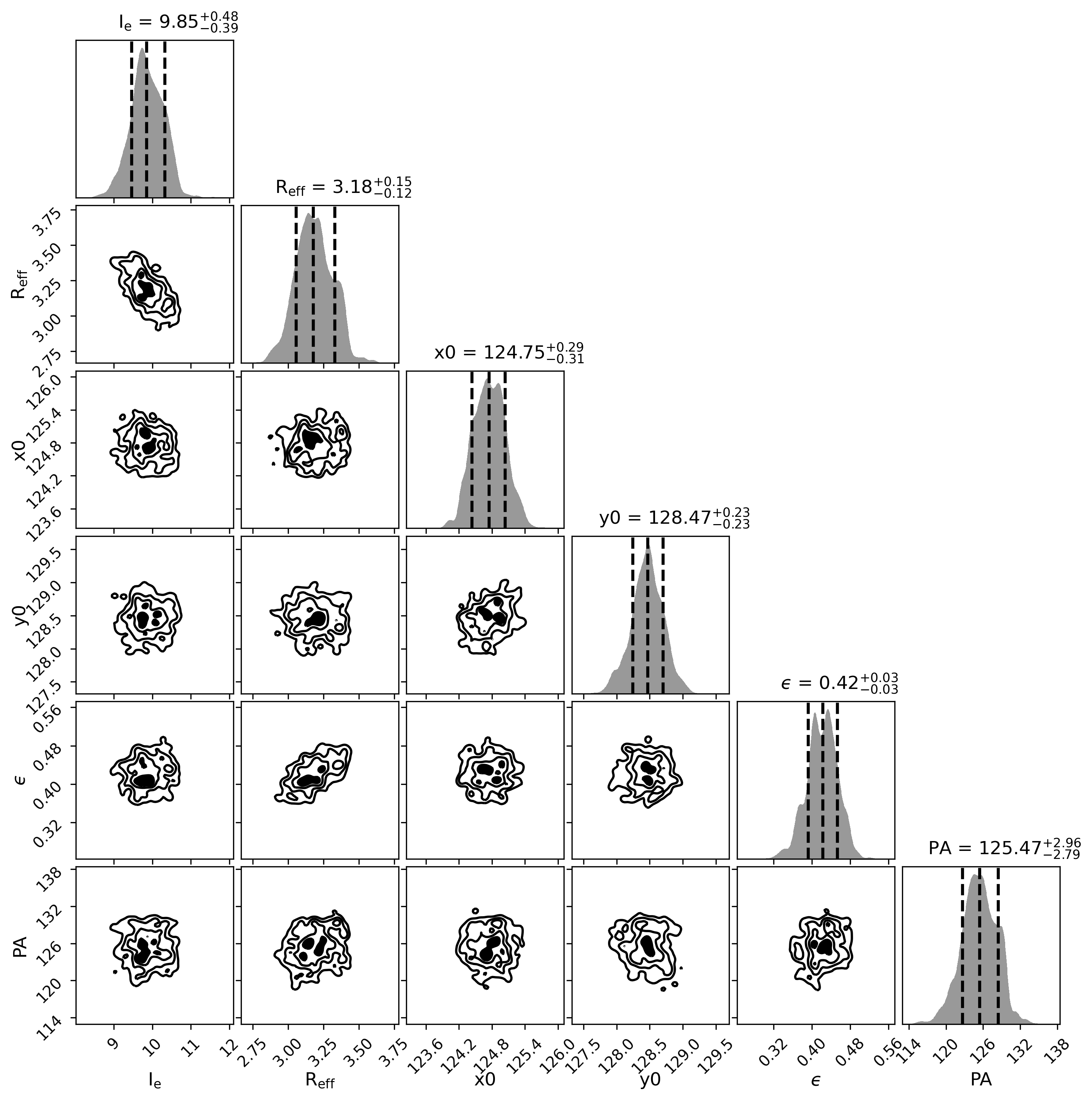}
\end{subfigure}\hfill
\begin{subfigure}{0.5\textwidth}
  \centering
    \includegraphics[width=\textwidth]{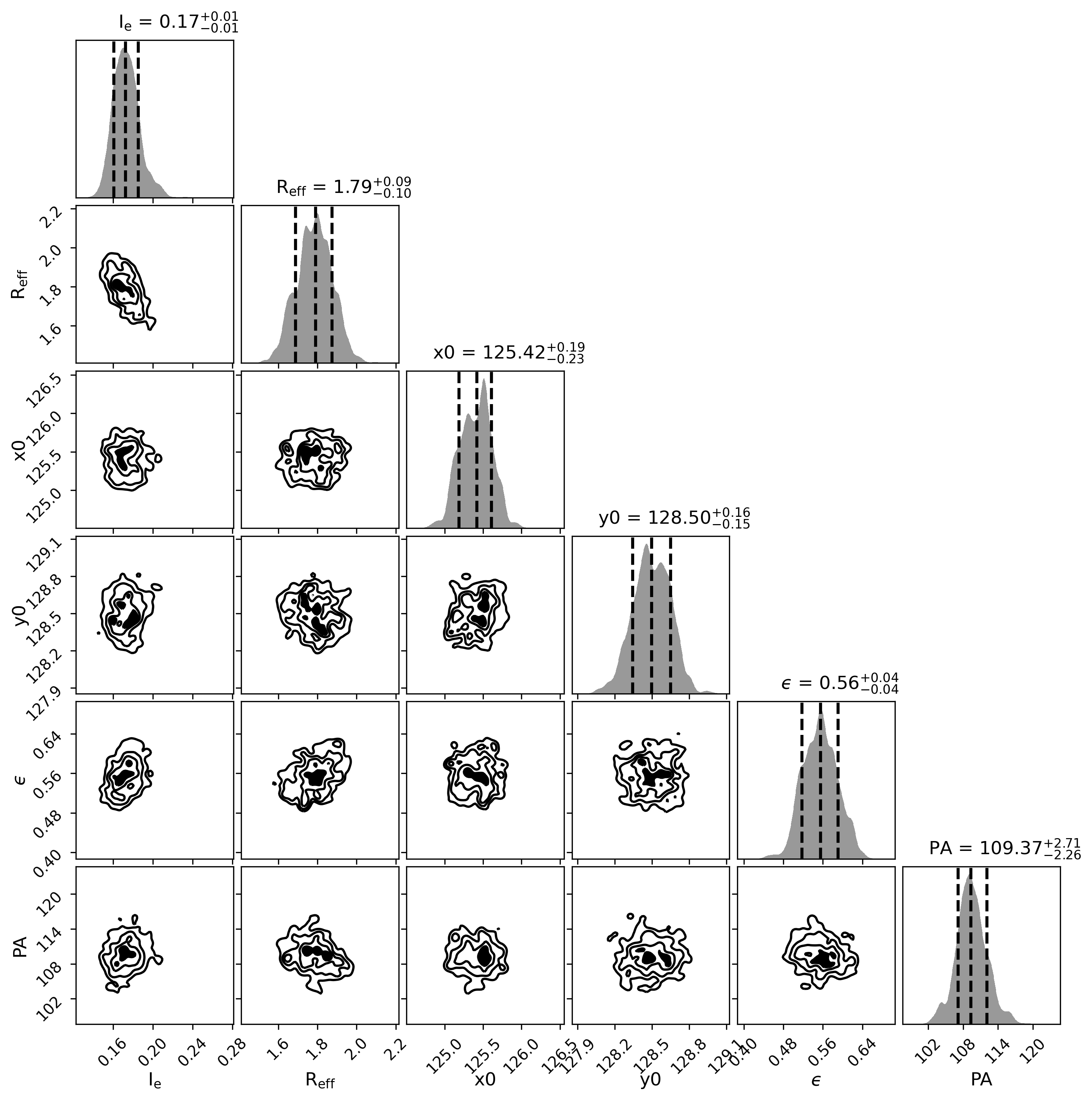}
\end{subfigure}
\caption{Posterior distributions of the surface brightness fits of the gas and dust distribution of the galaxies. Top left panel: [CII] emission of BRI1335-0417. Top right panel: Dust continuum emission of BRI1335-0417. Bottom left panel: [CII] emission of J081740. Bottom right panel: Dust continuum emission of J081740. All fits displayed have the S\'ersic index fixed to 1.0, except for the dust distribution of BRI1335-0417 in which a free S\'ersic index is preferred. We display the surface brightness at the effective radius in mJy/beam for the dust and mJy/beam km/s for the gas ($I_{\mathrm{e}}$), the effective radius in kpc ($R_{\mathrm{eff}}$), the galactic centre in pixels (x0, y0), the ellipticity ($\epsilon$), and the position angle in degrees (PA). The best values shown are the median followed by the lower and upper errors defined by the 16th and 84th percentiles, respectively.}
\label{fig:sb_bri_j}
\end{figure*}

\begin{figure*}
  \centering
  \includegraphics[width=\textwidth]{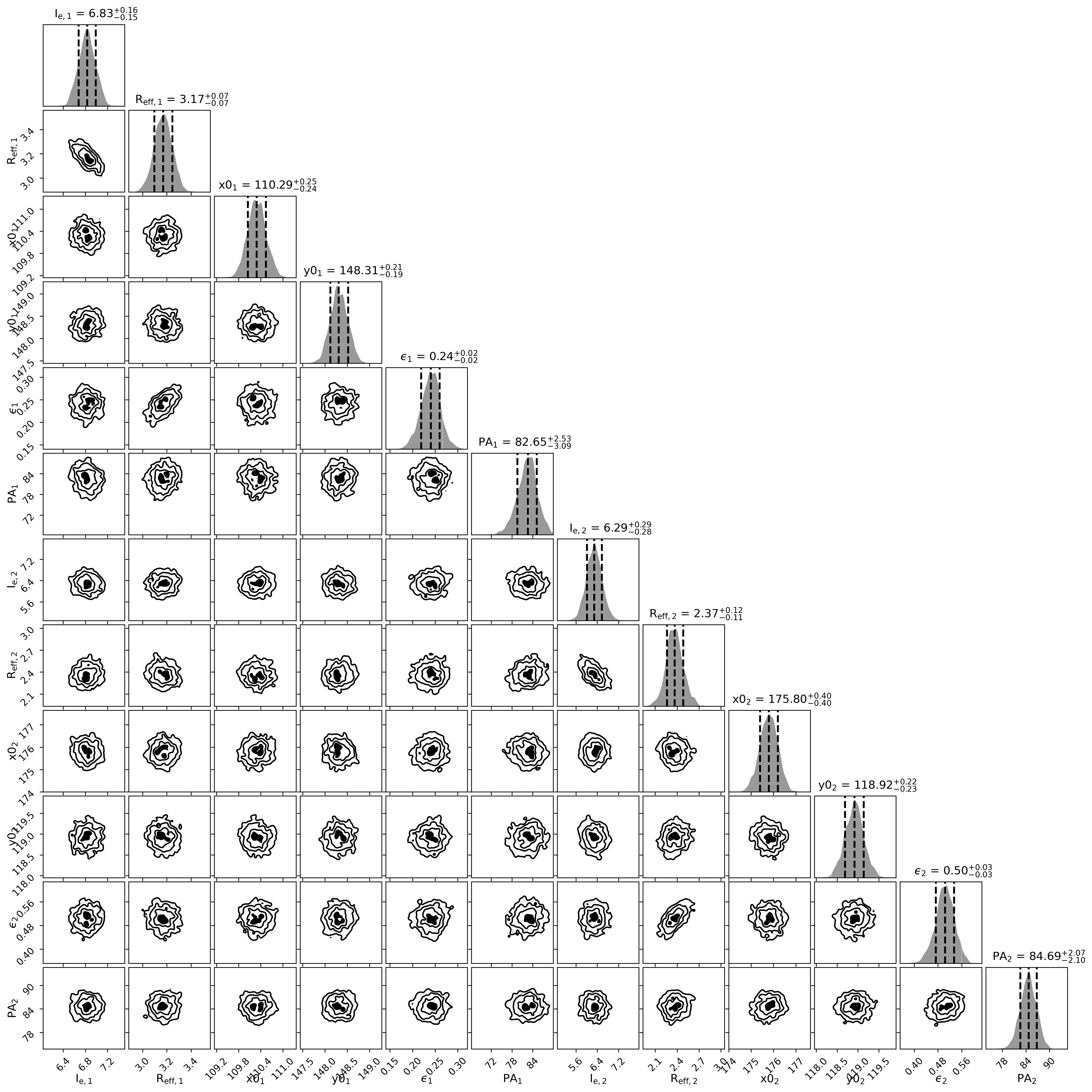}
\caption{Posterior distributions of the surface brightness fits of the [CII] emission of the SGP38326 system, we fit the galaxies SGP38326-1 and SGP38326-2 simultaneously. We display the surface brightness at the effective radius in mJy/beam for the dust and mJy/beam km/s for the gas ($I_{\mathrm{e}}$), the effective radius in kpc ($R_{\mathrm{eff}}$), the galactic centre in pixels (x0, y0), the ellipticity ($\epsilon$), and the position angle in degrees (PA). The best values shown are the median followed by the lower and upper errors defined by the 16th and 84th percentiles, respectively.}
\label{fig:sb_sgp_cii}
\end{figure*}

\begin{figure*}
  \centering
  \includegraphics[width=\textwidth]{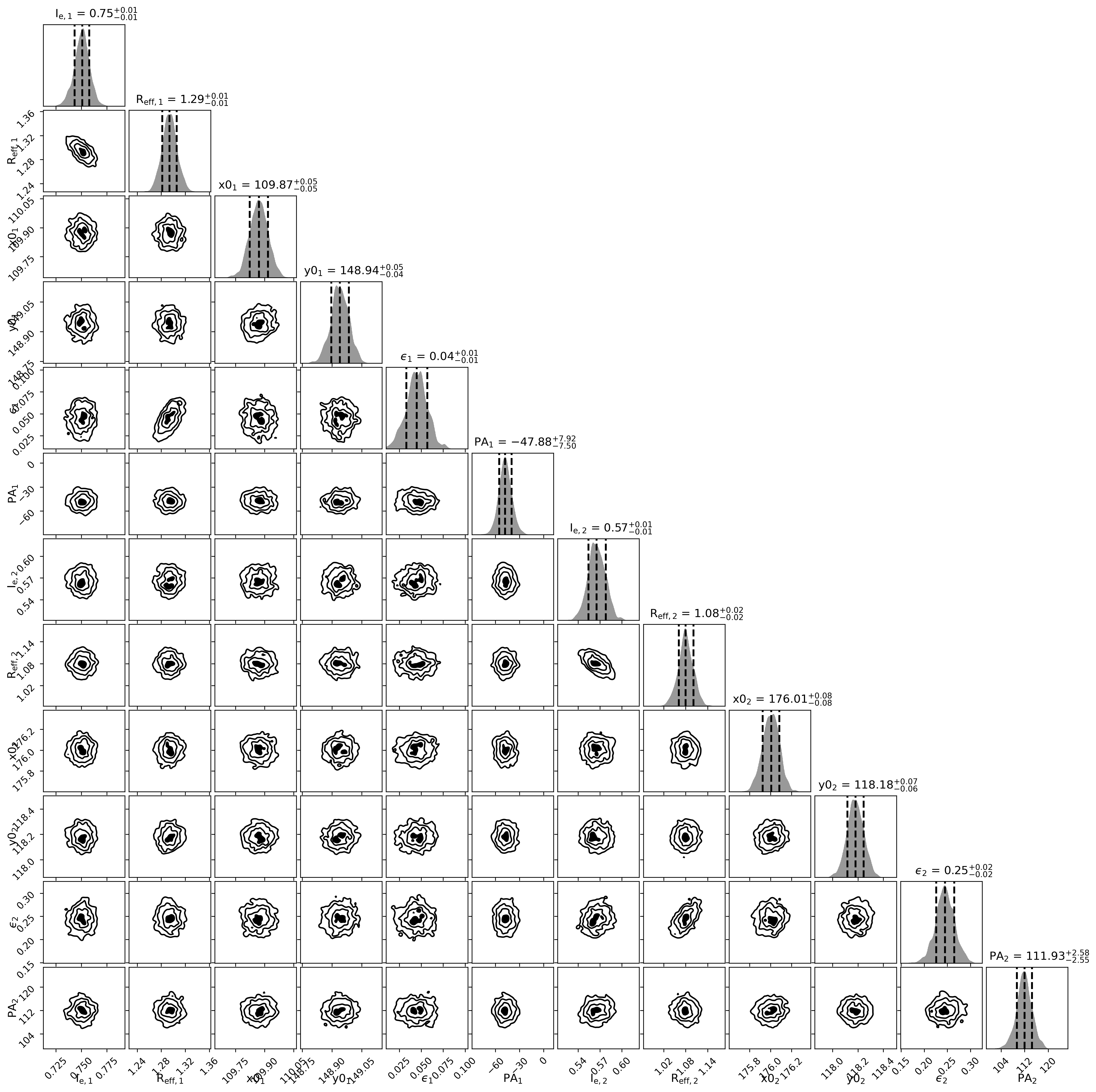}
  \caption{Posterior distributions of the surface brightness fits of the dust continuum emission of the SGP38326 system, we fit the galaxies SGP38326-1 and SGP38326-2 simultaneously. We display the surface brightness at the effective radius in mJy/beam for the dust and mJy/beam km/s for the gas ($I_{\mathrm{e}}$), the effective radius in kpc ($R_{\mathrm{eff}}$), the galactic centre in pixels (x0, y0), the ellipticity ($\epsilon$), and the position angle in degrees (PA). The best values shown are the median followed by the lower and upper errors defined by the 16th and 84th percentiles, respectively.}
  \label{fig:sb_sgp_dust}
\end{figure*}

\section{Dynamical Models}\label{ap:dynmod}

\subsection{Dynamical Models Best-fit Posteriors}
Here we show the posterior distributions of the best-fit dynamical models obtained as described in Section~\ref{sec:dynmod}. Our model contains five free parameters: stellar mass ($M_*$), stellar effective radius ($R_{\mathrm{eff,*}}$), stellar S\'ersic index ($n$), gas mass normalisation (gas\_norm) and baryon fraction ($f_{\mathrm{bar}}$), which allows us to derive the virial mass of the CDM halo. We note that for BRI1335-0417, we fix the baryon fraction to the cosmological baryon fraction, therefore we only have four free parameters. The best-fit parameters reproduce the observed rotation curves as shown in Figure~\ref{fig:dy_rotcur}.

\begin{figure*}

\begin{subfigure}{0.5\textwidth}
  \centering
    \includegraphics[width=\textwidth]{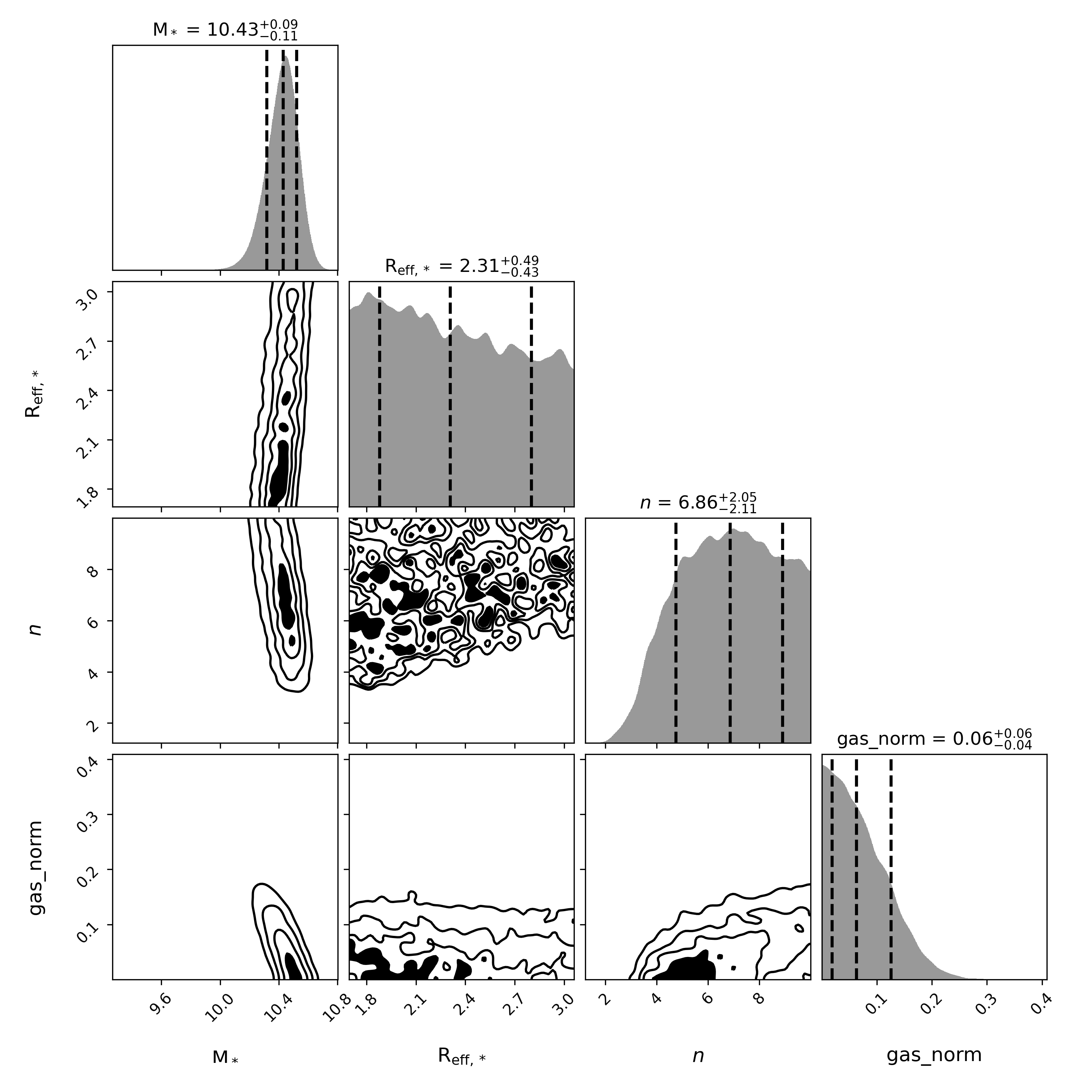}
  \caption{}
\end{subfigure}\hfill
\begin{subfigure}{0.5\textwidth}
  \centering
    \includegraphics[width=\textwidth]{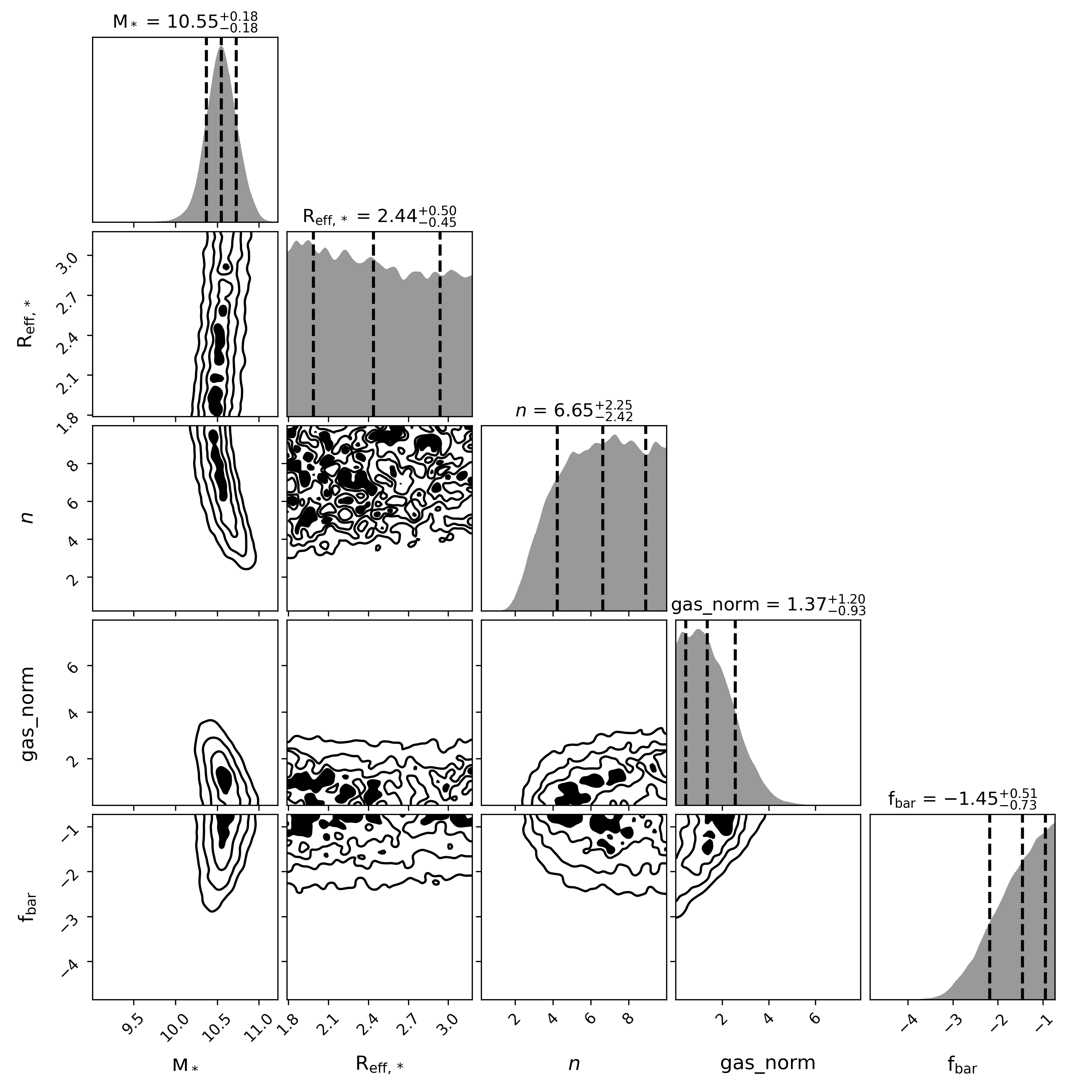}
  \caption{}
\end{subfigure}

\begin{subfigure}{0.5\textwidth}
  \centering
    \includegraphics[width=\textwidth]{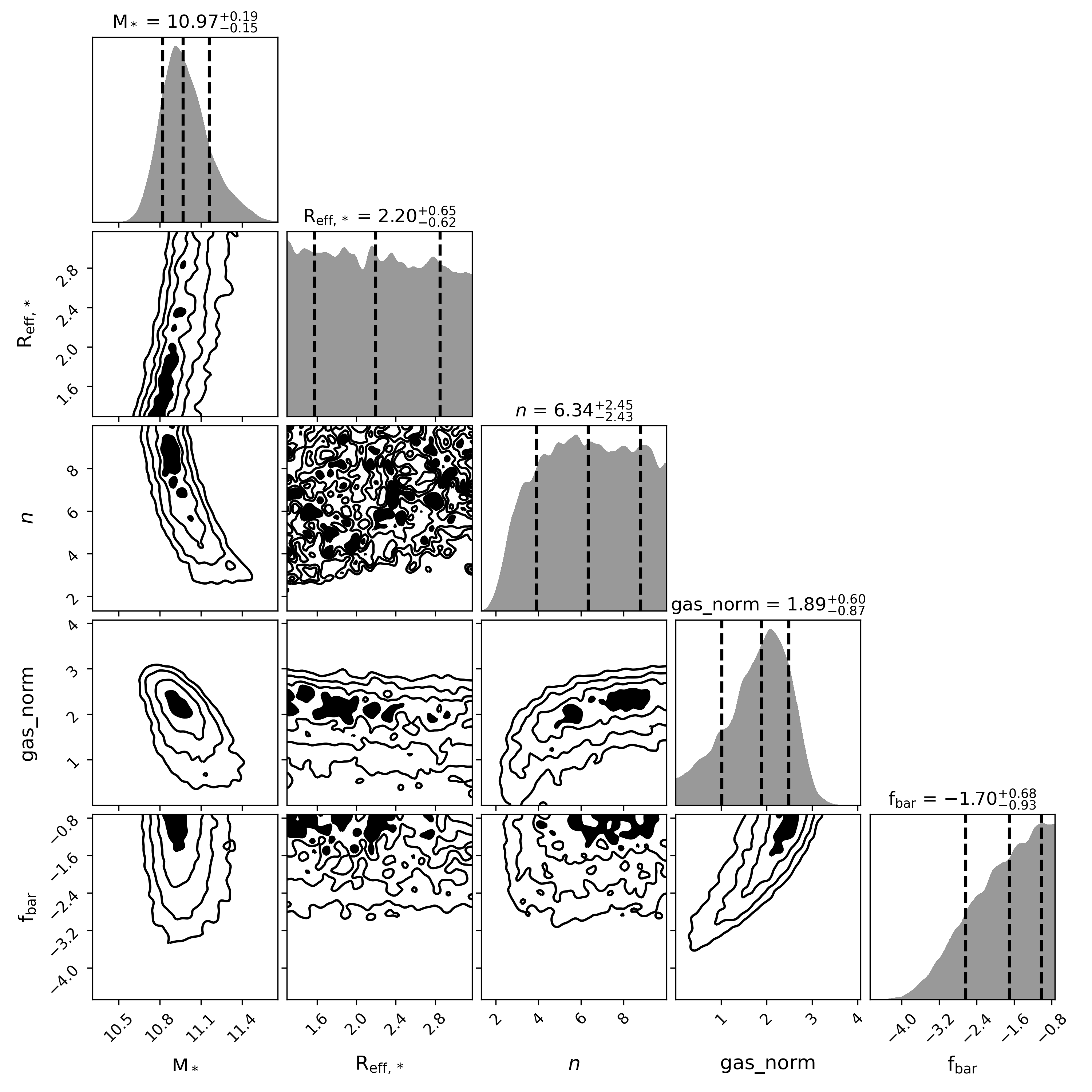}
  \caption{}
\end{subfigure}\hfill
\begin{subfigure}{0.5\textwidth}
  \centering
    \includegraphics[width=\textwidth]{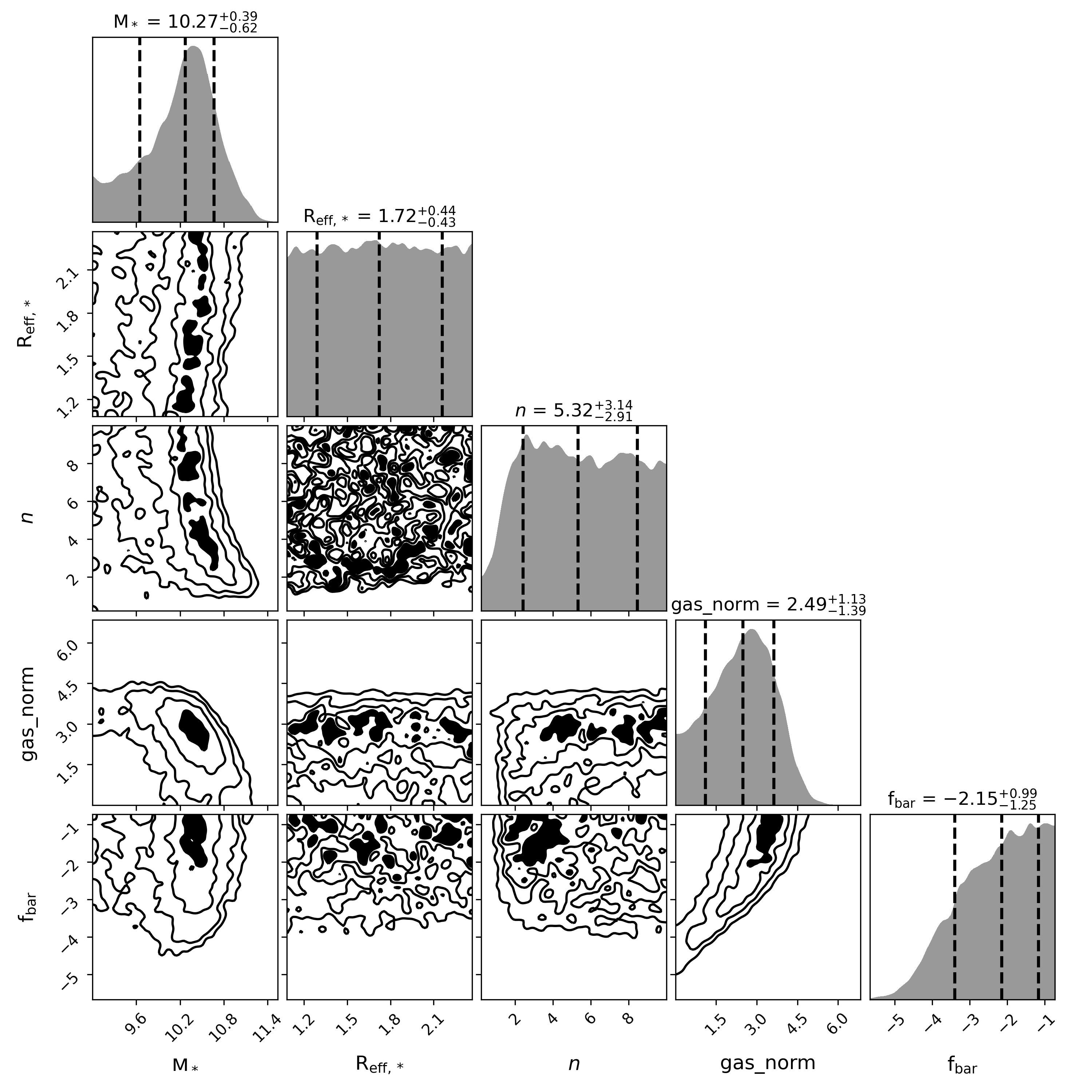}
  \caption{}
\end{subfigure}
\caption{Corner plot with the posteriors of the best-fit dynamical models for the galaxies. The columns show: (i) the stellar mass in solar masses (logspace of $M_{*}$); (ii) the effective radius of the stellar component in kpc ($R_{\mathrm{eff,*}}$); (iii) the S\'ersic index of the stellar component ($n$); (iv) the normalisation of the gas (gas\_norm); (v) the baryon fraction in logspace ($f_{\mathrm{bar}}$). Panels (a), (b), (c) and (d) are showing the results for the galaxies BRI1335-0417, J081740, SGP38326-1 and SGP38326-2, respectively. We note that for BRI1335-0417 we do not show the posterior for the baryon fraction as we fix it to the cosmological baryon fraction as discussed in Section~\ref{sec:res_dy}. We show dashed lines for the best values followed by the upper and lower errors defined by the 50th, 16th and 84th percentiles, respectively.}
\label{fig:a_dy}
\end{figure*}

\subsection{Further tests}
We have also performed a few more tests with different assumptions. Specifically, we test oversampling the rotation curves, both by modelling the kinematics with \texttt{3DBarolo} with more rings and by interpolation the rotation curves, and we find that neither retrieve significantly different results. This implies that, to retrieve better constrains on the parameters considered, the main limiting factor in the model is the information in the shape of the rotation curves that can only be achieved with higher resolution data and/or with more extended rotation curves (i.e. higher SNR). 
Moreover, we considered models with the DM mass completely free and found that the baryon fraction becomes higher than the cosmological fraction, implying that the rotation curves of these galaxies can be reproduced without a CDM halo component, albeit likely overestimating the gas mass in order to reproduce the outer rotation.
As mentioned in the main text, while the inner part of the rotation curve is sensitive to the stellar component, in the outer parts gas and DM tend to be degenerate.

Moreover, for BRI1335-0417, as explained in Section~\ref{sec:res_dy}, we fixed the baryon fraction to the cosmological baryon fraction when modelling the rotation curve. The main drawback from undertaking the same assumptions used for the dynamical models of the other galaxies is that BRI1335-0417 has significantly lower circular speed with respect with its expected baryonic mass.
Considering the disc inclination of 42$^{\circ}$ as derived in \citet{romanoliveira23}, the rotation velocity of BRI1335-0417 goes from 180 km s$^{-1}$ at 0.5 kpc to 124 km s$^{-1}$ at 4.6 kpc. After performing the asymmetric drift correction, the circular speeds become 191 km s$^{-1}$ in the centre and 153 km s$^{-1}$ in the outskirts. If we use these circular speeds to roughly estimate the total mass at each radius then we find that the innermost ring encloses a mass of $4.2 \times 10^9 M_\odot$ and the outermost ring encloses a mass of $2.5 \times 10^{10} M_\odot$.

If we take the CO luminosity observed for BRI1335-0417, presented in Table~\ref{tab:colum}, and consider a conservative case with the lowest CO luminosity reported in \citet{wagg14}, an $\alpha_{\mathrm{CO}} = 0.8$ and a line ratio (J=2-1/J=1-0) of r$_{\mathrm{l}} = 0.5$ \citep{boogaard20, yang17, kamenetzky16, dacunha13}, BRI1335-0417 should have a gas mass of $2.92 \pm 0.24 \times 10^{10} M_\odot$.
Just a simple comparison between the expected gas mass and the outer circular speed would imply that there is no room for neither the stellar component nor the CDM halo. Moreover, such a massive gas component could explain the outer rotation, however given the shape of the gas exponential disc it could never reproduce the inner circular speeds.
Therefore, for our fiducial model we considered the extreme case with the highest baryon fraction possible to derive the gas, stars and dark matter masses.

To explore whether we could potentially solve the discrepancy between the dynamical masses and the CO and stellar masses, in this Section we attempt at fitting the rotation curves with the disc inclination free and/or the NFW CDM halo concentration free. To include the inclination as a free parameter in our dynamical model we relate the circular speed $V_c$ to the disc inclination by
\begin{equation}
 V_c = \frac{V_{\mathrm{los}}}{\sin{(i)}},
\end{equation}
where, $V_{\mathrm{los}}$ is the line-of-sight rotation velocity and $i$ is the disc inclination. Therefore, if we were to consider a more face-on inclination of $i = 30^{\circ}$, the rotation of at the innermost radius would increase to 256 km s$^{-1}$ and the total mass enclosed within this radius would increase to $7.6 \times 10^{9} M_\odot$.

We perform a dynamical model keeping the inclination free to vary uniformly in the range of $20^{\circ} < i < 60^{\circ}$ while also keeping the baryon fraction and the gas normalisation free to vary in the same range as used for the other galaxies.
We find in this case that, despite potentially being able to increase the enclosed mass in the rotation curve to be modelled by decreasing the disc inclination, our dynamical model actually prefers higher inclinations ($i = 50^{\circ}$). Therefore, the resulting model is somewhat similar to the fiducial model presented in the body of this paper, with a stellar mass of $1.7 \times 10^{10} M_{\odot}$, 47\% lower than the one obtained with our fiducial model.
The best-fit baryon fraction is $f_{\mathrm{bar}} = 0.12$ and gas normalisation of gas\_norm = 0.04, however, both of these values concentrate at the upper and lower edge of the prior range, respectively.
This likely happens due to the flatness of the rotation curve of BRI1335-0417, since the innermost velocity is almost as high as the outskirt velocities any increase in the inner velocity translates to an even higher increase in the stellar mass. Ultimately, the main constrain is that the baryon fraction cannot exceed the cosmological baryon fraction.

Alternatively, we also test a dynamical model with a free DM concentration. In this case we keep the inclination fixed to 42 degrees, we use all the same priors for the remaining parameters and we let $c_{200}$ vary uniformly between 1 and 10. In this case, we find a stellar mass of $2.4 \times 10^{10} M_\odot$, a concentration $c_{200}$ of 2.6 while the problems with the gas normalisation and baryon fraction remain.

\end{appendix}

\end{document}